\documentclass[10pt,a4paper]{article}
\usepackage[numbers, square, comma, sort&compress]{natbib}
\usepackage{ifpdf}
\usepackage[latin1]{inputenc}
\usepackage{amsmath,amscd}
\usepackage{amssymb}
\usepackage{graphicx}
\usepackage[bookmarks = false, pdfstartview = FitV, colorlinks = false, linkcolor=green, urlcolor = blue]{hyperref} %LINKS IN PDF

\usepackage[small,sc]{caption}
\numberwithin{equation}{section}
\numberwithin{figure}{section}
\DeclareSymbolFont{symbolsC}{U}{txsyc}{m}{n}
\DeclareMathSymbol{\lambdabar}{\mathord}{symbolsC}{111}

\newcommand{\half}{\frac12}
\newcommand{\eps}{\varepsilon}
\newcommand{\ket}[1]{\left\vert{#1}\right\rangle}
\newcommand{\bra}[1]{\left\langle{#1}\right\vert}
\newcommand{\di}{\mathrm{d}}
\newcommand{\one}{\mbox{$1 \hspace{-1.0mm} {\bf l}$}}
\newcommand{\ex}[1]{\langle#1\rangle}
\newcommand{\bk}[2]{\langle#1|#2\rangle}
\newcommand{\m}[1]{\mathcal{#1}}
\newcommand{\ii}{\mathrm{i}}
\newcommand{\ee}{\mathrm{e}}

\begin{document}
\bibliographystyle{ieeetr}

\title{Localized qubits in curved spacetimes}
\author{\\\\
Matthew C. Palmer$^1$, Maki Takahashi$^1$, and Hans F. Westman$^{1,2,3}$\\
{\small \it $^{1}$School of Physics, The University of Sydney, Sydney, NSW 2006, Australia}\\%
{\small \it $^{2}$Centre for Time, The University of Sydney, Sydney, NSW 2006, Australia}\\%
{\small \it $^{3}$Perimeter Institute for Theoretical Physics, Waterloo, Ontario N2L 2Y5, Canada}\\[2mm]%
}

\date{{\small \today }}

\maketitle

%%%%%%%%%%%%%%%%%%%%%%%%%%%%%%%%%%%%%%%%%%%%%%%%%
\begin{abstract}
We provide a systematic and self-contained exposition of the subject of localized qubits in curved spacetimes. This research was motivated by a simple experimental question: if we move a spatially localized qubit, initially in a state $|\psi_1\rangle$, along some spacetime path $\Gamma$ from a spacetime point $x_1$ to another point $x_2$, what will the final quantum  state $|\psi_2\rangle$ be at point $x_2$? This paper addresses this question for two physical realizations of the qubit: spin of a massive fermion and polarization of a photon. Our starting point is the Dirac and Maxwell equations that describe respectively the one-particle states of localized massive fermions and photons. In the WKB limit we show how one can isolate a two-dimensional quantum state which evolves unitarily along $\Gamma$. The quantum states for these two realizations are represented by a left-handed 2-spinor in the case of massive fermions and a four-component complex polarization vector in the case of photons. In addition we show how to obtain from this WKB approach a fully general relativistic description of gravitationally induced phases. We use this formalism to describe the gravitational shift in the Colella--Overhauser--Werner 1975 experiment. In the non-relativistic weak field limit our result reduces to the standard formula in the original paper. We provide a concrete physical model for a Stern--Gerlach measurement of spin and obtain a unique spin operator which can be determined given the orientation and velocity of the Stern--Gerlach device and velocity of the massive fermion. Finally, we consider multipartite states and generalize the formalism to incorporate basic elements from quantum information theory such as quantum entanglement, quantum teleportation, and identical particles. The resulting formalism provides a basis for exploring precision quantum measurements of the gravitational field using techniques from quantum information theory.
\end{abstract}

%%%%%%%%%%%%%%%%%%%%%%%%%%%%%%%%%%%%%%%%%%%%%%%%%%%

\newpage
\tableofcontents

%%%%%%%%%%%%%%%%%%%%%%%%%%%%%%%%%%%%%%%%%%%%%%%%%%%

\newpage
\section*{Notation and conventions\label{notation}}
%%%%%%%%%%%%%%%%%%%%%%%%%%%%%%%%%%%%%%%%%%%%%%%%%%
We use the following index notation:
\begin{itemize}
\item[-] $\mu,\nu,\rho,\sigma,\ldots$ denote spacetime tensor indices
\item[-] $I,J,K,L,\ldots=0,1,2,3$ denote tetrad indices.
\item[-] $i,j,k,l,\ldots=1,2,3$ for spatial components of the tetrad (the `triad')
\item[-] $A,B,C,D,\ldots=1,2$ for spinor indices
\item[-] $A',B',C',D',\ldots=1,2$ for conjugate spinor indices
\end{itemize}
The Minkowski metric is defined as $\eta_{\mu\nu}=\mathrm{diag}(1,-1,-1,-1)$. We generally use natural units where $c=\hbar=1$, and in addition we set the charge of a proton to $e=1$.

We use the Weyl representation for the Dirac $\gamma$-matrices
\begin{eqnarray*}
\gamma^I=\left(\begin{array}{cc}0&\sigma^I\\ \bar{\sigma}^I&0\end{array}\right)
\end{eqnarray*}
where $\sigma^I=(1,\sigma^i)$ and $\bar{\sigma}^I=(1,-\sigma^i)$, and $\sigma^i$ are the usual Pauli matrices. Writing this object in spinor notation we have $\sigma^I=\sigma^I_{\ AA'}$ and $\bar{\sigma}^I=\bar{\sigma}^{IA'A}$. In order to interpret the spatial parts $\sigma^i_{\ AA'}$ and $\bar{\sigma}^{iA'A}$ as the Pauli matrices we use the convention in \cite{Bailin}: for $\bar{\sigma}^I=\bar{\sigma}^{IA'A}$ the primed index is the row index and the unprimed index is the column index, and the opposite assignment occurs for $\sigma^I=\sigma^I_{\ AA'}$. In spinorial notation $\sigma^I$ is not an operator, rather an operator $\hat{A}$ carries an index structure $A_{A}^{\ B}$ or $A_{\ B}^{A}$. Throughout this paper we will switch between the implicit index notation $\hat{A}$ and $A_{A}^{\ B}$ or $A_{\ B}^{A}$.

%\newpage

%%%%%%%%%%%%%%%%%%%%%%%%%%%%%%%%%%%%%%%%%%%%%%%%%%%%%%%%%%%%%%%%%
\section{Introduction}
%%%%%%%%%%%%%%%%%%%%%%%%%%%%%%%%%%%%%%%%%%%%%%%%%%%%%%%%%%%%%%%%%
This paper will provide a systematic and self-contained exposition of the subject of localized qubits in curved spacetimes with the focus on two physical realizations of the qubit: spin of a massive fermion and polarization of a photon. Although a great amount of research has been devoted to quantum field theory in curved spacetimes \cite{Birrell,WaldQFT,Mukhanov} and also more recently to relativistic quantum information theory in the presence of particle creation and the Unruh effect \cite{Alsing-DiracUnruh,Fuentes,FuentesBerry,Louko,Martinez}, the literature about localized qubits and quantum information theory in curved spacetimes is relatively sparse \cite{TerashimaUeda03,ASK,Pienaar11}. In particular, we are aware of only three papers, \cite{TerashimaUeda03,ASK,BDT11}, that deal with the following question: if we move a spatially localized qubit, initially in a state $|\psi_1\rangle$, along some spacetime path $\Gamma$ from a point $p_1$ in spacetime to another point $p_2$, what will the final quantum state $|\psi_2\rangle$ be at point $p_2$? This, and other relevant questions, were given as open problems in the field of relativistic quantum information by Peres and Terno in \cite[p.19]{PeresTerno04}. The formalism developed in this paper will be able to address such questions, and will also be able to deal with the basic elements of quantum information theory such as entanglement and multipartite states, teleportation, and quantum interference.

The basic object in quantum information theory is the qubit. Given a Hilbert space of some physical system, we can physically realize a qubit as any two-dimensional subspace of that Hilbert space. However, such physical realizations will in general not be localized in physical space. We shall restrict our attention to physical realizations that are well-localized in physical space so that we can approximately represent the qubit as a two-dimensional quantum state attached to a single point in space. From a spacetime perspective a localized qubit is then mathematically represented as a sequence of two-dimensional quantum states along some spacetime trajectory corresponding to the worldline of the qubit.

In order to ensure relativistic invariance it is then necessary to understand how this quantum state transforms under a Lorentz transformation. However, as is well-known, there are no finite-dimensional faithful unitary representations of the Lorentz group \cite{Wigner} and in particular no two-dimensional ones. The only faithful unitary representations of the Lorentz group are infinite dimensional (see e.g. \cite{KimNoz}).  Hence, these cannot be taken to mathematically represent a qubit, i.e. a two-level system. Naively it would appear that a formalism for describing localized qubits which is both relativistic and unitary is a mathematical impossibility.

In the case of flat spacetime the Wigner representations \cite{Wigner,Weinberg} provide unitary and faithful but infinite-dimensional representations of the Lorentz group. These representations make use of the symmetries of Minkowski spacetime, i.e. the full inhomogeneous Poincar\'e group which includes rotations, boosts, and translations. The basis states $|p,\sigma\rangle$ are taken to be eigenstates of the four momentum operators (the generators of spatio-temporal translations) $\hat{P}^\mu$, i.e. $\hat{P}^\mu|p,\sigma\rangle=p^\mu|p,\sigma\rangle$ where the symbol $\sigma$ refers to some discrete degree of freedom, perhaps spin or polarization. One strategy for obtaining a two-dimensional (perhaps mixed) quantum state $\rho_{\sigma\sigma'}$ for the discrete degree of freedom $\sigma$ would be to trace out the momentum degree of freedom. But as shown in \cite{PeresScudoTerno02,PeresTerno03b,PeresTerno04,BartlettTerno05} this density operator does not have covariant transformation properties.  The mathematical reason, from the theory presented in this paper, is that the quantum states for qubits with different momenta belong to {\it different} Hilbert spaces. Thus, the density operator $\rho_{\sigma\sigma'}$ is then a mixture of states which belong to different Hilbert spaces. The operation of `tracing out the momenta' is neither physically meaningful nor mathematically motivated.

Another strategy for defining qubits in a relativistic setting would be to restrict to momentum eigenstates $|p,\sigma\rangle$. The continuous degree of freedom $P$ is then fixed and the remaining degrees of freedom are discrete. In the case of a photon or fermion the state space is two dimensional and this can then serve as a relativistic realization of a qubit. This is the strategy in \cite{TerashimaUeda03,ASK} where the authors develop a theory of transport of qubits along worldlines. However, when we go from a flat spacetime to curved we lose the translational symmetry and thereby also the momentum eigenstates $\ket{p,\sigma}$. The only symmetry remaining is local Lorentz invariance which is manifest in the tetrad formulation of general relativity. Since the translational symmetry is absent in a curved spacetime it seems difficult to work with Wigner representations which rely heavily on the full inhomogeneous Poincar\'e group. The use of Wigner representations therefore needs further justification as they do not exist in curved spacetimes.

In this paper we shall refrain altogether from making use of the infinite-dimensional Wigner representation. Since our focus is on qubits physically realized as polarization of photons and spin of massive fermions our starting point will be the field equations that describe those physical systems, i.e. the Maxwell and Dirac equations in curved spacetimes. Using the WKB approximation we then show in detail how one can isolate a two-dimensional Hilbert space and determine an inner product, unitary evolution, and a quantum state. Our procedure reproduces the results of \cite{TerashimaUeda03,ASK}, and can be regarded as an independent justification and validation.

Notably, possible gravitationally induced global phases \cite{Stodolsky,Alsing,Anandan,AudretschLammerzahl,Sakurai,Werner}, which are absent in \cite{TerashimaUeda03,ASK}, are automatically included in the WKB approach. Such a phase is irrelevant if only single trajectories are considered. However, quantum mechanics allows for more exotic scenarios such as when a single qubit is simultaneously transported along a superposition of paths. In order to analyze such scenarios it is necessary to determine the gravitationally induced phase difference. We show how to derive a simple but fully general relativistic expression for such a phase difference in the case of spacetime Mach--Zehnder interferometry. Such a phase difference can be measured empirically \cite{Colella} with neutrons in a gravitational field. See \cite{Anandan,Mannheim,VarjuRyder} and references therein for further details and generalizations. The formalism developed in this paper can easily be applied to any spacetime, e.g. spacetimes with frame-dragging.

This paper aims to be self-contained and we have therefore included necessary background material such as the tetrad formulation of general relativity, the connection 1-form, spinor formalism and more (see \S\ref{sec-refframes} and \ref{secspinornotation}). For example, the absence of global reference frames in a curved spacetime has a direct bearing on how entangled states and quantum teleportation in a curved spacetime are to be understood conceptually and mathematically. We discuss this in section \ref{secQIinCST}.

%%%%%%%%%%%%%%%%%%%%%%%%%%%%%%%%%%%%%%%%%%%%%%%%%%%%%%%%%%%%%%%%
\section{An outline of methods and concepts}
%%%%%%%%%%%%%%%%%%%%%%%%%%%%%%%%%%%%%%%%%%%%%%%%%%%%%%%%%%%%%%%%
In this section we provide a general outline of the main ideas and concepts needed to understand the topic of localized qubits in curved spacetimes.

%%%%%%%%%%%%%%%%%%%%%%%%%%%%%%%%%%%%%%%%%%%%%%%%%%%%%%%%%%%%%%%%
\subsection{Localized qubits in curved spacetimes}\label{sec:localizedqubits}
%%%%%%%%%%%%%%%%%%%%%%%%%%%%%%%%%%%%%%%%%%%%%%%%%%%%%%%%%%%%%%%%
Let us now make precise the concept of a localized qubit. As a minimal characterization, a {\em localized qubit} is understood in this paper as any two-level quantum system which is spatially well-localized. Such a qubit is effectively described by a two-dimensional quantum state attached to a single point in space. From a spacetime perspective the history of the localized qubit is then a sequence (i.e. a one-parameter family) of two-dimensional quantum states $|\psi(\lambda)\rangle$ each associated with a point $x^\mu(\lambda)$ on the worldline of the qubit parameterized by $\lambda$. In this paper we will focus on qubits represented by the spin of an electron and the polarization of a photon and show how one can, by applying the WKB approximation to the corresponding field equation (the Dirac or Maxwell equation), extract a two-level quantum state associated with a spatially localized particle.

The sequence of quantum states $\ket{\psi(\lambda)}$ must be thought of as belonging to {\it distinct} Hilbert spaces $\m H_{x(\lambda)}$ attached to each point $x^\mu(\lambda)$ of our trajectory. The situation is identical to that in differential geometry where one must think of the tangent spaces associated with different spacetime points as mathematically distinct: since the parallel transport of a vector along some path from one point to another is path dependent there is no natural identification between vectors of one tangent space and the other. The parallel transport, for any type of object, is simply a sequence of infinitesimal Lorentz transformations acting on the object and it is this sequence that is in general path dependent. Thus, if we are dealing with a physical realization of a qubit whose state transforms non-trivially under the Lorentz group, as is the case for the two physical realizations that we are considering, we must also conclude that in general it is not possible to compare quantum states associated with distinct points in spacetime. As we shall see in sections \ref{secfermionQS} and \ref{sec-photonIP}, Hilbert spaces for different {\it momenta} $p^{\mu}(\lambda)$ of the particle carrying the qubit must also be considered distinct. The Hilbert spaces will therefore be indexed as $\mathcal{H}_{x,p}$, and so along a trajectory there will be a family of Hilbert spaces $\m H_{(x,p)(\lambda)}$.

\begin{figure}[h]
\begin{center}
\ifpdf
\includegraphics[width=5cm]{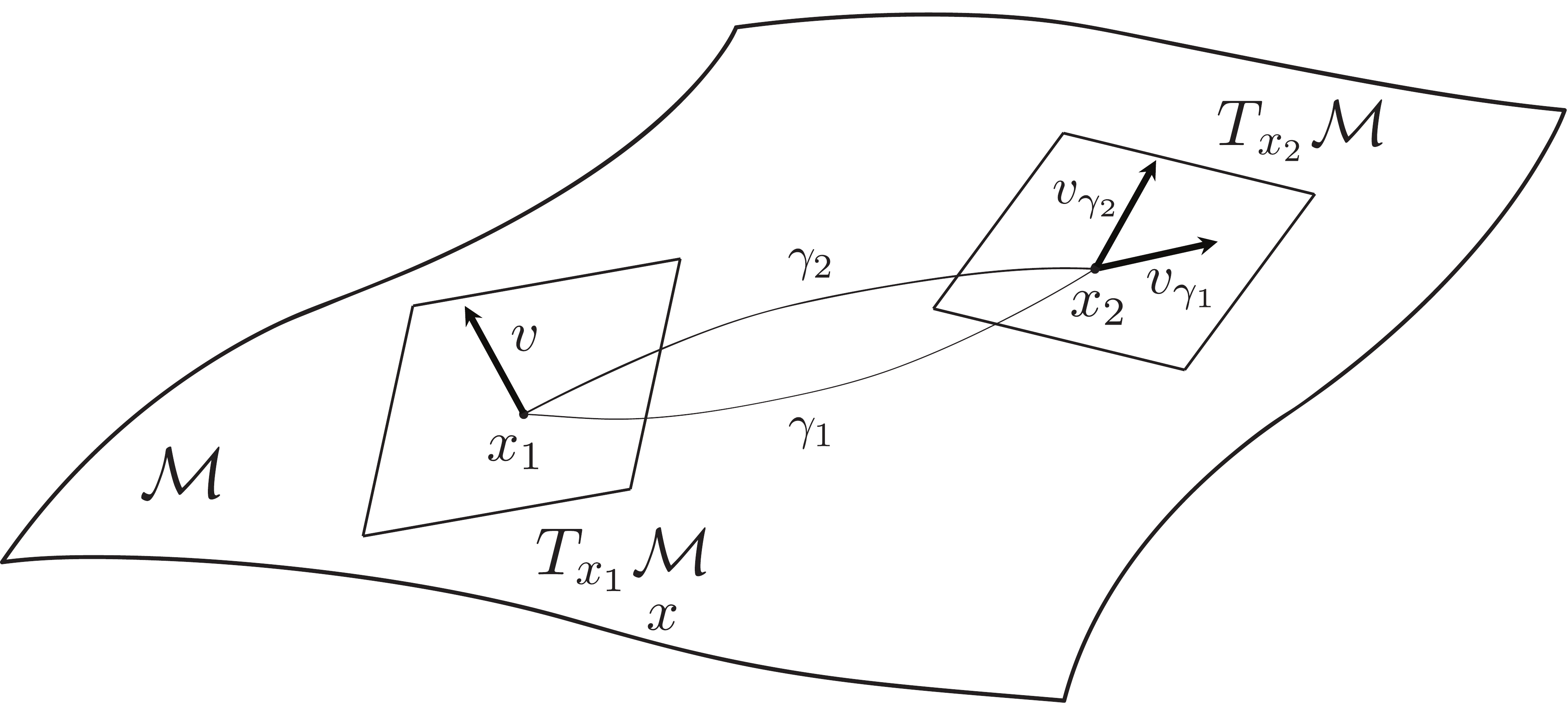}
\includegraphics[width=5cm]{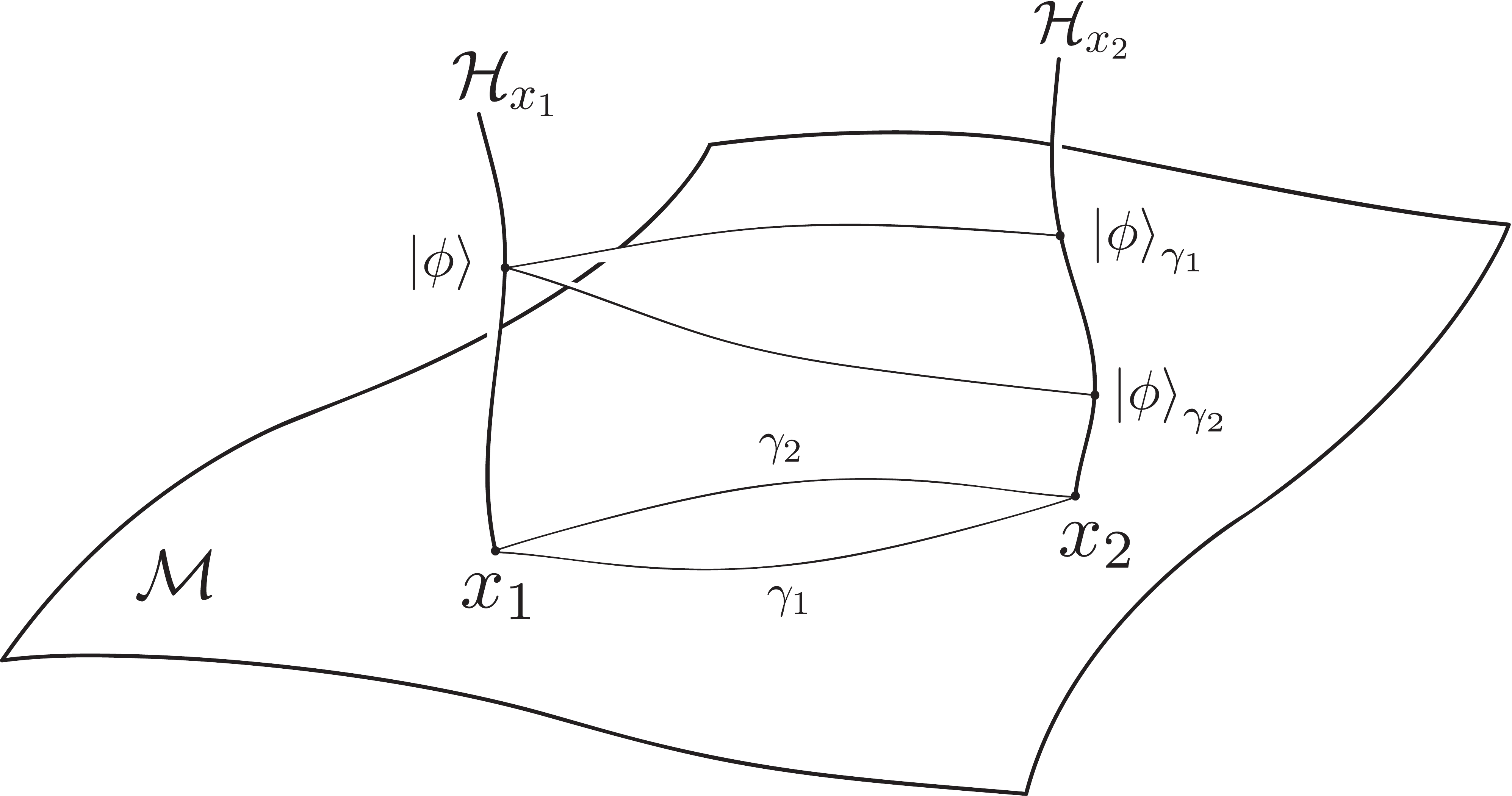}
\else
\includegraphics[width=5cm]{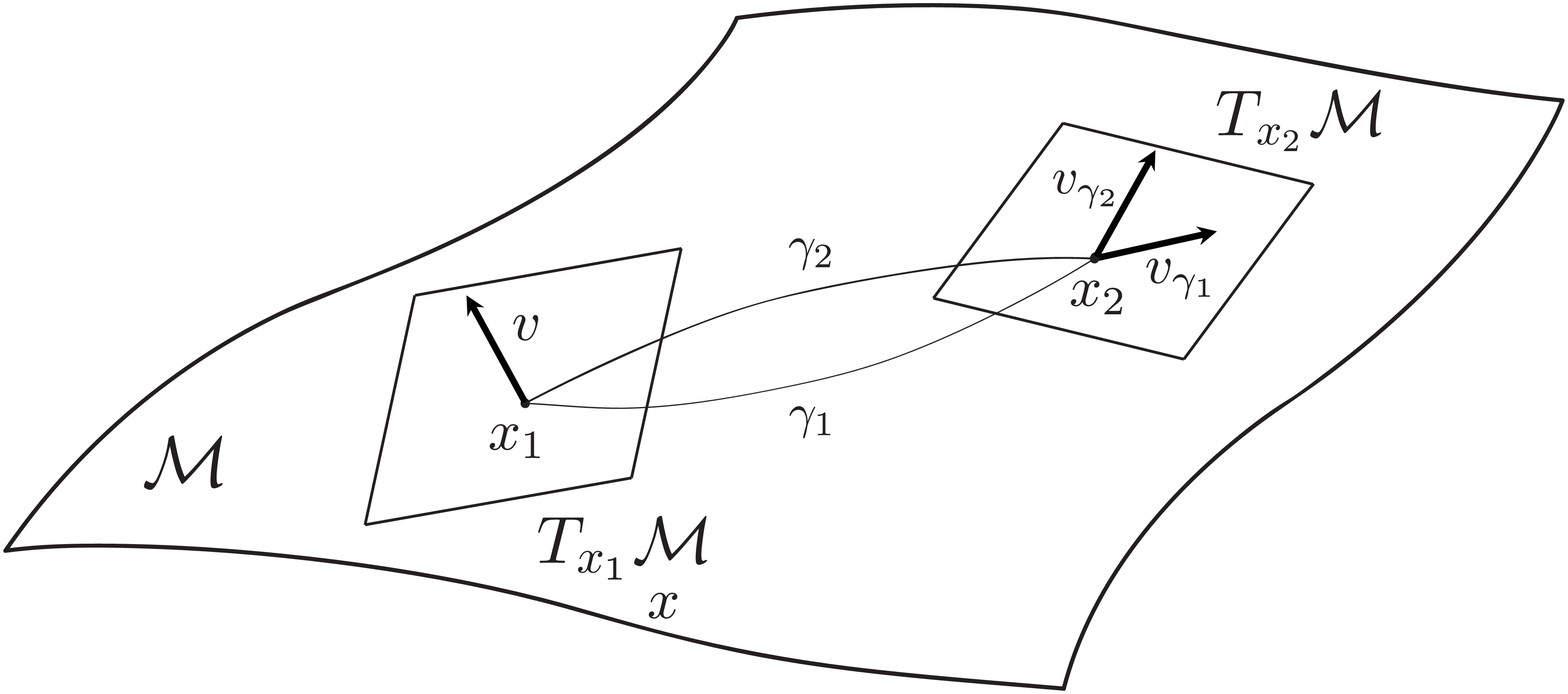}
\includegraphics[width=5cm]{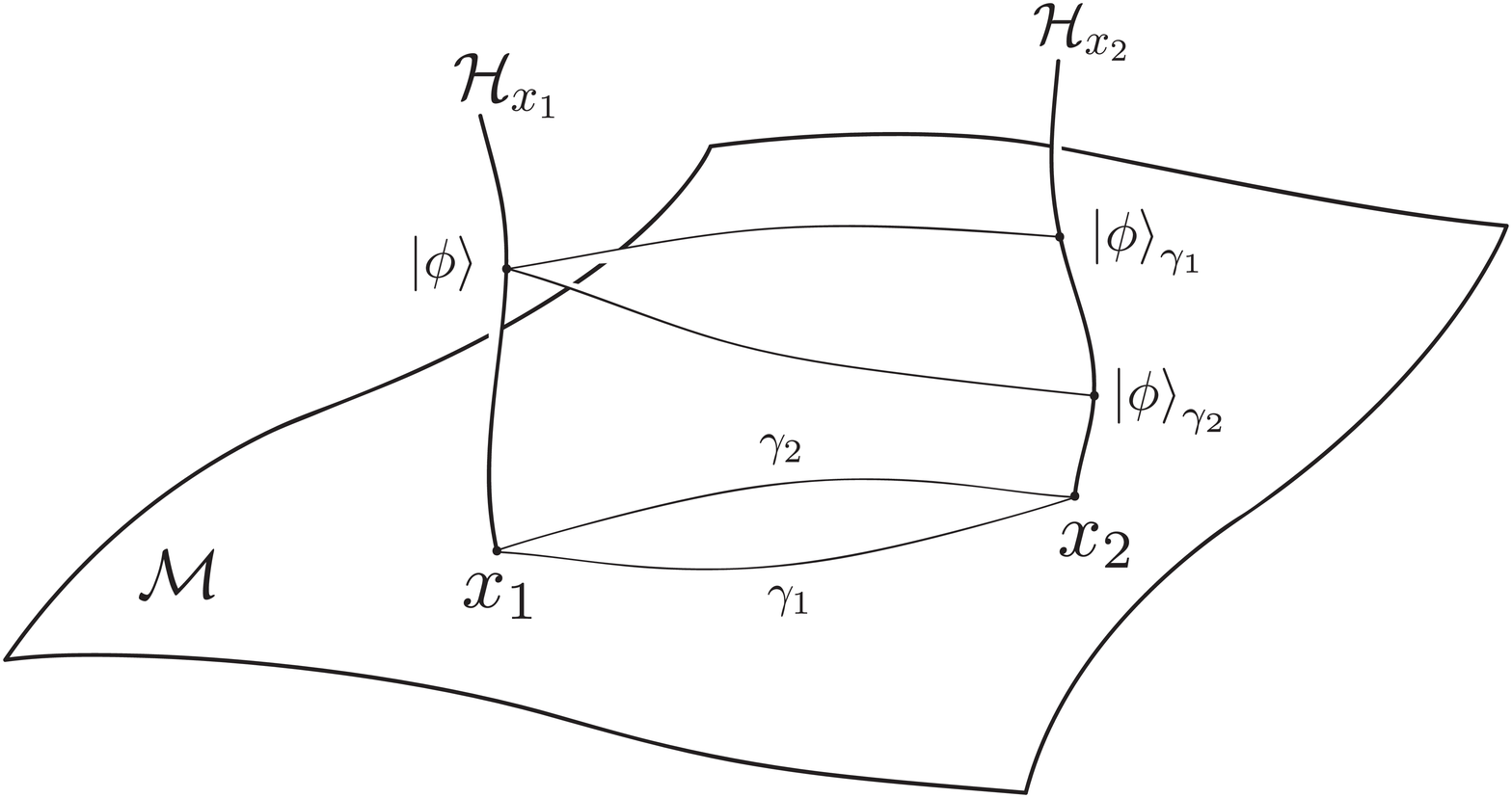}
\fi
\end{center}
\caption{If we parallel transport a vector $v$ from point $x_1$ to $x_2$ along two distinct trajectories $\Gamma_1$ and $\Gamma_2$ in curved spacetime we generally obtain two distinct vectors $v_{\Gamma_1}$ and $v_{\Gamma_2}$ at $x_2$. Thus, no natural identification of vectors of one tangent space and another exists in general and we need to associate a distinct tangent space for each point in spacetime. The same applies to quantum states and their Hilbert spaces: the state at a point $x_2$ of a qubit moved from a point $x_1$ would in general depend on the path taken, and hence the Hilbert space for each point is distinct.  The Hilbert spaces $\m H_{x_1}$ and $\m H_{x_2}$ are illustrated as vertical `fibres' attached to the spacetime points $x_1$ and $x_2$.}
\label{comparevector}
\end{figure}

The ambiguity in comparing separated states has particular consequences: It is in general not well-defined to say that two quantum states associated with distinct points in spacetime are the {\it same}. Nor is it mathematically well-defined to ask how much a quantum state has ``really" changed when moved along a path. Nevertheless, if two initially identical states are transported to some point $x$ but along two distinct paths, the difference between the two resulting states is well-defined, since we are comparing states belonging to the same Hilbert space (see figure \ref{comparevector}).

There are also consequences for how we interpret basic quantum information tasks such as
quantum teleportation: When Alice ``teleports" a quantum state over some distance to Bob we would like to say that it is the {\it same} state that appears at Bob's location. However, this will not have an unambiguous meaning. An interesting alternative is to instead use the maximally entangled state to {\it define} what is ``the same'' quantum state for Bob and Alice, at their distinct locations. We return to these issues in \S\ref{teleportation}.

In a strict sense a localized qubit can be understood as a sequence of quantum states attached to points along a worldline. We will however relax this notion of localized qubits slightly to allow for path superpositions as well. More specifically, we can consider scenarios in which a single localized qubit is split up into a spatial superposition, transported simultaneously along two or more distinct worldlines, and made to recombine at some future spacetime region so as to produce quantum interference phenomena. We will still regard these spatial superpositions as localized if the components of the superposition are each localized around well-defined spacetime trajectories.

%%%%%%%%%%%%%%%%%%%%%%%%%%%%%%%%%%%%%%%%%%%%%%%%%%%%%%%%%%
\subsection{Physical realizations of localized qubits\label{localqubits}}
%%%%%%%%%%%%%%%%%%%%%%%%%%%%%%%%%%%%%%%%%%%%%%%%%%%%%%%%%%
The concepts of a classical bit and a quantum bit (cbit and qubit for short) are abstract concepts in the sense that no importance is usually attached to the specific way in which we physically realize the cbit or qubit. However, when we want to manipulate the state of the cbit or qubit using external fields, the specific physical realization of the bit becomes important. For example, the state of a qubit, physically realized as the spin of a massive fermion, can readily be manipulated using an external electromagnetic field, but the same is not true for a qubit physically realized as the polarization of a photon.

The situation is no different when the external field is the gravitational one. In order to develop a formalism for describing transport of qubits in curved spacetimes it is necessary to pay attention to how the qubit is physically realized. Without knowing whether the qubit is physically realized as the spin of a massive fermion or the polarization of a photon, for example, it is not possible to determine how the quantum state of the qubit responds to the gravitational field. More precisely: gravity, in part, acts on a localized qubit through a sequence of Lorentz transformations which can be determined from the trajectory along which it is transported and the gravitational field, i.e. the connection one-form ${\omega_{\mu}}^I_{\ J}$. Since different qubits can constitute different representations under the Lorentz group, the  influence of gravity will be representation dependent. This is not at odds with the equivalence principle, which only requires that the qubits are acted upon with the same Lorentz transformation.

%%%%%%%%%%%%%%%%%%%%%%%%%%%%%%%%%%%%%%%%%%%%%%%%%%%%%%%%%
\subsection{Our approach}
%%%%%%%%%%%%%%%%%%%%%%%%%%%%%%%%%%%%%%%%%%%%%%%%%%%%%%%%%
Our starting point will be the one-particle excitations of the respective quantum fields. These one-particle excitations are fields $\Psi$ or $A_{\mu}$, which are governed by the classical Dirac or Maxwell equation, respectively. Our goal is to formulate a mathematical description for localized qubits in curved spacetime. Therefore we must find a regime in which the spatial degrees of freedom of the fields are suppressed so that the relevant state space reduces to a two-dimensional quantum state associated with points along some well-defined spacetime trajectory. Our approach is to apply the WKB approximation to these field equations (sections \ref{fermsemiclassapprox} and \ref{photongeoapprox}) and study spatially localized solutions. In this way we can isolate a two-dimensional quantum state that travels along a classical trajectory.

In the approach that we use for the two realizations, we start with a general wavefunction for the fields expressed as
\begin{eqnarray*}
\phi_A(x)=\psi_A(x)\varphi(x)\ee^{\ii\theta(x)}\ \mbox{ or }\ A_{\mu}(x)=\mathrm{Re}[\psi_{\mu}(x)\varphi(x)\ee^{\ii\theta(x)}]
\end{eqnarray*}
where the two-component spinor field $\phi_A(x)$ is the left-handed component of the Dirac field $\Psi$ and $x^{\mu}$ is some coordinate system. The decompositions for the two fields are similar: $\theta(x)$ is the phase, $\varphi(x)$ is the real-valued envelope, and $\psi_A(x)$ or $\psi_{\mu}(x)$ are fields that encode the quantum state of the qubit in the respective cases. These latter objects are respectively the normalized two-component spinor field and normalized complex-valued polarization vector field. Note that we are deliberately using the same symbol $\psi$ for both the two-component spinor $\psi_{A}$ and the polarization 4-vector $\psi_{\mu}$ as it is these variables that encode the quantum state in each case.

The WKB limit proceeds under the assumptions that the phase $\theta(x)$ varies in $x$ much more rapidly than any other aspect of the field and that the wavelength of the phase oscillation is much smaller than the spacetime curvature scale.  Expanding the field equations under these conditions we obtain:
\begin{itemize}
\item a field of wavevectors $k_{\mu}(x)$ whose integral curves satisfy the corresponding classical equations of motion;
\item a global phase $\theta$, determined by integrating $k_\mu$ along the integral curves;
\item transport equations that govern the evolution of  $\psi_A$ and $\psi_\mu$ along this family of integral curves;
\item a conserved current which will be interpreted as a quantum probability current.
\end{itemize}
The assumptions of the WKB limit by themselves do not ensure a spatially localized envelope $\varphi(x)$, and therefore do not in general describe localized qubits. In sections \ref{fermionlocalization} and \ref{photonlocalization} we add further assumptions that guarantee that the qubit is localized during its transport along the trajectory. The spatial degrees of freedom are in this way suppressed and we can effectively describe the qubit as a sequence of quantum states, encoded in the objects $\psi_A(\tau)$ or $\psi_\mu(\lambda)$. These objects constitute non-unitary representations of the Lorentz group. As we shall see in \S\ref{sec-unitarity}, unitarity is recovered once we have correctly identified the respective inner products. Notably, the Hilbert spaces $\m H_{(x,p)(\lambda)}$ we obtain are labelled with both the position and the momentum of the localised qubit.

Finally, since the objects $\psi_A(\tau)$ and $\psi_\mu(\lambda)$ have been separated from the phase $\ee^{\ii\theta(x)}$, the transport equations for these objects do not account for possible gravitationally induced global phases. We show how to obtain such phases in  \S\ref{secPhase} from the WKB approximation. Thus, with the inclusion of phases, we have provided a complete, Lorentz covariant formalism describing the transport of qubits in curved spacetimes. Hereafter it is straightforward to extend the formalism to several qubits in order to treat multipartite states, entanglement and teleportation (\S\ref{secQIinCST}), providing the basic ingredients of quantum information theory in curved spacetimes.

%%%%%%%%%%%%%%%%%%%%%%%%%%%%%%%%%%%%%%%%%%%%%%%%%%%%%%%%%%%%%%%%
\section{Issues from quantum field theory and the domain of applicability}
%%%%%%%%%%%%%%%%%%%%%%%%%%%%%%%%%%%%%%%%%%%%%%%%%%%%%%%%%%%%%%%%
The formalism describing qubits in curved spacetimes presented in this paper has its specific domain of applicability and cannot be taken to be empirically correct in all situations. One simple reason for this is that the current most fundamental theory of nature is not formulated in terms of localized qubits but instead involves very different objects such as quantum fields. There are four important issues arising from quantum field theory that restrict the domain of applicability:
\begin{itemize}
\item the problem of localization;
\item particle number ambiguity;
\item particle creation;
\item the Unruh effect.
\end{itemize}
Below we discuss these issues and indicate how they restrict the domain of applicability of the formalism of this paper.

%%%%%%%%%%%%%%%%%%%%%%%%%%%%%%%%%%%%%%%%%%%%%%%%%%%
\subsection{The localization problem}
%%%%%%%%%%%%%%%%%%%%%%%%%%%%%%%%%%%%%%%%%%%%%%%%%%%
The formalism of this paper concerns spatially localized qubits, with the wavepacket width being much smaller than the curvature scale. However, it is well-known from quantum field theory that it is not possible to localize one-particle states to an arbitrary degree. For example, localization of massive fermions is limited by the Compton wavelength $\lambda_c=h/mc$ \cite{Newtonwigner}. More precisely, any wavefunction constructed from exclusively positive frequency modes must have a tail that falls off with radius $r$ slower than $\ee^{-r/\lambda_c}$. However, this is of no concern if we only consider wavepackets with a width much larger than the Compton wavelength. This consequently restricts the domain of applicability of the material in this paper. In particular, since the width of the wavepacket is assumed to be much smaller than the curvature scale (see \S\ref{sec-particlenumberambiguity}), the localization theorem means that we cannot deal with extreme curvature scales of the order of the Compton wavelength.

A similar problem exists also for photons. Although the Compton wavelength for photons is ill-defined, it has also been shown that they must have non-vanishing sub-exponential tails \cite{Hegerfeldt,Birula}.

Given these localization theorems it is not strictly speaking possible to define a localized wavepacket with compact support. However, for the purpose of this paper we will assume that most of the wavepacket is contained within some region, smaller than the curvature scale, and the exponential tails outside can safely be neglected in calculations. We will assume from here on that this is indeed the case.

%%%%%%%%%%%%%%%%%%%%%%%%%%%%%%%%%%%%%%%%%%%%%%%%%%%%%%%%%%
\subsection{Particle number ambiguity\label{sec-particlenumberambiguity}}
%%%%%%%%%%%%%%%%%%%%%%%%%%%%%%%%%%%%%%%%%%%%%%%%%%%%%%%%%%
One important lesson that we have learned from quantum field theory in curved spacetimes is that a natural notion of particle number is in general absent; see e.g. \cite{WaldQFT}. It is only under special conditions that a natural notion of particle number emerges. Therefore, for arbitrary time-dependent spacetimes it is not in general possible to talk unambiguously about the spin of {\it one} electron or the polarization state of {\it one} photon as this would require an unambiguous notion of particle number. This is important in this paper because a qubit is realized by the spin of one massive fermion or polarization of one photon. %difference here

The particle number ambiguity can be traced back to the fact that the most fundamental mathematical objects in quantum field theory are the quantum field operators and not particles or Fock space representations. More specifically, how many particles a certain quantum state is taken to represent depends in general on how we expand the quantum field operators in terms of annihilation and creation operators $(\hat{a}_i,\hat{a}^\dagger_i)$:
\begin{eqnarray*}
\hat{\phi}(x)=\sum_i \bar{f}_i\hat{a}_i+f_i\hat{a}^\dagger_i
\end{eqnarray*}
which in turn depends on how the complete set of modes (which are solutions to the corresponding classical field equations) is partitioned into positive and negative frequency modes $(f_i,\bar{f}_i)$. In particular, the number operator $\hat{N}\equiv \sum_i\hat a^\dagger_i\hat a_i$ depends on the expansion of the quantum field operator $\hat{\phi}(x)$. This expansion can be done in an infinitude of distinct ways related by Bogoliubov transformations \cite{Birrell}. Particle number is therefore ill-defined. Since we base our approach on the existence of well-defined {\em one-particle} states for photons and massive fermions, the particle number ambiguity seems to raise conceptual difficulties.

We will now argue from the equivalence principle that the particle number ambiguity does not occur for spatially localized states. Consider first vanishing external fields and thus geodesic motion (we will turn to non-geodesics in the next section). In a pseudo-Riemannian geometry, for any sufficiently small spacetime region we can always find coordinates such that the metric tensor is the Minkowski metric $g_{\mu\nu}\stackrel{*}{=}\eta_{\mu\nu}$ and the affine connection is zero $\Gamma^{\rho}_{\mu\nu}\stackrel{*}{=}0$. However, this is true also for a sufficiently narrow strip around any extended spacetime trajectory, i.e. there exists an extended open region containing the trajectory such that $g_{\mu\nu}\stackrel{*}{=}\eta_{\mu\nu}$ and $\Gamma^{\rho}_{\mu\nu}\stackrel{*}{=}0$ \cite{WeinbergGC}. Thus, as long as the qubit wavepacket is confined to that strip it might as well be travelling in a flat spacetime. In fact, the usual free Minkowski modes $\ee^{\pm ip\cdot x}$ form a complete set of solutions to the wave equation for wavepackets localized within that strip. Using these modes we can then define positive and negative frequency and thus the notion of particle number becomes well-defined. Thus, if we restrict ourselves to qubit wavepackets that are small with respect to the typical length scale associated with the spacetime curvature, the particle number ambiguity is circumvented and it becomes unproblematic to think of the classical fields $\Psi(x)$ and $A_\mu(x)$ as describing one-particle excitations of the corresponding quantum field.

%%%%%%%%%%%%%%%%%%%%%%%%%%%%%%%%%%%%%%%%%%%%%%%%%%%%%%%%%%%%%%%%%%%%%
\subsection{Particle creation and external fields}
%%%%%%%%%%%%%%%%%%%%%%%%%%%%%%%%%%%%%%%%%%%%%%%%%%%%%%%%%%%%%%%%%%%%%

Within a strip as defined in the previous section, the effects of gravity are absent and therefore there is no particle creation due to gravitational effects for sufficiently localized qubits. If the trajectory $\Gamma$ along which the qubit is transported is non-geodesic, non-zero external fields need to be present along the trajectory. For charged fermions we could use an electromagnetic field. However, if the field strength is strong enough it might cause spontaneous particle creation and we would not be dealing with a single particle and thus not a two-dimensional Hilbert space. As the formalism of this paper presupposes a two-dimensional Hilbert space, we need to make sure that we are outside the regime where particle creation can occur.

When time-dependent external fields are present, the normal modes $\ee^{\pm ip\cdot x}$ are no longer solutions of the corresponding classical field equations and there will in general be no preferred way of partitioning the modes $(f_i,\bar f_i)$ into positive and negative frequency modes. Therefore, even when we confine ourselves to within the above mentioned narrow strip, particle number is ambiguous.

This type of particle number ambiguity can be circumvented with the help of asymptotic `in' and `out' regions in which the external field is assumed to be weak. In the scenarios considered in this paper there will be a spacetime region $\mathcal{R}_\text{\it prep.}$ in which the quantum state of the qubit is prepared, and a spacetime region $\mathcal{R}_\text{\it meas.}$ where a suitable measurement is carried out on the qubit. The regions are connected by one or many timelike paths along which the qubit is transported. The regions $\mathcal{R}_\text{\it prep.}$ and $\mathcal{R}_\text{\it meas.}$ are here taken to be macroscopic but still sufficiently small such that no tidal effects are detectable, and so special relativity is applicable. We allow for non-zero external fields in these regions and along the trajectory, though we assume that external fields (or other interactions) are weak in these end regions so that the qubit is essentially free there. This means that in $\mathcal{R}_\text{\it prep.}$ and $\mathcal{R}_\text{\it meas.}$ we can use the ordinary Minkowski modes $\ee^{ip\cdot x}$ and $\ee^{-ip\cdot x}$ to expand our quantum field. This provides us with a natural partitioning of the modes into positive and negative frequency modes and thus particle number is well-defined in the two regions $\mathcal{R}_\text{\it prep.}$ and $\mathcal{R}_\text{\it meas.}$. For our purposes we can therefore regard (approximately) the regions $\mathcal{R}_\text{\it prep.}$ and $\mathcal{R}_\text{\it meas.}$ as the asymptotic `in' and `out' regions of ordinary quantum field theory.

If we want to determine whether there is particle creation we simply `propagate' (using the wave equation with an external field) a positive frequency mode (with respect to the free Minkowski modes in $\mathcal{R}_\text{\it prep.}$) from region $\mathcal{R}_\text{\it prep.}$ to $\mathcal{R}_\text{\it meas.}$. In region $\mathcal{R}_\text{\it meas.}$ we then see whether the propagated mode has any negative frequency components (with respect to the free Minkowski modes in $\mathcal{R}_\text{\it meas.}$). If negative frequency components are present we can conclude that particle creation has occurred (see e.g.  \cite{Peskin}). This will push the physics outside our one-particle-excitation formalism and we need to make sure that the strength of the external field is sufficiently small so as to avoid particle creation.

One also has to avoid spin-flip transitions in photon radiation processes such as gyromagnetic emission, which describes radiation due to the acceleration of a charged particle by an external magnetic field, and the related Bremsstrahlung, which corresponds to radiation due to scattering off an external {\it electric} field \cite{Bordovitsyn,Melroseqpd1,Melroseqpd2}. For the former, a charged fermion will emit photons for sufficiently large accelerations and can cause a spin flip and thus a change of the quantum state of the qubit.   Fortunately, the probability of a spin-flip transition is much smaller than that of a spin conserving one, which does not alter the quantum state of the qubit \cite{Melroseqpd2}. In this paper we assume that the acceleration of the qubit is sufficiently small so that we can ignore such spin-flip processes.

%%%%%%%%%%%%%%%%%%%%%%%%%%%%%%%%%%%%%%%%%%%%%%%%%%%%%%%%%%%%%%%%
\subsection{The Unruh effect}
%%%%%%%%%%%%%%%%%%%%%%%%%%%%%%%%%%%%%%%%%%%%%%%%%%%%%%%%%%%%%%%%
Consider the case of flat spacetime. A violently accelerated particle detector could click (i.e. indicate that it has detected a particle) even though the quantum field $\hat{\phi}$ is in its vacuum state. This is the well-known Unruh effect \cite{Unruh,Birrell,Louko}. What happens from a quantum field theory point of view is that the term for the interaction between a detector and a quantum field allows for a process where the detector gets excited and simultaneously excites the quantum field. This effect is similar to that when an accelerated electron excites the electromagnetic field \cite{Akhmedov}. A different way of understanding the Unruh effect is by recognizing that there are two different timelike Killing vector fields of the Minkowski spacetime: one generates inertial timelike trajectories and the other generates orbits of constant proper acceleration. Through the separation of variables of the wave equation one then obtains two distinct complete sets of orthonormal modes: Minkowski modes and Rindler modes, corresponding respectively to each Killing field. The positive Minkowski modes have negative frequency components with respect to the Rindler modes and it can be shown that the Minkowski vacuum contains a thermal spectrum with respect to a Rindler observer.

In order to ensure that our measurement and preparation devices operate `accurately', their acceleration must be small enough so as not to cause an Unruh type effect.
%
%%%%%%%%%%%%%%%%%%%%%%%%%%%%%%%%%%%%%%%%%
\subsection{The domain of applicability}
%%%%%%%%%%%%%%%%%%%%%%%%%%%%%%%%%%%%%%%%%
Let us summarize. In order to avoid unwanted effects from quantum field theory we have to restrict ourselves to scenarios in which:
\begin{itemize}
\item the qubit wavepacket size is much smaller than the typical curvature scale (to ensure no particle number ambiguity);
\item in the case of massive fermions, because of the localization problem the curvature scale must be much larger than the Compton wavelength;
\item there is at most moderate proper acceleration of the qubit (to ensure no particle creation or spin-flip transition due to external fields);
\item there is at most moderate acceleration of preparation and measurement devices (to ensure negligible Unruh effect).
\end{itemize}
For the rest of the paper we will tacitly assume that these conditions are met.

%%%%%%%%%%%%%%%%%%%%%%%%%%%%%%%%%%%%%%%%%%%%%%%%%%%%%%%%%%%%%%%%%
\section{Reference frames and connection 1-forms\label{sec-refframes}}
%%%%%%%%%%%%%%%%%%%%%%%%%%%%%%%%%%%%%%%%%%%%%%%%%%%%%%%%%%%%%%%%%

The notion of a local reference frame, which is mathematically represented by a tetrad field $e^\mu_I(x)$, is essential for describing localized qubits in curved spacetimes. This section provides an introduction to the mathematics of tetrads with an eye towards its use for quantum information theory in curved spacetime. The hurried reader may want to skip to \S\ref{secFermion}. A presentation of tetrads can also be found in \cite[App. J]{Carroll}.

%%%%%%%%%%%%%%%%%%%%%%%%%%%%%%%%%%%%%%%%%%%%%%%%%%%%%%%%%%%%%%%%%
\subsection{The absence of global reference frames}
%%%%%%%%%%%%%%%%%%%%%%%%%%%%%%%%%%%%%%%%%%%%%%%%%%%%%%%%%%%%%%%%%
One main issue that arises when generalizing quantum information theory from flat to curved spaces is the absence of a global reference frame. On a flat space manifold one can define a global reference frame by first introducing, at an arbitrary point  $x_1$, some orthonormal reference frame, i.e. we associate three orthonormal spatial vectors $(\hat x_{x_1},\hat y_{x_1},\hat z_{x_1})$ with the point $x_1$. In order to establish a reference frame at some other point $x_2$ we can parallel transport each of the three vectors to that point. Since the manifold is flat the three resulting orthonormal directions are independent of the path along which they were transported. Repeating this for all points $x$ in our space we obtain a unique field of reference frames $(\hat{x}_x,\hat{y}_x,\hat{z}_x)$ defined for all points $x$ on the manifold.\footnote{In this paper we will implicitly always work in a topologically trivial open set. This allows us to ignore topological issues, e.g. the fact that not all manifolds will admit the existence of an everywhere non-singular field of reference frames.} Thus, from an arbitrarily chosen reference frame at a single point $x_1$ we can erect a unique {\em global reference frame}.

However, when the manifold is curved no unique global reference frame can be established in this way. The reference frame obtained at point $x_2$ by the parallel transport of the reference frame at $x_1$ is in general dependent on the path along which the frame was transported. Thus, in general there is no path-independent way of constructing global reference frames. Instead we have to accept that the choice of reference frame {\em at each point} on the manifold is completely arbitrary, leading us to the notion of local reference frames.

To illustrate this situation and its consequences in the context of quantum information theory in curved space, consider two parties, Alice and Bob, at separated locations. First we turn to the case where the space is flat and the entangled state is the singlet state. The measurement outcomes will be anticorrelated if Alice and Bob measure along the {\em same} direction. In flat space the notion of `same direction' is well-defined. However, in curved space, whether two directions are `the same' or not is a matter of pure convention, since the direction obtained from parallel transporting a reference frame from Alice to Bob is path dependent. Thus, the phrase `Alice and Bob measure along the {\em same} direction' does not have an unambiguous meaning in curved space.

With no natural way to determine that two reference frames at separated points have the same orientation, we are left with having to keep track of the arbitrary local choice of reference frame at each point. The natural way to proceed is then to develop a formalism that will be reference frame {\em covariant}, with the empirical predictions (e.g. predicted probabilities) of the theory required to be manifestly reference frame {\em invariant}. The formalism obtained in this paper meets these two requirements.

%%%%%%%%%%%%%%%%%%%%%%%%%%%%%%%%%%%%%%%%%%%%%%%%%%%%%%%%%%%%%%%%%%%%%%%%
\subsection{Tetrads and local Lorentz invariance \label{tetradllinvar}}
%%%%%%%%%%%%%%%%%%%%%%%%%%%%%%%%%%%%%%%%%%%%%%%%%%%%%%%%%%%%%%%%%%%%%%%%

The previous discussion was in terms of a curved space and a spatial reference frame consisting of three orthonormal spatial vectors. However, in this paper we consider curved spacetimes, and so we have to adjust the notion of a reference frame accordingly. We can do this by simply including the 4-velocity of the spatial reference frame as a {\em fourth} component $\hat t_x$ of the reference frame. Thus, in relativity a reference frame $(\hat{t}_x,\hat{x}_x,\hat{y}_x,\hat{z}_x)$ at some point $x$ consists of three orthonormal spacelike vectors and a timelike vector $\hat t_x$.

Instead of using the cumbersome notation $(\hat t_x,\hat{x}_x,\hat{y}_x,\hat{z}_x)$ to represent a local reference frame at a point $x$ we adopt the compact standard notation $e^\mu_I(x)$. Here $I=0,1,2,3$ labels the four orthonormal vectors of this reference frame such that $e^\mu_0\sim\hat{t},e^\mu_1\sim\hat{x},e^\mu_2\sim\hat{y}$, and $e^\mu_3\sim\hat{z}$, and $\mu$ labels the four components of each vector with respect to the coordinates on the curved manifold. The object $e^\mu_I(x)$ is called a {\em tetrad field}. This object represents a field of arbitrarily chosen orthonormal basis vectors for the tangent space for each point in the spacetime manifold $\m M$. This orthonormality is defined in spacetime by
\begin{eqnarray*}
g_{\mu\nu}(x)e^\mu_I(x)e^\nu_J(x)=\eta_{IJ}
\end{eqnarray*}
where $g_{\mu\nu}$ is the spacetime metric tensor and $\eta_{IJ}$ is the local flat Minkowski metric. Furthermore, orthogonality implies that the determinant $e=\det( e^\mu_I)$ of the tetrad as a matrix in $(\mu,I)$ must be non-zero. Thus there exists a unique inverse to the tetrad, denoted by $e^I_\mu$, such that $e^I_\mu e^\mu_J=\eta^I_J = \delta^I_J$ or $e^I_\mu e^\nu_I=g^{\nu}_{\mu}=\delta^{\nu}_{\mu}$. Making use of the inverse $e^I_\mu$ we obtain
\begin{eqnarray*}
g_{\mu\nu}(x)=e^I_\mu(x) e^J_\nu(x)\eta_{IJ}.
\end{eqnarray*}
Therefore, if we are given the inverse reference frame $e^I_\mu(x)$ for all spacetime points $x$ we can reconstruct the metric $g_{\mu \nu}(x)$. The tetrad $e^\mu_I(x)$ can therefore be regarded as a mathematical representation of the geometry.

As stressed above, on a curved manifold the choice of reference frame at any specific point $x$ is completely arbitrary. Consider then {\it local}, i.e. spacetime-dependent, transformations of the tetrad $e^\mu_I(x)\rightarrow e'^{\mu}_I(x)=\Lambda_I^{\ J}(x)e^\mu_J(x)$ that preserve orthonormality;
\begin{equation}
\eta_{IJ}=g_{\mu \nu}(x)e'^{\mu}_I(x) e'^{\nu}_J(x)\\=g_{\mu \nu}(x)\Lambda_I^{\ K}(x)e^\mu_K(x)\Lambda_J^{\ L}(x)e^\nu_L(x)=\eta_{KL}\Lambda_I^{\ K}(x)\Lambda_J^{\ L}(x)\label{etaLorentzinvariance}.
\end{equation}
The transformations $\Lambda_I^{\ J}(x)$ are recognized as local Lorentz transformations and leave $\eta_{IJ}$ invariant. Given that the matrices $\Lambda_I^{\ J}(x)$ are allowed to depend on $x^\mu$, so that different transformations can be performed at different points on the manifold,  the reference frames associated with different points are therefore allowed to be changed in an uncorrelated manner. However for continuity reasons we will restrict $\Lambda_I^{\ J}(x)$ to local {\em proper} Lorentz transformations, i.e. members of $SO^{+}(1,3)$.

The inverse tetrad $e^I_\mu$ transforms as $e^I_\mu\rightarrow e'^I_\mu=\Lambda^I_{\ J}e^J_\mu$ where $\Lambda^I_{\ K}\Lambda_J^{\ K}=\delta^I_J$. We now see that the gravitational field $g_{\mu\nu}$ is invariant under these transformations:
\begin{equation}
g'_{\mu\nu}=\eta_{IJ}e'^I_\mu e'^J_\nu=\eta_{IJ}\Lambda^I_{\ K} e^K_\mu \Lambda^J_{\ L}e^L_\nu=\eta_{IJ}\Lambda^I_{\ K}\Lambda^J_{\ L}e^K_\mu e^l_\nu=\eta_{KL} e^K_\mu e^L_\nu=g_{\mu\nu}.\label{gLorentzinvariance}
\end{equation}

Therefore, all tetrads related by a local Lorentz transformation $\Lambda^I_{\ J}(x)$ represent the same geometry $g_{\mu\nu}$. Thus, by switching from a metric representation to a tetrad representation we have introduced a new invariance: {\em local Lorentz invariance}.

As stated earlier it will be useful to formulate qubits in curved spacetime in a reference frame covariant manner. To do so we need to be able to represent spacetime vectors with respect to the tetrads and not the coordinates. A spacetime vector $V$ expressed in terms of the coordinates will carry the coordinate index $V^{\mu}$. However, the vector could likewise be expressed in terms of the tetrad basis, in this case  $V^{\mu} = V^{I}e^{\mu}_{I}$ where $V^{I}$ are the components of the vector in the tetrad basis given by $V^I=e^I_\mu V^\mu$. We  can therefore work with tensors represented either in the coordinate basis labelled by Greek indices $\mu, \nu, \rho,\text{ etc}$ or in the tetrad basis where tensors are labelled with capital Roman indices $I, J, K,\text{ etc}$. The indices are raised or lowered either with $g^{\mu\nu}$ or with $\eta^{IJ}$ depending on the basis \footnote{see the notation and conventions Section \ref{notation}}. We will switch between tetrad and coordinate indices freely throughout this paper.

%%%%%%%%%%%%%%%%%%%%%%%%%%%%%%%%%%%%%%%%%%%%%%%%%%%%%%%%%%%%%%%%
\subsection{The connection 1-form}
%%%%%%%%%%%%%%%%%%%%%%%%%%%%%%%%%%%%%%%%%%%%%%%%%%%%%%%%%%%%%%%%

In order to define a covariant derivative and parallel transport one needs a connection. When this connection is expressed in the coordinate basis, which is in general neither normalized nor orthogonal, this is referred to as the affine connection $\Gamma^\rho_{\mu\nu}$. Alternatively if the connection is expressed in terms of the orthonormal tetrad basis it is called the {\em connection one-form} $\omega_{\mu\ J}^{\ I}$. To see this, consider the parallel transport of a vector $V^\mu$ along some path $x^\mu(\lambda)$ given by the equation
\begin{eqnarray*}
\frac{DV^\mu}{D\lambda}\equiv\frac{\di V^\mu}{\di \lambda}+\frac{\di x^\nu}{\di\lambda} \Gamma^\mu_{\nu\rho}V^\rho\equiv0.
\end{eqnarray*}
where $\lambda$ is some arbitrary parameter.
The vector $V^\mu$ in the tetrad basis is expressed as $V^\mu=V^I e^\mu_I$. We can now re-express the parallel transport equation in terms of the tetrad components $V^I$:
\begin{align*}
\frac{D(e^\mu_IV^I)}{D\lambda}\equiv & \frac{\di(e^\mu_IV^I)}{\di\lambda}+ \frac{\di x^\nu}{\di\lambda}\Gamma^\mu_{\nu\rho}e^\rho_IV^I\\
=&e^\mu_I\left(\frac{\di V^I}{\di\lambda}+\frac{\di x^\nu}{\di\lambda}\left[e^I_\rho\partial_\nu e^\rho_J+\Gamma^\sigma_{\nu\rho}e^I_\sigma e^\rho_J\right]V^J\right).
\end{align*}
Thus, if we define
\begin{eqnarray*}
\omega_{\nu\ J}^{\ I} \equiv e^I_\rho\partial_\nu e^\rho_J+\Gamma^\sigma_{\nu\rho}e^I_\sigma e^\rho_J,
\end{eqnarray*}
the equation for the parallel transport of the tetrad components $V^I$ can be written as
\begin{eqnarray*}
\frac{DV^I}{D\lambda}\equiv\frac{\di V^I}{\di\lambda}+ \frac{\di x^\nu}{\di\lambda}\omega_{\nu\ J}^{\ I}V^J=0.
\end{eqnarray*}
The object $\omega_{\nu\ J}^{\ I}$ is called the {\em connection 1-form} or {\em spin-$1$ connection} and is merely the affine connection $\Gamma^\mu_{\nu\rho}$ expressed in a local orthonormal frame $e^\mu_I(x)$. It is also called a {\em Lie-algebra -valued 1-form} since, when viewed as a matrix $(\omega_\nu)^{I}_{\ J}$, it is a 1-form in $\nu$ of elements of the Lie algebra $\mathfrak{so}(1,3)$. The connection 1-form encodes the spacetime curvature  but unlike the affine connection it transforms in a covariant way (as a covariant vector, or in a different language, as a 1-form) under coordinate transformations, due to it having a single coordinate index $\nu$. However, as can readily be checked from the definition, it transforms {\em inhomogeneously} under a change of tetrad $e^I_\mu(x)\rightarrow \Lambda^I_{\ J}(x)e^J_\mu(x)$:
\begin{eqnarray}
\omega_{\mu\ J}^{\ I} \rightarrow\omega_{\mu\ J}^{\prime\ I}=\Lambda^I_{\ K}\Lambda_J^{\ L}\omega_{\mu\ L}^{\ K}+\Lambda^I_{\ K}\partial_\mu \Lambda_J^{\ K}.\label{connectiontransform}
\end{eqnarray}
The inhomogeneous term $\Lambda^I_{\ K}\partial_\mu \Lambda_J^{\ K}$ is present only when the rotations depend on the position coordinate $x^\mu$ and ensures that the parallel transport $\frac{DV^I}{D\lambda}$ transforms properly as a contravariant vector under local Lorentz transformations.

%%%%%%%%%%%%%%%%%%%%%%%%%%%%%%%%%%%%%%%%%%%%%%%%%%%%%%%%%%
\section{The qubit as the spin of a massive fermion\label{secFermion}}
%%%%%%%%%%%%%%%%%%%%%%%%%%%%%%%%%%%%%%%%%%%%%%%%%%%%%%%%%%

A specific physical realization of a qubit is the spin of a massive fermion such as an electron. An electron can be thought of as a spin-$\half$ gyroscope, where a rotation of $2\pi$ around some axis produces the original state but with a minus sign. Such an object is usually taken to be represented by a four-component Dirac field, which constitutes a reducible spin-$\half$ representation of the Lorentz group. However, given that we are after a qubit and therefore a two-dimensional object, we will work with a two-component Weyl spinor field $\phi_A(x)$, with $A=1,2$, which is the left-handed component of the Dirac field (see \ref{secspinornotation}). \footnote{We could work instead with the right-handed component, but this would yield the same results.} The Weyl spinor itself constitutes a finite-dimensional faithful -- and therefore {\it non-unitary} -- representation of the Lorentz group \cite{KimNoz} and one may therefore think that it could not mathematically represent a quantum state. As we shall see, unitarity is recovered by correctly identifying a suitable inner product.

We will begin by considering the Dirac equation in curved spacetime minimally coupled to an electromagnetic field. We rewrite this Dirac equation in second-order form (called the Van der Waerden equation) where the basic field is now a left-handed Weyl spinor $\phi_A$. This equation is then studied in the WKB limit which separates the spin from the spatial degrees of freedom. We then localize this field along a classical trajectory to arrive at a transport equation for the spin of the fermion which forms the physical realization of the qubit. We find that this transport equation corresponds to the Fermi--Walker transport of the spin along a non-geodesic trajectory plus an additional  precession of the fermion's spin due to the presence of local magnetic fields. We will see that from the WKB approximation a natural inner product for the two-dimensional vector space of Weyl spinors emerges. Furthermore, we will see in section \ref{secfermionQS} that in the rest frame of the qubit the standard notion of unitarity is regained. It is also in this frame where the transport equation is identical to the result obtained in \cite{TerashimaUeda03}.

%%%%%%%%%%%%%%%%%%%%%%%%%%%%%%%%%%%%%%%%%%%%%%%%%%%%
\subsection{The WKB approximation}\label{fermsemiclassapprox}
%%%%%%%%%%%%%%%%%%%%%%%%%%%%%%%%%%%%%%%%%%%%%%%%%%%%

Before we begin our analysis of the Dirac equation in the WKB limit we refer the reader to  \ref{secspinornotation} for notation and background material on spinors. This material is necessary for the relativistic treatment of massive fermions.

%%%%%%%%%%%%%%%%%%%%%%%%%%%%%%%%%%%%%%%%%%%%%%%%%
\subsubsection{The minimally coupled Dirac field in curved spacetime}
%%%%%%%%%%%%%%%%%%%%%%%%%%%%%%%%%%%%%%%%%%%%%%%%%
Fermions in flat spacetime are governed by the Dirac equation $\ii\gamma^\mu\partial_\mu\Psi=m\Psi$. Since we are dealing with curved spacetimes we must generalize the Dirac equation to include these situations. This is done as usual through minimal coupling by replacing the partial derivatives by covariant derivatives. The covariant derivative of a Dirac spinor is defined by \cite{Nakahara}
\begin{eqnarray}
\nabla_\mu\Psi=(\partial_\mu-\frac{\ii}{2}\omega_{\mu IJ}S^{IJ})\Psi\label{spinhalfcovarderiv}
\end{eqnarray}
where $S^{IJ}=\frac \ii4[\gamma^I,\gamma^J]$ are the spin-$\half$ generators of the Lorentz group and $\gamma^{I}$ are the Dirac $\gamma$-matrices which come with a tetrad rather than a tensor index. The gravitational field enters through the spin-1 connection $\omega_{\mu IJ}$. We assume that the fermion is electrically charged and include an electromagnetic field $F_{IJ}$  by minimal coupling so that we can consider accelerated trajectories. The Dirac equation in curved spacetime minimally coupled to an external electromagnetic field $A_\mu$ is then given by
\begin{eqnarray}
\ii\gamma^\mu D_\mu\Psi=m\Psi\label{curveddiraceq}
\end{eqnarray}
where we define the $U(1)$ covariant derivative as $D_\mu=\nabla_\mu-\ii eA_\mu$.

%%%%%%%%%%%%%%%%%%%%%%%%%%%%%%%%%%%%%%%%%%%%%%%%%%%%%%%%%
\subsubsection{The Van der Waerden equation: an equivalent second order formulation}
%%%%%%%%%%%%%%%%%%%%%%%%%%%%%%%%%%%%%%%%%%%%%%%%%%%%%%%%%
In order to proceed with the WKB approximation it is convenient to put the Dirac equation into a second-order form. This can be done by making use of the Weyl representation of the $\gamma$-matrices (see \ref{secspinornotation} for further details). In this representation the $\gamma$-matrices take on the form
\begin{eqnarray*}
\gamma^I=\begin{pmatrix}0&\sigma^I_{\ AA'}\\ \bar{\sigma}^{IA'A}&0\end{pmatrix}.
\end{eqnarray*}
The Dirac equation then splits into two separate equations
\begin{subequations}
\begin{eqnarray}
\ii\bar{\sigma}^{\mu A'A}D_\mu\phi_A&=m\chi^{A'}\label{left}\\
\ii\sigma^{\mu}_{\ AA'}D_\mu\chi^{A'}&=m\phi_A\label{right}
\end{eqnarray}
\end{subequations}
with $\bar{\sigma}^{\mu A'A}\equiv e^{\mu}_I\bar{\sigma}^{IA'A}$ and $\sigma^{\mu}_{\ A'A}\equiv e^{\mu}_I\bar{\sigma}^{I}_{\ A'A}$, and $\Psi=(\phi_A,\chi^{A'})$, where $\phi_A$ and $\chi^{A'}$ are left- and right- handed 2-spinors respectively. Solving for $\chi^{A'}$ in equation \eqref{left} and inserting the result into \eqref{right} yields a second-order equation called the {\it Van der Waerden equation} \cite{SakuraiAQM}
\begin{eqnarray*}
\sigma^\mu_{\ AA'}\bar{\sigma}^{\nu A'B}D_\mu D_\nu\phi_B+m^2\phi_A=0
\end{eqnarray*}
which is equivalent to the Dirac equation \eqref{curveddiraceq}. We can rewrite this equation in the following way
\begin{eqnarray}
0&=&\sigma^\mu_{\ AA'}\bar{\sigma}^{\nu A'B} D_\mu D_\nu\phi_B+m^2\phi_A\notag\\
&=&\sigma^\mu_{\ AA'}\bar{\sigma}^{\nu A'B}\left( D_{\{\mu} D_{\nu\}}+ D_{[\mu} D_{\nu]}\right)\phi_B+m^2\phi_A\notag\\
&=&g^{\mu\nu} D_\mu D_\nu\phi_A-\ii L^{\mu\nu\ B}_{\ \ A}(\mathfrak{R}_{\mu\nu B}^{\ \ \ \ C}-\ii e\delta_B^{\ C}F_{\mu\nu})\phi_C+m^2\phi_A\label{vdWexpanded}
\end{eqnarray}
where we have used that $2 D_{[\mu} D_{\nu]} = [D_\mu, D_\nu]$ and $2 D_{\{\mu} D_{\nu\}} = \{D_\mu, D_\nu\}$, and $\sigma^{\{ \mu}\bar\sigma^{\nu\} } = g^\mu\nu$. We identify $F_{\mu\nu}\equiv2\nabla_{[\mu} A_{\nu]}$ as the electromagnetic tensor and $\mathfrak{R}_{\mu\nu A}^{\ \ \ \ B}\phi_{B}:=2\nabla_{[\mu}\nabla_{\nu]}\phi_{A}$ as a spin-$\half$ curvature 2-form associated with the left-handed spin-$\half$ connection $\frac{\ii}{2}\omega_{\mu IJ}L^{IJ\ B}_{\ \ A}$, where $\hat{L}^{\mu\nu} = e^{\mu}_I e^{\nu}_J\hat{L}^{IJ} = \frac \ii2\sigma^{[\mu},\bar{\sigma}^{\nu]}$ are the left-handed spin-$\half$ generators related to the Dirac four-component representation by  $\hat{S}^{\mu\nu}=\hat{L}^{\mu\nu}\oplus \hat{R}^{\mu\nu}$. We have tacitly assumed here that the connection is torsion-free. Torsion can be included (at least in the case of vanishing electromagnetic field) and will slightly modify the way the spin of the qubit changes when transported along a trajectory. We refer the reader to \cite{Audretsch,Anandan94,Bergmann} for further details on torsion.

%%%%%%%%%%%%%%%%%%%%%%%%%%%%%%%
\subsubsection{The basic ansatz}
%%%%%%%%%%%%%%%%%%%%%%%%%%%%%%%
The starting point of the WKB approximation is to write the left-handed two-spinor field $\phi_A$ as
\begin{eqnarray*}
 \phi_A(x)=\varphi_A(x) \ee^{\ii\theta(x)/\epsilon}
\end{eqnarray*}
and study the Van der Waerden equation in the limit $\epsilon\rightarrow0$, where $\epsilon$ is a convenient expansion parameter. Physically this means that we are studying solutions for which the phase is varying much faster than the complex amplitude $\varphi_A$. In the high frequency limit $\epsilon\rightarrow0$ the fermion will not `feel' the presence of a finite electromagnetic field. We are therefore going to assume that as the frequency increases the strength of the electromagnetic field also increases. We thus assume that the electromagnetic potential is given by $\frac1\epsilon A_\mu$. $\epsilon$ is to be thought of as a `dummy' parameter whose only role is to identify the different orders in an expansion. Once the different orders have been identified the value of $\epsilon$ in any equation can be set to 1.

%%%%%%%%%%%%%%%%%%%%%%%%%%%%%%%%%%%%%%%%%%%%%%%%%%%%%%%%%%%%%%%%%%%%%%%%
\subsubsection{The Van der Waerden equation in the WKB limit}
%%%%%%%%%%%%%%%%%%%%%%%%%%%%%%%%%%%%%%%%%%%%%%%%%%%%%%%%%%%%%%%%%%%%%%%%
Rewriting the Van der Waerden equation in terms of the new variables $\varphi_A$ and $\theta$, and collecting terms of similar order in $\frac{1}{\epsilon}$, yields
\begin{equation}
g^{\mu\nu}\nabla_\mu \nabla_\nu\varphi_A-\ii L^{\mu\nu\ B}_{\ \ A}\mathfrak{R}_{\mu\nu B}^{\ \ \ \ C}\varphi_C+\frac{\ii}{\epsilon}(2k^\mu\nabla_\mu\varphi_A+\varphi_A\nabla_\mu k^\mu+\ii eF_{\mu\nu}L^{\mu\nu\ B}_{\ \ A}\varphi_B)
-\frac{1}{\epsilon^2}k_\mu k^\mu\varphi_A+m^2\varphi_A=0\label{vdWWKB}
\end{equation}
where we define the momentum/wavevector as the gauge invariant quantity $k_\mu=\nabla_\mu\theta-eA_\mu$.

If we assume that both the typical scale $\ell$ over which $\varphi_A$ varies and the curvature scale $\m R$ are large compared to the scale $\lambdabar$ over which the phase varies (which is parameterized by $\epsilon$), the first two terms of \eqref{vdWWKB} can be neglected. In the WKB limit the mass term represents a large number and is therefore treated as a ${1}/{\epsilon^2}$ term. The remaining equations are then
\begin{subequations}
\begin{align}
&2k^\mu\nabla_\mu\varphi_A+\varphi_A\nabla_\mu k^\mu+\ii eF_{\mu\nu}L^{\mu\nu\ B}_{\ \ A}\varphi_B =0\label{transncons}\\
&k^\mu k_\mu-m^2=0\label{disprel}.
\end{align}
\end{subequations}
%
%%%%%%%%%%%%%%%%%%%%%%%%%%%%%%%%%%%%%%%%%%%%%%%%%%%%%%%%%%%
\subsubsection{Derivation of the spin transport equation and conserved current \label{spinortransportderiv}}
%%%%%%%%%%%%%%%%%%%%%%%%%%%%%%%%%%%%%%%%%%%%%%%%%%%%%%%%%%%
The dispersion relation \eqref{disprel} implies that $k$ is timelike. Furthermore, by taking the covariant derivative of the dispersion relation and assuming vanishing torsion
\begin{eqnarray*}
\nabla_\nu(k^\mu k_\mu-m^2)&=&2k^\mu\nabla_\nu k_\mu=2k^\mu\nabla_\nu(\nabla_\mu\theta-eA_\mu)\\
&=&2(k^\mu\nabla_\mu k_\nu+ek^\mu F_{\mu\nu})=0
\end{eqnarray*}
we readily see that the integral curves of $u^\mu(x)\equiv k^\mu(x)/m$, defined by $\frac{\di x^\mu}{\di\tau}=u^\mu$, satisfy the classical Lorentz force law
\begin{eqnarray}
m\frac{D^2x^\mu}{D\tau^2}+e\frac{\di x^\nu}{\di\tau} F_\nu^{\ \mu}=0\label{FWvelocity}.
\end{eqnarray}
where $a^\mu\equiv\frac{D^2x^\mu}{D\tau^2}=\frac{\di x^\nu}{\di\tau}\nabla_\nu u^\mu$ and $u^\mu u_\mu=1$. Thus, the integral curves of $k^\mu$ are classical particle trajectories.

To see the implications of the first equation \eqref{transncons} we contract it with $k_\mu\bar{\sigma}^{\mu A'A}\bar{\varphi}_{A'}$ and add the result to its conjugate. Simplifying this sum with the use of \eqref{FWvelocity} and the identity (\cite[Eqn (2.85) p19]{DHM2010})

\begin{eqnarray*}
\bar{\sigma}^{KA'A}L^{IJ\ B}_{\ \ A}=\frac{\ii}{2}(\eta^{KI}\bar{\sigma}^{JA'B}-\eta^{KJ}\bar{\sigma}^{IA'B}-\ii\epsilon^{KIJ}_{\ \ \ \ L}\bar{\sigma}^{LA'B})
\end{eqnarray*}
yields
\begin{eqnarray}
\nabla_\mu (\varphi^2)k^\mu+\varphi^2\nabla_\mu k^\mu=0\label{contracted}
\end{eqnarray}
where $\varphi^2\equiv u_\mu\bar{\sigma}^{\mu A'A}\bar{\varphi}_{A'}\varphi_A$. Eq.\eqref{contracted}  can also be rewritten as
\begin{eqnarray}
\nabla_\mu(\varphi^2k^\mu)=0\label{diraccurrent}
\end{eqnarray}
which tells us that we have a conserved energy density $j^\mu\equiv\sqrt{-g}\varphi^2k^\mu$ \footnote{$\varphi^2$ has dimension $L^{-3}$.}, with $g=\det g_{\mu\nu}$.

Secondly, \eqref{contracted} yields $\nabla_\mu k^\mu=-(2k^\mu\nabla_\mu\varphi)/\varphi$ and when this is inserted back into \eqref{transncons} we obtain
\begin{eqnarray*}
2k^\mu\nabla_\mu\psi_A+\ii eF_{\mu\nu}L^{\mu\nu\ B}_{\ \ A}\psi_B =0.
\end{eqnarray*}
By making use of the integral curves $x^\mu(\tau)$ we obtain the ordinary differential equation
\begin{eqnarray}
\frac{D\psi_A}{D\tau}+\ii\frac e{2m}F_{IJ}L^{IJ\ B}_{\ \ A}\psi_B =0\label{fermionundecipheredtransport}
\end{eqnarray}
where $\frac{D\psi_A}{D\tau}=\frac{\di\psi_A}{\di\tau}-\frac \ii2u^\mu\omega_{\mu IJ}L^{IJ\ B}_{\ \ A}\psi_B$ is the spin-$\half$ parallel transport. Equation \eqref{fermionundecipheredtransport} governs the evolution of the normalized spinor $\psi_A\equiv\varphi_A/\varphi$ along integral curves. Below $\psi_A$ will assume the role of the qubit quantum state.

%%%%%%%%%%%%%%%%%%%%%%%%%%%%%%%%%%%%%%%%%%%%%%%%%%%%%%%%%%%%%%
\subsection{Qubits, localization and transport}\label{fermionlocalization}
%%%%%%%%%%%%%%%%%%%%%%%%%%%%%%%%%%%%%%%%%%%%%%%%%%%%%%%%%%%%%%
The aim of this paper is to obtain a formalism for localized qubits. However, the WKB approximation does not guarantee that the fermion is spatially localized, i.e. the envelope $\varphi(x)$ need not have compact support in a small region of space. In addition, even if the envelope initially is well-localized there is nothing preventing it from distorting and spreading, and becoming delocalized. We therefore need to make additional assumptions beyond the WKB approximation to guarantee the initial and continued localization of the qubit. As pointed out in \S\ref{sec-particlenumberambiguity}, by restricting ourselves to localized envelopes we avoid the particle number ambiguity and can interpret the Dirac field as a one-particle quantum wavefunction.

%%%%%%%%%%%%%%%%%%%%%%%%%%%%%%%%%%%%%%
\subsubsection{Localization\label{fermionlocalizationsubsec}}
%%%%%%%%%%%%%%%%%%%%%%%%%%%%%%%%%%%%%%
Before we begin let us be a bit more precise as to what it means for a qubit to be `localized'. In order to avoid the particle number ambiguity we know that the wavepacket size $\m L$ has to be much less than the curvature scale $\m R$. We also know from quantum field theory that it is not possible to localize a massive fermion to within its Compton wavelength $\lambda_\text{com}\equiv h/mc$ using only positive frequency modes. Mathematically we should then have $\lambda_\text{com}<\m L \ll \m R$ where $\m L$ is the packet length in the rest frame of the fermion. If $\lambda_\text{com}\sim \m R$ the formalism of this paper will not be empirically correct.

How well-localized a wavepacket is, is determined by the support of the envelope. Strictly speaking we know from quantum field theory that a localized state will always have exponential tails which cannot be made to vanish using only positive frequency modes. However, the effects of such tails are small and for the purpose of this paper we will neglect them and assume that the wavepacket has compact support.

The equation that governs the evolution of the envelope within the WKB approximation is the continuity equation \eqref{diraccurrent}
\begin{eqnarray*}
\nabla_\mu (u^\mu \varphi^2(x))=0.
\end{eqnarray*}
If we assume that the divergence of the velocity field $u^\mu$ is zero, i.e. $\nabla_\mu u^\mu=0$, the continuity equation reduces to
\begin{eqnarray*}
\nabla_\mu (u^\mu \varphi^2(x))=u^\mu\nabla_\mu \varphi^2+\varphi^2\nabla_\mu u^\mu=u^\mu\nabla_\mu \varphi^2=0,
\end{eqnarray*}
or, using the integral curves of $u^\mu$,
\begin{eqnarray*}
\frac{\di \varphi^2}{\di\tau}=0.
\end{eqnarray*}
Thus, the shape of the envelope in the qubit's rest frame remains unchanged during the evolution. However, because of the uncertainty principle \cite{PeresQM}, if the wavepacket has finite spatial extent it cannot simultaneously have a sharp momentum, and therefore the divergence in velocity cannot be exactly zero. We can then relax the assumption, since the only thing that we need to guarantee is that the final wavepacket is not {\em significantly} distorted compared to the original one. Since $\nabla_\mu u^\mu$ measures the rate of change of the rest-frame volume $\frac1V\frac{\di V}{\di\tau}$ \cite{MTW}  we should require that
\begin{eqnarray*}
\langle\nabla_\mu u^\mu\rangle \ll\frac{1}{\tau_{_\Gamma}}
\end{eqnarray*}
where $\ex{\nabla_\mu u^\mu}$ is the typical value of $|\nabla_\mu u^\mu|$, and $\tau_{_\Gamma}$ the proper time along some path $\Gamma$  assumed to have finite length. If we combine this assumption of negligible divergence with the assumption that the envelope is initially localized so that the wavepacket size is smaller than the curvature scale, we can approximately regard the envelope as being rigidly transported while neither distorting nor spreading during its evolution.

To further suppress the spatial degrees of freedom we need also an assumption about the two-component spinor $\psi_A(x)$. This variable could vary significantly within the localized support of the envelope $\varphi(x)$. However, as we want to attach a single qubit quantum state to each point along a trajectory we need to assume that $\psi_A(x)$ only varies along the trajectory and not spatially. More precisely, we assume that $\psi_A(t,\vec x)=\psi_A(t)$ when we use local Lorentz coordinates $(t,\vec x)$ adapted to the rest frame of the particle. This implies that the wavepacket takes on the form
\begin{eqnarray*}
\phi_A(t,\vec{x})=\psi_A(t) \varphi(t,\vec x)\ee^{\ii\theta(t,\vec x)}.
\end{eqnarray*}
This form is not preserved for all reference frames since in other local Lorentz coordinates $\psi_A$ will have spatial dependence. Nevertheless, if the packet is sufficiently localized and $\psi_A$ varies slowly the wave-packet will approximately be separable in spin and position for most choices of local Lorentz coordinates. With these additional assumptions we have effectively `frozen out' the spatial degrees of freedom of the wavepacket. The spinor $\psi_A$ can now be thought of not as a function of spacetime $\psi_A(x)$ satisfying a partial differential equation, but rather as a spin state $\psi_A(\tau)$ defined on a classical trajectory $\Gamma$ satisfying an ordinary differential equation \eqref{fermionTP}. We can therefore effectively characterize the fermion for each $\tau$ by a position $x^\mu(\tau)$, a 4-velocity $\di x^\mu/\di\tau=u^\mu(\tau)$, and a spin $\psi_A(\tau)$. Once we have identified the spin as a quantum state this will provide the realization of a localized qubit.

%%%%%%%%%%%%%%%%%%%%%%%%%%%%%%%%%%%%%%%%%%%%%%%%%%%%%%%%%%%%%%%%%
\subsubsection{The physical interpretation of WKB equations\label{sec-spinorphysinterp}}
%%%%%%%%%%%%%%%%%%%%%%%%%%%%%%%%%%%%%%%%%%%%%%%%%%%%%%%%%%%%%%%%%
As discussed in section \ref{sec-particlenumberambiguity}, if we restrict ourselves to localized wavepackets we can interpret $\phi_A(x)$ as one-particle excitations of the quantum field. This allows us to interpret the conserved current $j^\mu/m=\sqrt{g}\varphi^2u^\mu$ as the probability current of a single particle. In this way we can provide a physical interpretation of the classical two-component spinor field  $\phi_A(x)$ as a quantum wavefunction of a single particle.

Next, let us examine the transport equation \eqref{fermionundecipheredtransport}. The electromagnetic tensor $F_{IJ}$ that appears in the term $\ii eF_{IJ}\hat L^{IJ}/m$ can be decomposed into a component parallel to the timelike 4-velocity $u^I$ and a spacelike component perpendicular to $u^I$ using a covariant spatial projector $h^I_J =\delta^I_J-u^{I} u_{J}$.
 We can then rewrite $F_{IJ}L^{IJ\ B}_{\ \ A}$ as $(2u_Iu^KF_{KJ}+h_I^{\ K}h_J^{\ L}F_{KL})L^{IJ\ B}_{\ \ A}$.
The first term corresponds to the electric field as defined in the rest frame, $u_Iu^KF_{KJ}$. This will produce an acceleration $u^\mu\nabla_\mu u_I=a_I=-\frac emu^JF_{JI}$ of the fermion as described by the Lorentz force equation \eqref{FWvelocity}.  The second term is recognized as the magnetic field experienced by the particle, i.e. the magnetic field as defined in the rest frame of the particle, $B^\text{rest}_{IJ}$.
We thus obtain the transport equation for $\psi_A$;
\begin{eqnarray}
\frac{D\psi_A}{D\tau}-\ii u_Ia_JL^{IJ\ B}_{\ \ A}\psi_B+   \ii\frac e{2m} B^\text{rest}_{IJ}L^{IJ\ B}_{\ \ A}\psi_B=0\label{fermionTP}.
\end{eqnarray}
This has a simple physical interpretation. The third term represents the magnetic precession which is induced by the torque that the magnetic field exerts on the spin. This takes the usual form $\frac \ii2\frac emF_{ij}L^{ij}=\frac \ii2\frac emB_{ij}\half\eps^{ij}_{\ \ k}\sigma^k=-\frac{\ii}{2}\frac em\mathbf B\cdot\mathbf\sigma$ if we express it in a tetrad co-moving with the particle, i.e. $e_0^\mu=u^\mu$.

The two first terms represent the spin-half version of the Fermi--Walker derivative:
\begin{eqnarray}
\frac{D^{FW}\psi_A}{D\tau}\equiv\frac{D\psi_A}{D\tau}-\ii u_Ia_JL^{IJ\ B}_{\ \ A}\psi_B\label{spinhalfFW}.
\end{eqnarray}
The presence of a Fermi--Walker derivative can be understood directly from physical considerations. Heuristically we understand the electron as a spin-$\half$ object, i.e. loosely as a quantum gyroscope. The transport of the orientation of an ordinary classical gyro is not governed by the parallel transport equation but rather, it is governed by a Fermi--Walker transport equation. The Fermi--Walker equation arises when we want to move a gyroscope along some spacetime path without applying any external torque \cite{MTW}. \footnote{At first one might think that this is just what the parallel transport equation achieves. However, this is only true for geodesic motion  ($a_I=0$), where the Fermi--Walker and parallel transport equations agree.} We thus identify \eqref{spinhalfFW} as describing torque-free transport of the electron, resulting in the usual Thomas precession of the spin \cite{WeinbergGC}. Finally, the parallel transport term $D\psi_A/D\tau$ encodes the influence of gravity on the qubit, governed by the spin-1 connection $\omega_{\mu\ J}^{\ I}$.

\subsubsection{A summary of the WKB limit}
Let us summarize the results from the previous section.
\begin{itemize}
\item The full wavepacket is written as $\phi_{A}(x) = \psi_{A}(x)\varphi(x)\ee^{\ii\theta(x)}$.
\item The current $j^\mu/m=\sqrt{g}\varphi^2u^ \mu$ is a conserved  probability density.
\item The phase $\theta$ and the vector potential $A_\mu$ define a field of 4-velocities $u_\mu=\frac1m(\nabla_{\mu}\theta-eA_\mu)$.
\item The integral curves of $u^\mu$ are timelike and satisfy the classical Lorentz equation $ma_\mu=eu^\nu F_{\mu\nu}$.
\item The two-component spinor $\psi_A(\tau)$ defined along some integral curve of $u^\mu$ satisfies the transport equation
\begin{eqnarray}
\frac{D\psi_A}{D\tau}-\ii u_Ia_JL^{IJ\ B}_{\ \ A}\psi_B+ \ii\frac e{2m}h_I^{\ K}h_J^{\ L}F_{KL}L^{IJ\ B}_{\ \ A}\psi_B=0\label{spinhalfFWandBprecession}
\end{eqnarray}
which dictates how the spin is influenced by the presence of an electromagnetic and gravitational field.
\end{itemize}
%

%%%%%%%%%%%%%%%%%%%%%%%%%%%%%%%%%%%%%%%%%%%%%%%%%%%
\subsection{The quantum Hilbert space}\label{fermionQHS}
%%%%%%%%%%%%%%%%%%%%%%%%%%%%%%%%%%%%%%%%%%%%%%%%%%%

The spinor $\psi_A\in W$ (where $W$ is a two dimensional complex vector space) could potentially encode a two dimensional quantum state. However, given that $\psi_{A}$ constitutes a faithful and therefore {\em non-unitary} representation of the Lorentz group this identification might seem problematic. This issue is resolved by identifying a velocity-dependent inner product on the space $W$. In doing so we are able to promote $W$ to a Hilbert space and so regard $\psi_A$ as a quantum state. Let us now show how the two-component spinor $\psi_A$ can be taken as a representation of the quantum state for a qubit, and that it does indeed evolve unitarily.

%%%%%%%%%%%%%%%%%%%%%%%%%%%%%%%%%%%%%%%%%%%%%%%%%%%%%%%%%%%%%%%%%%%%%
\subsubsection{The quantum state and inner product\label{newinnerproduct}\label{secfermionQS}}
%%%%%%%%%%%%%%%%%%%%%%%%%%%%%%%%%%%%%%%%%%%%%%%%%%%%%%%%%%%%%%%%%%%%%
Although the space of two-component spinors $W$ is a two-dimensional complex vector space, it is not a Hilbert space as there is no positive definite sesquilinear inner product defined {\em a priori}. However, in the above analysis of the Dirac  field in the WKB limit the object $I_{u}^{A'A}\equiv u_I\bar{\sigma}^{IA'A}$ emerged naturally. Note that this object is simply the inner product for the Dirac field in the WKB limit and has the appropriate index structure of an inner product for a spinor space (see \ref{appIP}). Thus we take the inner product between two spinors $\psi_A^{1}$ and  $\psi_A^{2}$ to be given by
\begin{eqnarray}
\bk{\psi^{1}}{\psi^{2}}=I_{u}^{A'A}\bar{\psi}^{1}_{A'}\psi^{2}_A=u_I\bar{\sigma}^{IA'A}\bar{\psi}^{1}_{A'}\psi^{2}_A
\label{IP}
\end{eqnarray}
which in the rest frame $u^I=(1,0,0,0)$ takes on the usual form $u_{0}\sigma^{0 A'A}\bar{\psi}^{1}_{A'}\psi^{2}_A = \delta^{A'A}\bar{\psi}^{1}_{A'}\psi^{2}_A$. The connection between Dirac notation and spinor notation can therefore be identified as
\begin{eqnarray*}
|\phi\rangle\sim\phi_A\qquad\langle\phi|\sim I_{u}^{A'A}\bar{\phi}_{A'}.
\end{eqnarray*}

First note that the inner product \eqref{IP} is manifestly Lorentz invariant. This follows immediately from the fact that all indices have been contracted.\footnote{Lorentz invariance can be verified explicitly by making use of $\Lambda^I_{\ J}(x)\bar{\sigma}^{JB'B}(x)\bar\Lambda^{\ A'}_{B'}(x)\Lambda^{\ A}_{B}(x)=\bar{\sigma}^{IA'A}$ \cite{DHM2010}.}  Secondly, $I^{A'A}_u$  satisfies all the criteria for an inner product on a complex vector space $W$: Sesquilinearity\footnote{Sesquilinearity is the property that the inner product is linear in its second argument and antilinear in its first.} is immediate, and the positive definiteness follows if $u^I$ is future causal and timelike, since the eigenvalues $\lambda_{\pm}=u^0(1\pm v)$ of $I_{u}^{A'A}$ are strictly positive, where $u^0\equiv(1-v^2)^{-\half}$ and $v$ denotes the speed of the particle as measured in the tetrad frame. Thus, in the WKB limit, $I_{u}^{A'A}$ can be taken to define an inner product on the spinor space $W$ which therefore becomes a Hilbert space. The spinor $\psi_A$ is then a member of a Hilbert space and thus it plays the role of a quantum state. A qubit is then characterized by its trajectory $\Gamma$ and the quantum states $\psi_A(\tau)$ attached to each point along the trajectory.

In \S \ref{sec:localizedqubits} we saw that we need a separate Hilbert space for each spacetime point $x$. However, the inner product is also velocity dependent, or equivalently momentum dependent. Thus, we must also regard states corresponding to qubits with different momenta as belonging to different Hilbert spaces. In particular, we cannot compare or add quantum states with different 4-momenta $p_1\neq p_2$ even if the quantum states are associated with the same position in spacetime. Consequently the Hilbert space of the qubit is labelled not only with its spacetime position but also with its 4-momentum. We therefore denote the Hilbert space as $\m H_{x,p}$.

%%%%%%%%%%%%%%%%%%%%%%%%%%%%%%%%%%%%%%%%%%%%%%%%%%%%%%%%%%%%%%%%%%%%%
\subsubsection{Wigner rotation \label{fermionwignerrot}}
%%%%%%%%%%%%%%%%%%%%%%%%%%%%%%%%%%%%%%%%%%%%%%%%%%%%%%%%%%%%%%%%%%%%%
In order to establish a relation to the Wigner representations and Wigner rotations \cite{Weinberg} we first note that the basis
\begin{eqnarray*}
\xi_A = \begin{pmatrix}1\\ 0\end{pmatrix} \qquad \chi_A = \begin{pmatrix}0\\1\end{pmatrix}
\end{eqnarray*}
in which the quantum state is expanded, $\psi_A=\psi_1\xi_A+\psi_2\chi_A$, is an oblique basis and not orthonormal with respect to the inner product $I^{A'A}_u$, i.e. $\bk\xi\chi\neq0$ and $\bk\xi\xi\neq1\neq\bk \chi\chi$. One consequence of this is that the transport equation \eqref{spinhalfFW} appears non-unitary as it contains both terms that look Hermitian (e.g. $\hat{L}^{ij}=\half\eps^{ij}_{\ \ k}\hat{\sigma}^k$), and terms that look anti-Hermitian (e.g. $\hat{L}^{0j}=-\frac \ii2 \hat{\sigma}^j$). However, as will shall see in \S\ref{sec-unitarity} the transport is unitary with respect to the inner product $I_u^{A'A}$.

The connection to the Wigner formalism is seen by re-expressing the quantum state in an orthonormal basis. This is given by
\begin{eqnarray*}
\tilde\xi_A = \Lambda^{\ B}_{A}\xi_B \qquad \tilde\chi_A = \Lambda^{\ B}_{A}\chi_B
\end{eqnarray*}
where $\Lambda_A^{\ B}$ is the spin-$\half$ representation corresponding to the Lorentz transformation defined by $u_I=\Lambda_{I}^{\ J}\delta^0_J$. Orthonormality follows from the Lorentz invariance of $\bar{\sigma}^{JB'B}$ and the fact that $\xi_A$ and $\chi_A$ are orthonormal with respect to the inner product $\delta^{A'A}$. For example, $\tilde\xi$ and $\tilde\chi$ are orthogonal which can be seen by making use of the invariance of $\bar\sigma^{IA'A}$:
\begin{eqnarray*}
\bk{\tilde\xi}{\tilde\chi}&=&\bar\Lambda_{A'}^{\ \ C'}\bar\xi_{C'}\Lambda_B^{\ D}\chi_Du_I\bar\sigma^{IA'B}
\nonumber\\
&=&\bar\xi_{C'}\chi_D\bar\Lambda_{A'}^{\ \ C'}\Lambda_B^{\ D}\Lambda^{J}_{\ I}\delta_J^0\bar\sigma^{IA'B}=\bar\xi_{C'}\chi_D\delta^{BC'}=0
\end{eqnarray*}
and we can in a similar way demonstrate that $\bk{\tilde\xi}{\tilde\xi}=\bk{\tilde\chi}{\tilde\chi}=1$. The components $(\tilde\psi_1,\tilde\psi_2)$ are defined by $\psi_A=\tilde\psi_1\tilde\xi_A+\tilde\psi_2\tilde\chi_A$ and can now be understood as the components $\tilde\psi_A ={ \Lambda^{-1}}^{\ B}_{A}\psi_B$ of the spinor in the particle's rest frame.

Given that in the rest frame the basis $(\tilde\xi_A,\tilde\chi_A)$ is indeed orthonormal, it is instructive to also express the Fermi--Walker transport in such a basis. By doing so we will not only see that the evolution is indeed unitary, but in addition we will make contact with the transport equation identified by \cite{TerashimaUeda03} in which the authors made use of infinite-dimensional representations and the Wigner rotations.

Explicitly the spin-$\half$ Lorentz boost as defined above takes the form \cite{DHM2010}
\begin{eqnarray}
\Lambda_{A}^{\;B} = \sqrt{\frac{\gamma+1}{2}}\sigma^{0\ B}_{\ A}+\sqrt{\frac{\gamma-1}{2\beta^2}} \beta_{i}\sigma^{i\ B}_{\ A}\label{spinhalfLboost}
\end{eqnarray}
where $\beta^{i}$ is the boost velocity,  $\gamma=(1-\beta^2)^{-\half}$ is its Lorentz factor, and the Pauli operators are given by $\sigma^{I\ B}_{\ A}=\sigma^0_{\ AA'}\sigma^{IA'B}$. The corresponding spin-1 boost (acting on a contravariant vector) is
\begin{eqnarray}
\Lambda_{\ J}^{I}=
\begin{pmatrix}
\gamma&\gamma\beta_j\\
\gamma\beta^i&\delta^i_j+\frac{\gamma^2\beta^i\beta_j}{\gamma+1}
\end{pmatrix}\label{spin1Lboost}
\end{eqnarray}
where $\beta_j=\delta_{ij}\beta^i$. Substituting $\psi_{A} =  \Lambda_{A}^{\;B}\tilde{\psi}_{B}$ into the Fermi--Walker derivative \eqref{spinhalfFW} yields
\begin{align*}
\frac{D^{FW}\psi_{A}}{D\tau} &=\frac{\di\psi_{A}}{\di\tau}-\frac \ii2 u^{\mu} \omega_{\mu\;IJ} L^{IJ\ B}_{\ \ A}\psi_{B} - \ii u_{I} a_{J} L^{IJ\ B}_{\ \ A}\psi_{B} \\
&=\Lambda_{A}^{\;B} \frac{\di\tilde{\psi}_B}{\di\tau}+ \frac{\di \Lambda_{A}^{\;B}}{\di\tau}\tilde{\psi}_B - \ii\left( \frac 12 u^{\mu} \omega_{\mu\;IJ} + u_{I} a_{J}\right) L^{IJ\ B}_{\ \ A}\Lambda_{B}^{\;C}\tilde{\psi}_C=0.
\end{align*}
The latter expression can be rearranged to give an evolution equation for the rest-frame spinor
\[
\frac{\di\tilde{\psi}_A}{\di\tau} = \left[-{\Lambda^{-1}}_{A}^{\;B} \frac{\di \Lambda_{B}^{\;D}}{\di\tau} + \ii\left( \frac 12 u^{\mu} \omega_{\mu\;IJ} +u_{I} a_{J}\right) {\Lambda^{-1}}_A^{\ B}L^{IJ\ C}_{\ \ B} \Lambda_{C}^{\;D}\right]\tilde{\psi}_D.
\]
One can then simplify this using the identities $ \Lambda_{A}^{\;B}\Lambda_{\ C}^{D} {L^{IJ}}_{B}^{\;\;C}=  \Lambda^{I}_{\;K}\Lambda^{J}_{\;L}{L^{KL}}_{A}^{\;D}$ \footnote{This can be shown using the Lorentz invariance of $\bar{\sigma}^{IA'A}$ and the definition of ${L^{IJ}}_A^{\ B}$ in terms of $\sigma^I$: see \S\ref{spinhalfLG}.}
and ${\Lambda^{-1}}_A^{\ B}=\Lambda_{\ A}^B$ \cite[p9]{Bailin} to obtain spin-1 boosts to the terms involving $L^{IJ}$. Using now explicit expressions of the spin-$\half$ and spin-1 boosts \eqref{spinhalfLboost} and \eqref{spin1Lboost}, one can cancel many of the terms to yield the result
\begin{equation}
 \frac{\di\tilde{\psi}_A}{\di\tau} = \frac{\ii\gamma^{2}}{2(\gamma+1)}\beta_{i}\frac{\di\beta_{j}}{\di\tau}\epsilon^{ij}_{\;\;k}\sigma^{k\;B}_{\;A} \tilde{\psi}_{B} +\ii u^{\mu}\left(\frac12  \omega_{\mu\;ij}+\gamma  \omega_{\mu\;0j}\beta_{i}+\frac{\gamma^{2}}{\gamma+1} \omega_{\mu\;il}\beta^{l}\beta_{j} \right)L^{ij\ B}_{\ \ A}\tilde{\psi}_{B}.\label{restframetrans}
\end{equation}
This is the transport equation for the quantum state $\psi_{A}$ expressed in terms of the rest-frame spinor $\tilde{\psi}_{A}$. First we note that the transport is unitary with respect to the standard inner product $\delta^{A'A}$ as it only contains terms proportional to  $\hat{L}^{ij}=\half\eps^{ij}_{\ \ k}\hat{\sigma}^k $. It is however not manifestly Lorentz invariant. Secondly, it is also equivalent to the transport equation derived by \cite{TerashimaUeda03} who used the infinite-dimensional Wigner representations \cite{Weinberg}. We have thus re-derived their result using the Dirac equation in the WKB limit. In addition, we have done so while avoiding the use of momentum eigenstates $|p,\sigma\rangle$, which are strictly speaking not well-defined in a curved spacetime as no translational invariance is present.

Notice that there is no term proportional to the identity $\delta_A^{\ B}$ which would correspond to an accumulation of global phase.  In fact, global phase is missing in \cite{TerashimaUeda03}. On the other hand, as we shall see in \S\ref{secPhase}, these phases are automatically included in the WKB approach adopted in this paper.

Although the unitarity of the transport becomes manifest when written in terms of the rest-frame spinor it is not necessary to work with equation \eqref{restframetrans}. Once we generalize the notion of unitarity in \S\ref{sec-unitarity} we will see that we can treat the evolution of the quantum state in terms of the manifestly Lorentz covariant Fermi--Walker transport \eqref{spinhalfFW}.

%%%%%%%%%%%%%%%%%%%%%%%%%%%%%%%%%%%%%%%%%%%%%%%%%%%%%%%%%%%%%%%%%%%%%
\section{The qubit as the polarization of a photon}
%%%%%%%%%%%%%%%%%%%%%%%%%%%%%%%%%%%%%%%%%%%%%%%%%%%%%%%%%%%%%%%%%

Another specific physical realization of a qubit is the polarization of a single photon. This is an important example since it lends itself easily to physical applications. We obtain this realization via the WKB limit of Maxwell's equations in curved spacetime \cite{MTW,Woodhouse}. The polarization of a photon is described by a unit spacelike 4-vector $\psi_\mu$ called the polarization vector \cite{MTW,Rindler,Woodhouse}. Restricting ourselves to localized wavepackets we obtain the description of a photon with definite 4-momentum/wavevector $k^\mu$ and polarization vector $\psi_\mu(\lambda)$ which is parallel transported along a null geodesic $x^{\mu}(\lambda)$. We will see that in fact $\psi_{\mu}$ contains only two gauge invariant degrees of freedom and thus can be taken to encode the quantum state of a photonic qubit.

Although we consider only geodesic trajectories in this paper it is possible to consider non-geodesic trajectories. We refer the reader to \ref{jerk} for a discussion of approaches to this problem.  A physically motivated way to obtain non-geodesic trajectories would be to introduce a medium in Maxwell's equations through which the photon propagates. Nevertheless, even without explicitly including a medium, it is easy to include optical elements such as mirrors, prisms, and other unitary transformations as long as their effect on polarization can be considered separately to the effect of transport through curved spacetime.

%%%%%%%%%%%%%%%%%%%%%%%%%%%%%%%%%%%%%%%%%%%%%%%%%%%%%%%%%%%%%%%%%%%%%
\subsection{Parallel transport from the WKB approximation}\label{photongeoapprox}
%%%%%%%%%%%%%%%%%%%%%%%%%%%%%%%%%%%%%%%%%%%%%%%%%%%%%%%%%%%%%%%%%%%%%
In this section we shall see that the parallel transport equation for the polarization vector emerges directly from the WKB approximation \cite{MTW,Woodhouse}. Gauge invariance and gauge fixing in the WKB approach are important for properly isolating the quantum state and we have therefore paid attention to this issue.

%%%%%%%%%%%%%%%%%%%%%%%%%%%%%%%%%%%%%%%%%%%%%%%%%%%%%%%%%%%%%%%%%%%%%%%%%%%%%%%%%%%%
\subsubsection{The basic ansatz}
%%%%%%%%%%%%%%%%%%%%%%%%%%%%%%%%%%%%%%%%%%%%%%%%%%%%%%%%%%%%%%%%%%%%%%%%%%%%%%%%%%%%

The WKB approximation for photons follows a procedure similar to that for the Dirac field. First we write the vector potential $A_{\mu}$ as
\begin{eqnarray}
A_\mu=\mathrm{Re}[\varphi_\mu \ee^{\ii\theta/\epsilon}]
\label{EMpolarform}.
\end{eqnarray}
As in the case for the Dirac field, the WKB limit is where the phase $\theta$ is oscillating rapidly compared to the slowly varying complex amplitude $\varphi_\mu$. As before, this is expressed through the expansion parameter $\epsilon$. Maxwell's equations can then be studied in the limit $\epsilon\rightarrow 0$. Although we omit taking the real part of $\varphi_\mu \ee^{\ii\theta/\epsilon}$ in this section it is implicitly understood that this is done.

%%%%%%%%%%%%%%%%%%%%%%%%%%%%%%%%%%%%%%%%%%%%%%%%%%%%%%%%%%%%%%%%%%%%%%%%%%%%%%%%%%%%
\subsubsection{Gauge transformations in the WKB limit}
%%%%%%%%%%%%%%%%%%%%%%%%%%%%%%%%%%%%%%%%%%%%%%%%%%%%%%%%%%%%%%%%%%%%%%%%%%%%%%%%%%%%
Let us now study the $U(1)$ gauge transformations in terms of the new variables $\theta$ and $\varphi_\mu$. It is clear that not all gauge transformations $A_\mu\rightarrow A_\mu+\nabla_\mu \lambda$ will preserve the basic form $A_\mu=\varphi_\mu \ee^{\ii\theta/\eps}$. We therefore consider gauge transformations of the form $\lambda=\zeta \ee^{\ii\theta/\eps}$ where $\zeta$ is a slowly varying function. This class of gauge transformations can be written in the polar form of \eqref{EMpolarform} as
\[
A_\mu\rightarrow A_\mu+\nabla_\mu\lambda= A_\mu+\nabla_\mu(\zeta \ee^{\ii\theta/\epsilon})=\left(\varphi_\mu+\nabla_\mu \zeta+\frac{\ii}{\epsilon}k_\mu\zeta\right)\ee^{\ii\theta/\epsilon}
\]
and so $\varphi_\mu\rightarrow \varphi_\mu+\nabla_\mu \zeta+\frac{\ii}{\epsilon}k_\mu\zeta$.

In the limit $\epsilon\rightarrow 0$ note that $\varphi_\mu$ does not behave properly under the gauge transformations of the type that we are considering since the second term blows up. This has no physical significance and is just an artefact of describing the vector potential as being of the specific form \eqref{EMpolarform}. Such a gauge transformation leaves the physics unchanged but will no longer preserve the form of the solution \eqref{EMpolarform} where we have a slowly varying envelope and rapid phase. Because of this it is necessary to further restrict the space of gauge transformations to ``small" gauge transformations $\zeta=-\ii\epsilon\xi$. In that limit we then have
\begin{eqnarray}
\varphi_\mu\rightarrow \varphi_\mu-\ii\epsilon\nabla_\mu \xi+k_\mu\xi \label{varphigauge}
\end{eqnarray}
and so $\varphi_\mu\rightarrow \varphi_\mu+k_\mu\xi+\mathcal{O}(\epsilon)$. %%NOTE!!!
However, as we shall see below, in order to maintain gauge invariance of the equations in all orders of $\epsilon$ it is important to keep both orders of $\epsilon$ in the gauge transformation \eqref{varphigauge}.

%%%%%%%%%%%%%%%%%%%%%%%%%%%%%%%%%%%%%%%%%%%%%%%%%%%%%%%%%%%%%%%%%%%%%%%%%%%%%%%%%%%%
\subsubsection{The gauge condition}
%%%%%%%%%%%%%%%%%%%%%%%%%%%%%%%%%%%%%%%%%%%%%%%%%%%%%%%%%%%%%%%%%%%%%%%%%%%%%%%%%%%%

In the literature we find two suggestions for imposing a gauge. For example, in \cite{MTW} the Lorenz gauge is used, $\nabla_\mu A^\mu=(\nabla_\mu \varphi^\mu+\frac{\ii}{\epsilon}k^\mu \varphi_\mu)\ee^{\ii\theta/\epsilon}=0$, and in  \cite{Woodhouse} the gauge $k^\mu \varphi_\mu=0$ is imposed so that the complex amplitude $\varphi_\mu$ is always orthogonal to the wavevector $k^\mu$. However, for our purposes neither of these gauge conditions turns out to be suitable. Rather we will work in a gauge where $k_\mu$ and $\varphi_\mu$ are orthogonal up to first-order terms in $\epsilon$, i.e.
\begin{eqnarray*}
\varphi_\mu k^\mu=\epsilon \alpha(x)
\end{eqnarray*}
where $\alpha$ is taken to be some arbitrary function of $x^\mu$.

%%%%%%%%%%%%%%%%%%%%%%%%%%%%%%%%%%%%%%%%%%%%%%%%
\subsubsection{Maxwell's equations in the WKB limit}
%%%%%%%%%%%%%%%%%%%%%%%%%%%%%%%%%%%%%%%%%%%%%%%%
Let us now turn to Maxwell's equations in vacuum:
\begin{eqnarray}
\nabla_\mu F^\mu_{\ \ \nu}=g^{\rho\mu}\nabla_\rho(\nabla_\mu A_\nu-\nabla_\nu A_\mu)=0\label{MW}.
\end{eqnarray}
The equations $\nabla_{[\rho}F_{\mu\nu]}=0$ are mere identities when we work with a vector potential $A_\mu$ rather than the gauge invariant $F_{\mu\nu}\equiv\nabla_\mu A_\nu-\nabla_\nu A_\mu$. If we substitute the ansatz $A_\mu=\varphi_\mu \ee^{\ii\theta/\epsilon}$ into \eqref{MW} we obtain
\begin{equation}
\square \varphi_\nu-\nabla^\mu \nabla_\nu \varphi_\mu+\frac{\ii}{\epsilon}(2k^\mu\nabla_\mu \varphi_\nu+\varphi_\nu\nabla_\mu k^\mu-k_\nu\nabla_\mu \varphi^\mu-\nabla_\nu(\varphi_\mu k^\mu))-\frac{1}{\epsilon^2}(k^2\varphi_\nu-k_\nu \varphi_\mu k^\mu)=0.\label{MW1}
\end{equation}
Gauge invariance can be a bit subtle in this context so let us make a few remarks. Eq.\eqref{MW1} is of course invariant under gauge transformations $\varphi_\mu\rightarrow \varphi_\mu-\ii\epsilon\nabla_\mu\xi+k_\mu\xi$ as this is nothing but Maxwell's equations \eqref{MW} rewritten in different variables. However, note that the terms of zeroth, first, and second order (in $1/\epsilon$) of Eq. \eqref{MW1}:
\begin{subequations}\label{orders}
\begin{align}
&\square \varphi_\nu-\nabla^\mu \nabla_\nu \varphi_\mu\label{zero}\\
&2k^\mu\nabla_\mu \varphi_\nu+\varphi_\nu\nabla_\mu k^\mu-k_\nu\nabla_\mu \varphi^\mu-\nabla_\nu(\varphi_\mu k^\mu)\label{first}\\
&k^2\varphi_\nu-k_\nu \varphi_\mu k^\mu\label{second}
\end{align}
\end{subequations}
are not {\em separately} gauge invariant. This is so because the gauge transformation $\varphi_\mu\rightarrow \varphi_\mu-\ii\epsilon\nabla_\mu \xi+k_\mu\xi$ contains terms of different orders in $\epsilon$. Thus, after a gauge transformation of the second-order term \eqref{second} we end up with first-order terms in $\epsilon$, which then belong to \eqref{first}. Similarly first-order terms in $\epsilon$ in \eqref{first} end up in \eqref{zero}. It is then easy to verify that the entire equation \eqref{MW1} is gauge invariant although the separate terms in \eqref{orders} are not.
\subsubsection{Equations of motions in the gauge $\varphi_\mu k^\mu=\epsilon\alpha$}
Imposing the gauge condition $k^\mu \varphi_\mu=\epsilon\alpha$ on \eqref{MW1} yields the equation
\begin{equation}
\left[\square \varphi_\nu-\nabla^\mu \nabla_\nu \varphi_\mu-\nabla_\nu\alpha+\frac{\ii}{\epsilon}(2k^\mu\nabla_\mu \varphi_\nu+\varphi_\nu\nabla_\mu k^\mu-k_\nu(\nabla_\mu \varphi^\mu-\alpha))-\frac{1}{\epsilon^2}k^2\varphi_\nu\right]\ee^{\ii\theta/\epsilon}=0.
\end{equation}
We now demand that the solutions for $\varphi_\mu$ be independent of $\epsilon$ in the limit when $\epsilon$ is small. Physically this means that for high frequencies the form of the solutions should be independent of the frequency (parameterized by $\epsilon$). Consequently, each separate order of $\frac{1}{\epsilon}$ in the expansion must be zero. The equations corresponding to the first and second orders then read
\begin{subequations}\label{orders1}
\begin{align}
&2k^\mu\nabla_\mu \varphi_\nu+\varphi_\nu\nabla_\mu k^\mu-k_\nu(\nabla_\mu \varphi^\mu-\alpha)=0\label{firstorder}\\
&k^\mu k_\mu=0\label{secondorder}
\end{align}
\end{subequations}
for $\varphi_\nu\neq0$. The zeroth-order equation is to be thought of as `small' in comparison to the higher order terms in $1/\epsilon$ and is therefore ignored and not imposed as an equation of motion. The second equation \eqref{secondorder} is trivially gauge invariant since $k_\mu$ does not transform. The first equation is only gauge invariant up to first-order terms in $\epsilon$. This can be seen by letting $\alpha$ transform as $\alpha\rightarrow\alpha+k^\mu\nabla_\mu\xi$ under a gauge transformation, making use of \eqref{second}, and the fact that $k_\mu$ satisfies the geodesic equation as shown in \eqref{photonintegralcurves}.

%%%%%%%%%%%%%%%%%%%%%%%%%%%%%%%%%%%%
\subsubsection{The derivation of parallel transport and conserved currents}
%%%%%%%%%%%%%%%%%%%%%%%%%%%%%%%%%%%%

Equation \eqref{secondorder} tells us that the wavevector $k_\mu$ is a null vector, and that its integral curves $x^\mu(\lambda)$ defined by $\di x^\mu/\di\lambda\propto k^\mu$ lie on a light cone. Taking the derivative of Eq.\eqref{secondorder} yields
\begin{equation}
\nabla_\nu (k^\mu k_\mu)=2k^\mu\nabla_\nu k_\mu\equiv2k^\mu\nabla_\nu\nabla_\mu\theta= 2k^\mu\nabla_\mu\nabla_\nu\theta=2k^\mu\nabla_\mu k_\nu=0 \label{photonintegralcurves}
\end{equation}
which tells us that the integral curves are null geodesics.\footnote{We have assumed in \eqref{zero} that the spacetime torsion is zero. A non-zero torsion field could possibly influence the polarization (see \cite{Bergmann}).} These are expected since we have considered Maxwell's equations in vacuum. Non-geodesic trajectories can be obtained by introducing a medium through which the photon propagates. See \ref{jerk} for a discussion.

Contracting equation \eqref{firstorder} with $\bar\varphi_v$ and adding to it its complex conjugate yields the continuity equation
\begin{eqnarray}
2\bar\varphi^\nu k^\mu\nabla_\mu \varphi_\nu+2\varphi^\nu k^\mu\nabla_\mu \bar\varphi_\nu+2\bar\varphi^\nu \varphi_\nu\nabla_\mu k^\mu=-2\nabla_\mu(\varphi^2k^\mu)=0\label{photoncontinuity}
\end{eqnarray}
where $\varphi^2\equiv-g^{\mu\nu}\bar\varphi_\mu \varphi_\nu$. Note that $\varphi^2$ is gauge invariant up to first-order terms in $\epsilon$, i.e. $\varphi^2\rightarrow \varphi^2+\mathcal{O}(\epsilon)$, and therefore also $j^\mu\equiv\sqrt{g}\varphi^2k^\mu$ is gauge invariant to first order. This means that $j^\mu$ is a conserved current in the WKB limit. Since $j^0$ has the units of a probability density\footnote{We recall that $A_\mu$ has dimensions $L^{-1}$ in natural units with $e=1$.} we can interpret $j^\mu$ as a conserved probability density current.

We can also deduce that $\nabla_\mu k^\mu=-\frac{2}{\varphi}k^\mu\nabla_\mu \varphi$ and if we insert this in equation \eqref{firstorder} and define the polarization vector $\psi_\nu$ through $\varphi_\nu\equiv \varphi\psi_\nu$, we obtain
\begin{align*}
2k^\mu\nabla_\mu \varphi_\nu+\varphi_\nu\nabla_\mu k^\mu-k_\nu(\nabla_\mu \varphi^\mu-\alpha)&=2k^\mu\nabla_\mu \varphi_\nu-\varphi_\nu\frac{2}{\varphi}k^\mu\nabla_\mu \varphi-k_\nu(\nabla_\mu \varphi^\mu-\alpha)\\&=2\varphi k^\mu\nabla_\mu \psi_\nu-k_\nu(\nabla_\mu \varphi^\mu-\alpha)=0
\end{align*}
which implies that
\begin{eqnarray*}
k^\mu\nabla_\mu \psi_\nu=\left(\frac{\nabla_\mu \varphi^\mu-\alpha}{2\varphi}\right)k_\nu.
\end{eqnarray*}
However, since $\alpha$ is arbitrary the whole right-hand side is arbitrary and we can write
\begin{eqnarray}\label{betatrans}
k^\mu\nabla_\mu \psi_\nu=\beta k_\nu.
\end{eqnarray}
The right-hand side is proportional to the wavevector $k_\nu$ and represents an arbitrary infinitesimal gauge transformation of $\psi_{\mu}$. Let us now introduce the integral curves of $u^\mu=k^\mu/E$ given by $\di x^{\mu}/\di \lambda=u^{\mu}$ where $E$ is an arbitrary constant with dimensions of energy. We can then write equation \eqref{betatrans} as
\begin{eqnarray}\label{photonPT}
\frac{D\psi_{\mu}}{D\lambda}=\beta u_{\mu}.
\end{eqnarray}
%s
Thus the transport of the polarization vector $\psi_\mu$ is given by the parallel transport along the null geodesic integral curves of $u_{\mu}$, with an arbitrary infinitesimal gauge transformation at each instant.

%%%%%%%%%%%%%%%%%%%%%%%%%%%%%%%%%%%%%%%%%%%%%%%%
\subsection{Localization of the qubit}\label{photonlocalization}
%%%%%%%%%%%%%%%%%%%%%%%%%%%%%%%%%%%%%%%%%%%%%%%%

As in the fermion case, the WKB approximation is not enough to guarantee either that the wavepacket is localized or that it stays localized under evolution, and again it is not possible to achieve strict  localization. Indeed it can be proved that a photon must have non-vanishing sub-exponential tails \cite{Hegerfeldt,Birula}. As in the case of fermions, \S\ref{fermionlocalization}, we are going to ignore these small tails and treat the wavepacket as effectively having compact support within some small region much smaller than the typical curvature scale.

The continuity equation \eqref{photoncontinuity} dictates the evolution of the envelope $\varphi(x)$. Divided by the energy as measured in some arbitrary frame it becomes $\nabla_\mu(\varphi^2u^\mu)=0$. Again we see that the assumption $\nabla_{\mu}u^{\mu}=0$ simplifies this equation. However, the interpretation of $\nabla_{\mu}u^{\mu}$ is a bit different. Instead of quantifying how much a spatial volume element is changing (as in \S\ref{fermionlocalization}), it quantifies how much an area element, transverse to $u^{\mu}$ in some arbitrary reference frame, changes \cite{Poisson}:
\begin{eqnarray*}
\nabla_{\mu}u^\mu=\frac1A\frac{\di A}{\di\lambda}
\end{eqnarray*}
where $\lambda$ is an affine parameter defined by $\di x^{\mu}/\di \lambda = u^{\mu}$. In this case we require that $\ex{\nabla_{\mu}u^\mu} \ll 1/\lambda_{\Gamma}$, where ${\lambda_{\Gamma}}$ is the affine length of the trajectory $\Gamma$. Thus, it gives us a measure of the transverse distortion of a wavepacket. For photons there can be no longitudinal distortion since all components, regardless of frequency, travel with the speed of light. Initial localization and the assumption that $\nabla_{\mu}u^{\mu}\approx0$ therefore guarantee that the wave-packet is rigidly transported along the trajectory.

Once we assume that the polarization vector $\psi_{I}$ does not vary spatially within the wavepacket we can effectively describe the system as a polarization vector $\psi_\mu(\lambda)$ for each $\lambda\in\Gamma$. Having effectively suppressed the spatial degrees of freedom of the wavepacket, the polarization $\psi_\mu$ can thus be thought of as a function defined on a classical trajectory $\Gamma$, satisfying an ordinary differential equation \eqref{photonPT}. A photonic qubit can then be characterized by a position $x^\mu(\lambda)$, a wavevector $k^\mu(\lambda)$, and a spacelike complex-valued polarization vector $\psi_\mu(\lambda)$.

%%%%%%%%%%%%%%%%%%%%%%%%%%%%%%%%%%%%%%%%%%%%%%%%%%%%%%%%
\subsection{A summary of WKB limit}
%%%%%%%%%%%%%%%%%%%%%%%%%%%%%%%%%%%%%%%%%%%%%%%%%%%%%%%%

To summarize, the WKB approximation yields the following results and equations:
\begin{itemize}
\item The integral curves $x^\mu(\lambda)$ of $u_\mu$ are null geodesics
\item The vector $j^\mu=\sqrt{g}\varphi^2 k^\mu$ is a conserved probability density current.
\item The polarization vector $\psi_\mu$ satisfies $\psi_{\mu} u^\mu=0$ and transforms as $\psi_{\mu}\rightarrow\psi_{\mu}+\upsilon u_\mu$ under gauge transformation up to first-order terms in $\epsilon$.
\item The transport of $\psi_\mu$ is governed by \eqref{photonPT} which is simply the parallel transport along integral curves of $u^\mu$ modulo gauge transformations.
\end{itemize}
We have now established a formalism for the quantum state of a localized qubit which is invariant under $\psi_\mu\rightarrow \psi_\mu+\upsilon u_\mu$ and $\psi_\mu u^\mu=0$ up to first-order terms in $\epsilon$. We shall from this point on  neglect the small terms of order $\epsilon$.

%%%%%%%%%%%%%%%%%%%%%%%%%%%%%%%%%%%%%%%%%%%%%%%%%%%%%%%%%%%%%%%%%%%%%%%%%%
\subsection{The quantum state\label{secphotonQS}}
%%%%%%%%%%%%%%%%%%%%%%%%%%%%%%%%%%%%%%%%%%%%%%%%%%%%%%%%%%%%%%%%%%%%%%%%%%
We now show that the polarization 4-vector has only two complex degrees of freedom and in fact it can be taken to encode a two-dimensional quantum state. We do this first with a tetrad adapted to the velocity of the photon for simplicity and then with a general tetrad. It is convenient and more transparent to work with tetrad indices instead of the ordinary tensor indices and we shall do so here.

%%%%%%%%%%%%%%%%%%%%%%%%%%%%%%%%%%%%%%%%%%%%%%%5%
\subsubsection{Identification of the quantum state with an adapted tetrad\label{sec-adaptedphotonQS}}
%%%%%%%%%%%%%%%%%%%%%%%%%%%%%%%%%%%%%%%%%%%%%%%%%

Recall from the previous section that we partially fixed the gauge to $u_{I} \psi^{I}=0$. The remaining gauge transformations are of the form $\psi^{I}\rightarrow \psi^{I}+\upsilon u^{I}$. Indeed, if $u_{I} \psi^{I}=0$ we also have that  $u_{I} (\psi^{I}+\upsilon u^{I})=0$ for all complex-valued functions $\upsilon$, since $u^I$ is null.

To illustrate in more detail what effect this gauge transformation has on the polarization vector we adapt the tetrad reference frame $e^\mu_I$ to the direction of the photon so that $u^\mu\propto e^\mu_0+e^\mu_3$. Notice that there are several choices of tetrads that put the photon 4-velocity into this standard form. The two-parameter family of transformations relating these different tetrad choices are (1) spatial rotations around the $z$-axis and (2) boosts along the $z$-axis.

With a suitable parameterization of the photon trajectory such that $e^0_\mu(\di x^\mu/\di\lambda)=1$ we can eliminate the proportionality factor and we have $u^\mu=e^\mu_0+e^\mu_3$. In tetrad components $u^I=(1,0,0,1)$ and we see that the tetrad $z$-component $e^\mu_3$ is aligned with the photon's 3-velocity. Since $0=u_{I}\psi^{I}=\psi^{0}-\psi^{3}$ it follows that $\psi^{0}=\psi^{3}=\nu$ and the polarization vector can be written as
\begin{eqnarray*}
\psi^{I}=\begin{pmatrix}\nu\\\psi^1\\ \psi^2\\\nu\end{pmatrix}.
\end{eqnarray*}
It is clear that a gauge transformation
\begin{eqnarray*}
\psi^I=\begin{pmatrix}\nu\\\psi^1\\ \psi^2\\\nu\end{pmatrix}\rightarrow \psi^I+\upsilon u^I=\begin{pmatrix}\nu+\upsilon\\\psi^1\\ \psi^2\\\nu+\upsilon\end{pmatrix}
\end{eqnarray*}
leaves the two middle components unchanged and changes only the zeroth and third components. The two complex components $\psi^1$ and $\psi^2$, which form the Jones vector \cite{Hecht}, therefore represent gauge invariant true degrees of freedom of the polarization vector whereas the zeroth and third components represent pure gauge.

We can now identify the quantum state as the two gauge invariant middle components $\psi^1$ and $\psi^2$, where $\psi^1$ is the horizontal and $\psi^2$ the vertical component of the quantum state in the linear polarization basis:
\begin{eqnarray*}
|1\rangle\sim\left(\begin{array}{c}0\\1\\ 0\\0\end{array}\right),\qquad |2\rangle\sim \left(\begin{array}{c}0\\0\\ 1\\0\end{array}\right)
\end{eqnarray*}
or simply $|A\rangle\sim \delta^I_{A}$ with $A=1,2$. The quantum state is then
\begin{eqnarray*}
\ket{\psi}\sim\psi^{A} = \delta^{A}_{I}\psi^{I}  = \begin{pmatrix}\psi^1\\\psi^2\end{pmatrix}.
\end{eqnarray*}
Note we have deliberately used a notation similar to that used for representing spinors; however, $\psi^{A}$ should not be confused with an $SL(2,\mathbb C)$ spinor.  In order to distinguish $\psi^A$ from $\psi^I$ we will refer to the former as the Jones vector and the latter as the polarization vector.

%%%%%%%%%%%%%%%%%%%%%%%%%%%%%%%%%%%%%%%%%%%%%%%5%
\subsubsection{Identification of the quantum state with a non-adapted tetrad\label{sec-photonqsident}}
%%%%%%%%%%%%%%%%%%%%%%%%%%%%%%%%%%%%%%%%%%%%%%%%%

\begin{figure}[h]
\begin{center}
\ifpdf
\includegraphics[height=30mm]{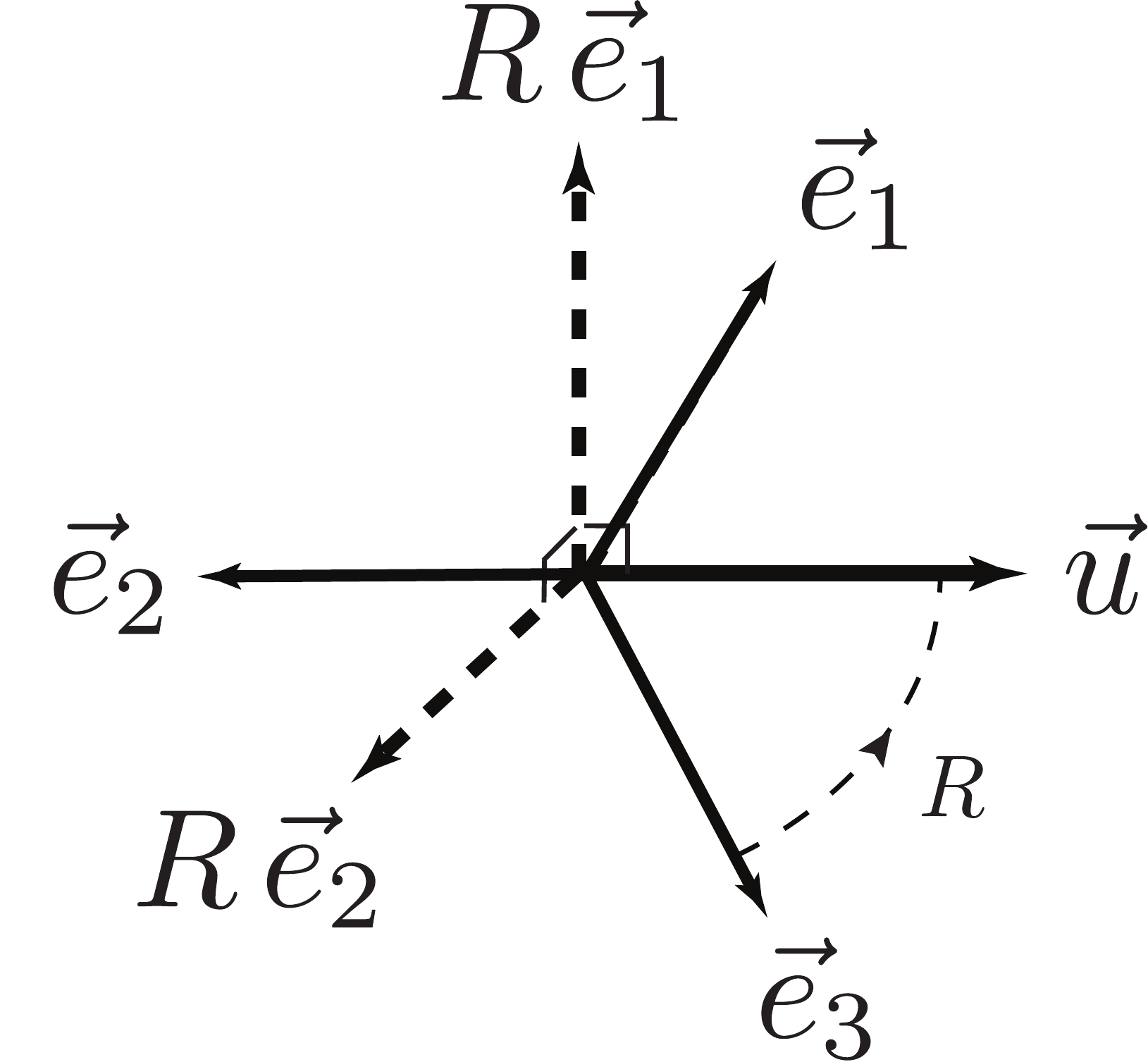}
\else
\includegraphics[height=30mm]{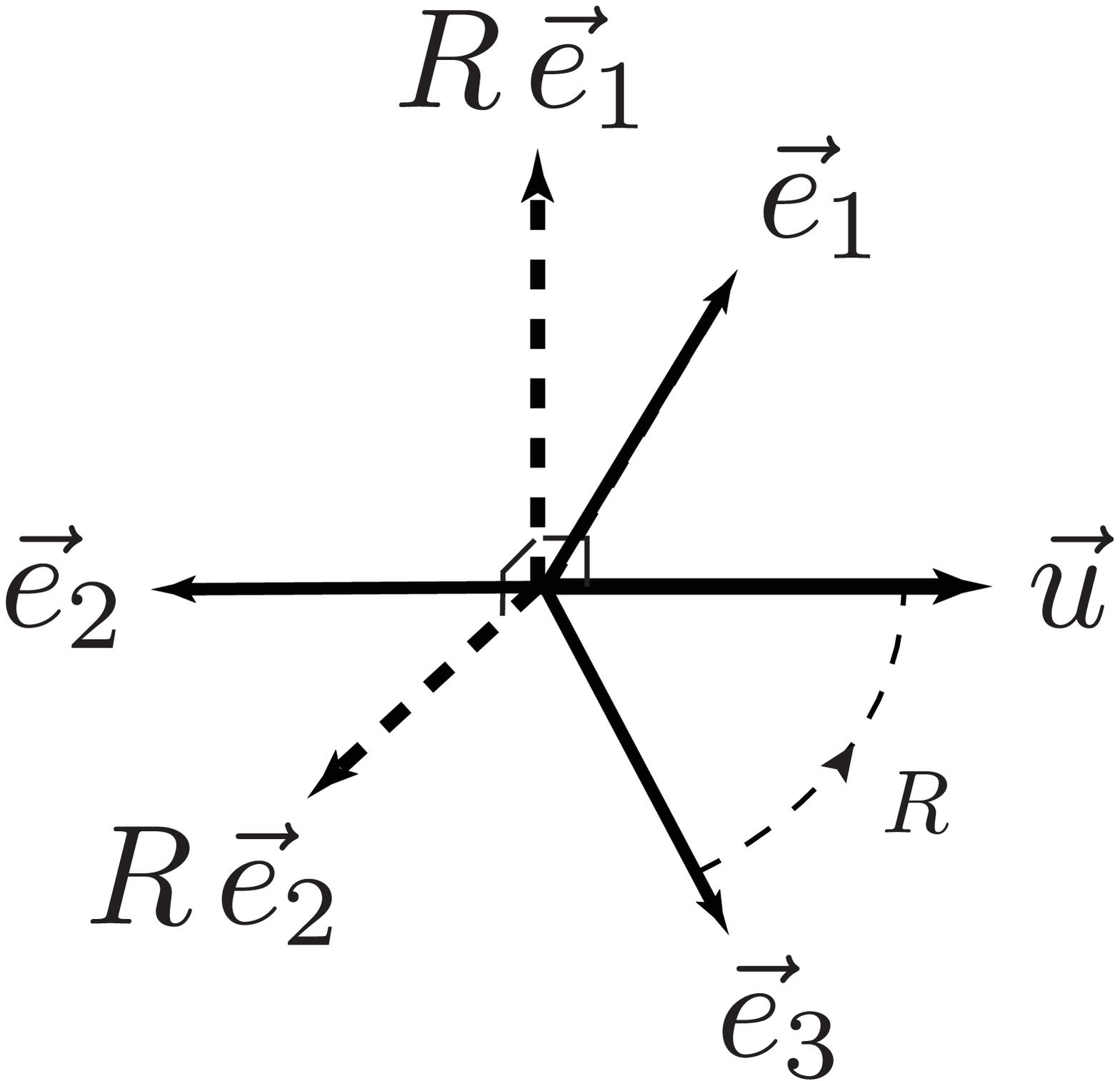}
\fi
\caption{The rotation $R$ adapts the spacelike vectors of the tetrad $\vec e_i$ so that the z-axis $\vec e_{3}$ is aligned to the 3-velocity of the photon $\vec u$. A polarization vector is then in the plane spanned by $R\vec e_1$ and $R\vec e_2$.\label{adaption}}
\end{center}
\end{figure}

In the above discussion we have used an adapted tetrad in order to identify the quantum state. We can write a map for this adaption explicitly, which will provide a generic non-adapted formalism. To adapt one simply introduces a rotation which takes the 4-velocity $u^{I}$ to the standard form \cite{Weinberg}
\begin{eqnarray*}
u^I\to u^{\prime I} = R^{I}_{\ J}u^{J}=\begin{pmatrix}1\\0\\0\\1\end{pmatrix}
\end{eqnarray*}
which results in the tetrad being aligned with the photon's 3-velocity, as illustrated in figure \ref{adaption}.

Such a rotation is explicitly given by
\begin{eqnarray*}
R^{I}_{\;J}(u)=\delta_0^I\delta_J^0 -\hat r^{I}\hat r_{J}-\frac{u_3}u(P^{I}_{\;J}+\hat r^I\hat r_J) -\sqrt{1-\left(\frac{u_3}u\right)^2}\eps^{I}_{KJ0}\hat r^K
\end{eqnarray*}
where $u_3/u=u_\mu e^\mu_3/u=-\cos\theta$ is the angle between the direction of the photon and the z-component of the tetrad, $r^I\equiv \eps^I_{J30}u^J$ is the spatial axis of rotation with $\hat r^{I} \equiv r^{I}/|r|$, and $P^I_{\ J}\equiv \delta^I_J-e^I_0e^0_J$ is the projector onto the spacelike hypersurface orthogonal to the tetrad time axis (see Fig.\ref{adaption}).

It is important to stress that there are several other possible choices for this spatial rotation corresponding to different conventions for the linear polarization basis. Furthermore, the rotation matrix above becomes undefined for $\theta=\pi$ which is unavoidable for topological reasons.

The rotation $R^{I}_{\;J}$ induces a linear polarization basis $\delta^{A}_{I}R^{I}_{\;J}$. We can now extract the components of the quantum state expressed in this basis as
\begin{eqnarray}
\psi^A= f^A_J\psi^J\text{,\quad with\quad }f^A_J\equiv\delta^{A}_{I}R^{I}_{\;J}.\label{qs-adapt-pol}
\end{eqnarray}
It is clear that the specific linear polarization basis used here depends on how we have adapted the tetrad to the velocity of the photon. However, regardless of what convention one chooses, the quantum state $\psi^A$ is gauge invariant. Alternatively we could think of the quantum state directly in terms of an equivalence class of polarization vectors $\psi^I\sim \psi^I+\upsilon u^I$ orthogonal to photon velocity $u^I$.  The advantage of this approach is that once one has developed a gauge invariant formalism one need not work with the cumbersome two component Jones vector, but instead can work solely with the gauge covariant polarization vector $\psi^I$. This will be addressed below and in Section \ref{sec-photonunitaritymmt}.

$f^A_I$ from \eqref{qs-adapt-pol} turns out to provide a `diad' frame: The two vectors $f^1_I$ and $f^2_I$ span the two-dimensional space orthogonal to both the photon's 4-velocity $u_I$ and the time component of a tetrad $e^t_I$. If we let $f^I_A$ be the inverse of $f^A_I$ we have that $f^I_Af^A_J = \delta^I_J$ and $f^I_Au_I = f^I_Ae^t_I=0$. In fact, if we define $w_I$ to be a null vector defined by $e^t_I=\half(u_I+w_I)$ \cite{Poisson}, the vectors $u_I$, $w_I$, $f^1_I$ and $f^2_I$ span the full tangent space. This decomposition will be useful when identifying unitary operations in Section \ref{sec-photonunitaritymmt}.

%%%%%%%%%%%%%%%%%%%%%%%%%%%%%%%%%%%%%%%%%%%%%%%
\subsection{The inner product\label{sec-photonIP}}
%%%%%%%%%%%%%%%%%%%%%%%%%%%%%%%%%%%%%%%%%%%%%%%

We must identify an inner product on the complex vector space for polarization so that it can be promoted to a Hilbert space. In the analysis of the WKB limit we found that $j^{I} = -\sqrt{g}k^{I}\varphi^2\eta_{JK}\psi^{J}\bar{\psi}^{K}$ corresponded to a conserved 4-current which was physically interpreted as a conserved probability density current. A natural inner product between two polarization 4-vectors $\psi^{I}$ and $\phi^{J}$ is then given by
\begin{eqnarray}\label{photonIP}
-\eta_{IJ}\bar\phi^{I}\psi^{J}.
\end{eqnarray}
This form is clearly sesquilinear and positive definite for spacelike polarization vectors.\footnote{There is no primed index for conjugate terms because the vector representation of the Lorentz group is real.} Unlike the case for fermions, the inner product $\eta_{IJ}$ is not explicitly dependent on the photon 4-velocity. However, if we consider the gauge transformation $\psi^{I}\to\psi^{I}+\upsilon_1 u^{I}_{1}$ and $\phi^{I}\to\phi^{I}+\upsilon_2 u^{I}_{2}$ it is clear that unless $u^{I}_{1}=u^{I}_{2}$, i.e. $k^{I}_{1}\propto k^{I}_{2}$, the inner product \eqref{photonIP} is not gauge invariant. We conclude that two polarization vectors corresponding to two photons with non-parallel null velocities do not lie in the same Hilbert space. Furthermore, in order to be able to {\em coherently} add two polarization states it is  also necessary to have $k_1^I=k_2^I$, i.e. the two photons must have the same frequency. Under such conditions the inner product is both Lorentz invariant and gauge invariant. With the inner product \eqref{photonIP} the complex vector space of polarization vectors is promoted to a Hilbert space which is notably labelled again with both position and 4-momentum $p^I=\hbar k^I$.

The above inner product \eqref{photonIP} reduces to the standard inner product for a two-dimensional Hilbert space. This is best seen through the use of an adapted tetrad. In an adapted frame the inner product of $\psi^I=(\nu,\psi^1,\psi^2,\nu)$ with some other polarization vector $\phi^I=(\mu,\phi^1,\phi^2,\mu)$ is given by
\begin{eqnarray*}
-\eta_{IJ}\bar{\phi}^I\psi^J=-\bar{\mu}\nu+\bar{\phi}_{1}\psi^1+\bar{\phi}_{2}\psi^2+ \bar{\mu}\nu=\bar{\phi}_1\psi^1+\bar{\phi}_{2}\psi^2=\langle\phi|\psi\rangle.
\end{eqnarray*}
Thus, the standard inner product $\langle\phi|\psi\rangle=\bar{\phi}_{1}\psi^1+\bar{\phi}_{2}\psi^2$ is simply given by $\bk\phi\psi=-\eta_{IJ}\bar{\phi}^I\psi^J$, where we associate
\begin{eqnarray*}
\ket{\psi} \sim \psi^{I}\quad \mbox{and}\quad \bra{\phi} \sim -\bar\phi_{I} = -\eta_{IJ}\bar\phi^{J}.
\end{eqnarray*}
We can now work directly with the polarization 4-vector $\psi^{I}$ which transforms in a manifestly Lorentz covariant and gauge covariant manner.

%%%%%%%%%%%%%%%%%%%%%%%%%%%%%%%%%%%%%%%%%%%%%%%%%%%%%%%
\subsection{The relation to the Wigner formalism\label{secphotonWigner}}
%%%%%%%%%%%%%%%%%%%%%%%%%%%%%%%%%%%%%%%%%%%%%%%%%%%%%%%

The Wigner rotation $W_A^{\ B}(k,\Lambda)$ on the quantum state represented by the Jones vector which results from the transport of the polarization vector can be identified in the same way as was done in \S\ref{fermionwignerrot} for fermions. Specifically, this is achieved by determining the evolution of the Jones vector $\psi^{A}$ that is induced by the transport of the quantum state represented by the polarization 4-vector $\psi^{I}$. Substituting $\psi^{I} = f^{I}_{B} \psi^{B}$ in the transport equation \eqref{photonPT}, $\frac{\di \psi^{I}}{\di \lambda} + u^{\mu}\omega_{\mu \ J}^{\ I}\psi^{J}= \beta u^{I}$, and multiplying by $f^A_I$, we obtain
\begin{eqnarray}
\frac{\di \psi^{A}}{\di \lambda} = - \left(u^{\mu}e_{I}^{A}\omega^{\ I}_{\mu\ J} f^{J}_{B} +f^A_I\frac{\di f^I_B}{\di\lambda}\right)\psi^B + \beta f_{I}^{A} u^{I}.\label{photonqstransport}
\end{eqnarray}
The last term is zero, as $f^A_I$ is the diad frame defined to be orthogonal to $u^{I}$. If we first consider \eqref{photonqstransport} in an adapted tetrad as in \S\ref{sec-adaptedphotonQS} we see that the derivative $\frac{\di f_B^I}{\di\lambda}$ vanishes and the remaining term on the right-hand side can be simplified to
\begin{eqnarray}
\frac{\di \psi^{A}}{\di \lambda} = \ii u^{\mu}\omega_{\mu 12}{\sigma^y}^A_{\ B} \psi^{B}\label{adaptedPQST}
\end{eqnarray}
where we have made use of the antisymmetry of the spin-1 connection in order to introduce the antisymmetric  Pauli Y matrix ${\sigma^y}^A_{\ B}$.\footnote{Note again that this should not be confused with the  $\sigma$-matrices encountered when working with spinors.} Eq.\eqref{adaptedPQST} is then clearly unitary and helicity preserving as it is proportional to ${\sigma^y}^A_{\ B}$ in the linear polarization basis. We can now readily identify the infinitesimal Wigner rotation as $W^A_{\ B} = \ii u^{\mu}\omega_{\mu 12} {\sigma^y}^A_{\ B} $, where the rotation angle is $u^{\mu}\omega_{\mu 12}$. In a non-adapted tetrad frame the map $f^I_A=\delta^J_AR_J^{\ I}$ can be seen to put \eqref{photonqstransport} in the form \eqref{adaptedPQST} with a modified spin-1 connection $\omega'^{\ I}_{\mu\ J}$. The Wigner rotation for non-adapted tetrads is then
\begin{eqnarray}
W^A_{\ B}=\ii u^\mu(R^{\ I}_1\partial_\mu R^{2}_{\ I}+R_1^{\ I}\omega_{\mu I}^{\ \ J} R^{2}_{\ J}){\sigma^y}^A_{\ B}.\label{eq-genphotonWigner}
\end{eqnarray}
A general Wigner rotation is understood as the composition of maps $W^A_{\ B}\equiv f_{B}^{I}(\Lambda u) \Lambda_{I}^{\ J}f_{J}^{A}(u)$. It is therefore no surprise that the transport of the polarization vector induces a Wigner rotation: The action of the gravitational field along a trajectory is simply a sequence of infinitesimal Lorentz transformations which are given by $u^{\mu}\omega_{\mu\ J}^{\ I}$ (\S\ref{localqubits}). The transport of the Jones vector is therefore described by a sequence of infinitesimal Wigner rotations given by \eqref{eq-genphotonWigner}. Notice that the Wigner rotation takes on a form which is not manifestly Lorentz covariant. This is because the Wigner rotation describes a {\it spatial} rotation. This should be contrasted with the manifestly Lorentz covariant representation in terms of parallel transported polarization vectors. Furthermore, after we have developed a measurement formalism in section \ref{polarizationmeasurement} it will become clear that there is no need to work with the cumbersome Wigner rotations.

%%%%%%%%%%%%%%%%%%%%%%%%%%%%%%%%%%%%%%%%%%%%%%%%%%%%%%%%%%%%%%%%%%%
\section{Phases and interferometry}\label{secPhase}
%%%%%%%%%%%%%%%%%%%%%%%%%%%%%%%%%%%%%%%%%%%%%%%%%%%%%%%%%%%%%%%%%%%%

So far we have determined the transport of the quantum state of a single qubit along one spacetime trajectory. If we inspect the transport equations \eqref{fermionTP} and \eqref{betatrans} we see that neither one contains a term proportional to the identity ($\delta_A^B$ for fermions and $\delta^I_J$ for photons). Such a term would lead to an overall accumulation of global phase $\ee^{\ii\theta}\psi_A$ or $\ee^{\ii\theta}\psi^I$. This leads one to suspect that not all of the possible contributions to the global phase have been taken into account in these transport equations. Indeed this is the case, as can be seen immediately by considering the full wavepacket in the WKB approximation
\begin{eqnarray*}
\Psi_\sigma(x)= \varphi(x)\psi_\sigma(x)\ee^{\ii\theta(x)}
\end{eqnarray*}
where $\varphi(x)$ is a real-valued envelope and $\sigma=1,2$ for fermions or $\sigma=0,1,2,3$ for photons.\footnote{In the case of a scalar particle (and thus not a qubit) there are still gravitational phases, and here the index $\sigma$ can just be removed.} Clearly there is an additional phase $\theta(x)$ which is not included in $\psi_\sigma(x)$.

Since global phase is unobservable this is of course of no concern if we restrict ourselves to a qubit moving along a single trajectory. However, quantum mechanics allows for more exotic experiments where a single qubit is split up into a spatial superposition, simultaneously transported along multiple distinct paths, and recombined so as to produce quantum interference phenomena. Here it becomes necessary to keep track of the phase difference between the components of the spatial superposition in order to be able to predict the measurement probabilities at the detectors.

In this section we will extend the formalism in this paper to include gravitationally induced phase difference in experiments involving path superpositions. The formalism will be derived from equations of the WKB approximation together with the assumptions of localization. With these assumptions, the details of the spatial profile of the qubit become irrelevant, and we can satisfactorily describe the experiment solely in terms of a phase difference $\Delta \theta$ between two quantum states. This phase difference will depend on the spacetime geometry $g_{\mu\nu}$ and the trajectories along which the qubit is simultaneously transported. We show how the various sources for the phase difference can be understood from a wave-geometric picture. We lastly apply the formalism of this paper to gravitational neutron interferometry \cite{Colella,Werner,Sakurai} and obtain an exact general relativistic expression for the phase difference which in various limits reproduces the results in \cite{Anandan,AudretschLammerzahl,VarjuRyder} in which higher order corrections to the non-relativistic result were proposed.

%%%%%%%%%%%%%%%%%%%%%%%%%%%%%%%%%%%%%%%%%%%%%%%%%%%%%%%%%%%%%%%%%%%%
\subsection{Spacetime Mach--Zehnder interferometry}\label{phasedef}
%%%%%%%%%%%%%%%%%%%%%%%%%%%%%%%%%%%%%%%%%%%%%%%%%%%%%%%%%%%%%%%%%%%%

\begin{figure}[h]
\begin{center}
\ifpdf
\includegraphics[height=60mm]{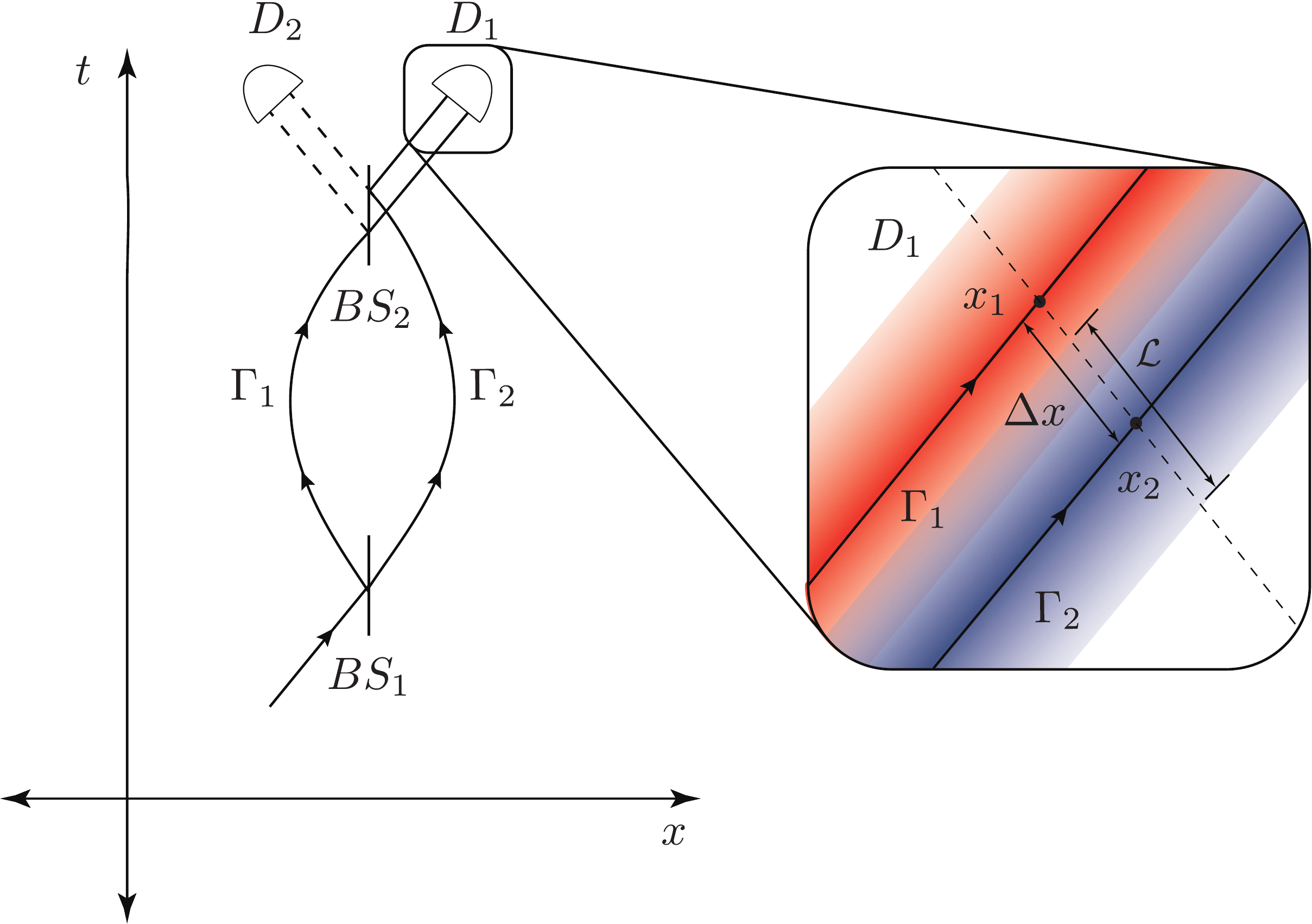}
\else
\includegraphics[height=60mm]{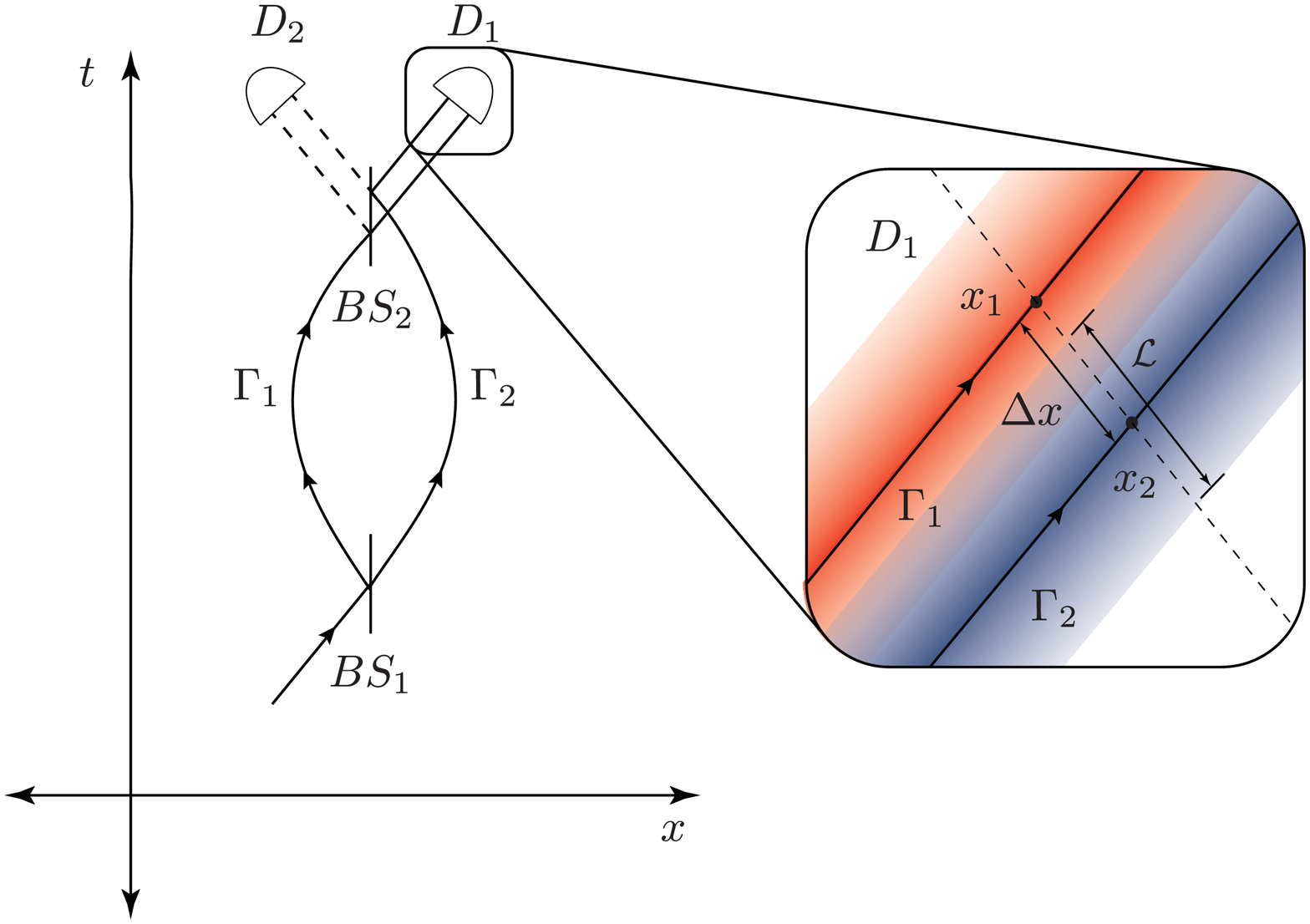}
\fi
\caption{Spacetime figure of a Mach--Zehnder type interferometer, illustrating a single qubit subjected to a beam splitter $BS_1$ resulting in a superposition of the qubit travelling along two distinct spacetime paths $\Gamma_1$ and $\Gamma_2$. In some future spacetime region containing the beam splitter $BS_2$ the components of the spatial superposition are assumed to recombine to produce possible interference phenomena in the detector regions $D_1$ and $D_2$. These regions contain two trajectories, indicating that the times of arrival at the second beam splitter $BS_2$ are not in general the same. The variables $x_{1,2}$ are arbitrary spacetime points in the region $D_1$ along trajectories $\Gamma_1$ and $\Gamma_2$ and are useful for calculating the total phase difference. The red and blue strips represent the spatial extents of the wavepackets along $\Gamma_{1}$ and $\Gamma_{2}$ respectively. These correspond to the length $\m L$ of the wavepacket as measured along the line joining the points $x_{1}$ and $x_{2}$.
\label{fig-spacetimeMZ}}
\end{center}
\end{figure}

We consider, as a concrete example of an interference experiment, standard Mach--Zehnder interferometry. As usual, there is a qubit incident on a beam splitter (e.g. a half-silvered mirror) which creates a spatial superposition of the qubit. The two components of the spatially superposed state (each assumed to be spatially well-localized) are then transported along two different paths and later made to interfere using another beam splitter. This produces two output rays each incident on a particle detector, as illustrated in figure \ref{fig-spacetimeMZ}.

Let us now focus our attention on a small region $D_1$ situated on the right output arm.
There are then two classical paths $\Gamma_1$ and $\Gamma_2$ which arrive at $D_1$, as illustrated in Fig.\ref{fig-spacetimeMZ}. Note that in order for us to derive a formalism in terms of quantum states, the wavevectors $k^\mu_1$ and $k^\mu_2$ of $\Gamma_1$ and $\Gamma_2$ must be approximately equal in this region, i.e. $k^\mu_1=k^\mu_2=k^\mu$. This is because the Hilbert space of a quantum state is labelled with momentum, as explained in \S\ref{secfermionQS} and \S\ref{sec-photonIP}.

 In general the times of arrival of the two paths $\Gamma_1$ and $\Gamma_2$ at the second beam splitter $BS_{2}$ will differ for the two paths. As we shall see this contributes to the total phase difference between the two packets.

Let $x$ be some suitable local Lorentz coordinate system in region $D_1$. The wavepacket in region $D_1$ is then given by the superposition
\begin{eqnarray}
a\Psi^{(1)}_\sigma(x)+b\Psi^{(2)}_\sigma(x)\label{additionofwavepackets}
\end{eqnarray}
where $\Psi^{(1)}_\sigma(x)$ and $\Psi^{(2)}_\sigma(x)$ are the packets propagated along $\Gamma_1$ and $\Gamma_2$ respectively. $a$ and $b$ are determined from the reflection and transmission coefficients of the various beam splitters in the experiment. In the case of 50-50 beam splitters, $a=b=\frac{\ii}{\sqrt{2}}$ in region $D_1$ (see Fig.\ref{fig-spacetimeMZ}). We will ignore any overall global phase factor resulting from reflections.

%%%%%%%%%%%%%%%%%%%%%%%%%%%%%%%%%%%%%%%%%%%%%%%%%%%%%%%%%%%%%%%%%%%%%%%%%%%%%%%%%%%%%%%%%%%%%%%%%%%%%
\subsection{The phase difference from the WKB approximation \label{addingstates}}\label{phasediff}
%%%%%%%%%%%%%%%%%%%%%%%%%%%%%%%%%%%%%%%%%%%%%%%%%%%%%%%%%%%%%%%%%%%%%%%%%%%%%%%%%%%%%%%%%%%%%%%%%%%%%
In order to make empirical predictions in a Mach--Zehnder type interference experiment we must determine explicitly the forms of $\Psi^{(1)}_\sigma(x)$  and $\Psi^{(2)}_\sigma(x)$ in \eqref{additionofwavepackets} in the detector region $D_{1}$. By making use of the field equations in the WKB limit and the localization assumptions we will see that $\Psi^{(1)}_\sigma(x)$  and $\Psi^{(2)}_\sigma(x)$ will differ by a phase accumulated along the trajectory and a rigid translation/displacement, resulting in an overall phase difference. The derivations differ in the cases of fermions and photons and we will treat them separately.

%%%%%%%%%%%%%%%%%%%%%%%%
\subsubsection{Fermions}
%%%%%%%%%%%%%%%%%%%%%%%%
In the small region $D_1$ the wavepacket in the WKB approximation is given by
\begin{equation}
a\phi^{(1)}_A(x)+b\phi^{(2)}_A(x)=a\varphi_1(x)\psi_A^{(1)}(x)\ee^{\ii\theta_1(x)}+ b\varphi_2(x)\psi_A^{(2)}(x)\ee^{\ii\theta_2(x)}\label{superposwavefunctions}
\end{equation}
where $x$ is some local Lorentz coordinate system, and $a$ and $b$ are real-valued coefficients. The functions $\phi_i(x)$, $\psi^{(i)}_{A}(x)$ and $\theta_i(x)$ ($i=1,2$) are defined in section \ref{fermsemiclassapprox}. We are now going to successively make use of the equations of the WKB approximation and the localization assumptions to simplify the expression \eqref{superposwavefunctions} and thereby extract the relative phase difference between the two components in the superposition.

First we use the fact that under the mathematical assumptions detailed in \S\ref{fermionlocalizationsubsec} the envelope will be transported rigidly and will not distort. Therefore, $\varphi_1(x)$ and $\varphi_2(x)$ will differ at most up to a rigid translation and rotation. We assume that the packet is `cigar shaped' and is always oriented in the direction of motion. The final envelopes will then differ at most up to a translation and we can write $\varphi_i(x)=\varphi(x-x_i)$ for some suitable function $\varphi(x)$ and an arbitrary choice of spacetime points $x^\mu_1,x^\mu_2\in D_1$ situated on the trajectories $\Gamma_1,\Gamma_2$ respectively (see figure \ref{fig-spacetimeMZ}).

If we now assume that
\begin{eqnarray*}
\frac{(x^\mu-x_i^\mu)\nabla_\mu\varphi(x)}{\varphi(x)}\ll1,\qquad i=1,2
\end{eqnarray*}
for all points $x\in\m D_1$, the difference in the envelopes $\varphi_1(x)$ and $\varphi_2(x)$ is negligible. The translational difference in the envelopes can then be neglected and factored out:
\begin{eqnarray}
a\phi^{(1)}_A(x)+b\phi^{(2)}_A(x)\approx\varphi(x)\left(a\psi_A^{(1)}(\Gamma_1)\ee^{\ii\theta_1(x)}+ b\psi_A^{(2)}(\Gamma_2)\ee^{\ii\theta_2(x)}\right)\label{locsuperposwavefunctions}
\end{eqnarray}
where $\psi_A^{(i)}(\Gamma_i)$ are determined by integrating the transport equation \eqref{fermionTP} to the points $x_1$, $x_2$, respectively. Thus for the purpose of interferometry the details of the envelope become irrelevant and can be ignored.

We now focus on the phase $\theta_1$ and $\theta_2$. As we pointed out in section \ref{newinnerproduct}, in order to coherently add two quantum states it is necessary to assume that the wavevectors of the packets are the same, $k_1^\mu=k_2^\mu=k^\mu$. Therefore in the region $D_1$ we have from the WKB approximation that the phases $\theta_1(x)$ and $\theta_2(x)$ both satisfy the equation
\begin{eqnarray}\label{fermphaseq}
\nabla_\mu\theta=k_\mu+e A_\mu.
\end{eqnarray}
Within the small region  $D_1$ we regard $k_\mu(x)$ and $A_\mu(x)$ as constant and so the partial differential equation \eqref{fermphaseq} has the solution $\theta_i(x)=(k_\mu+e A_\mu)(x^\mu-x^\mu_i)+\theta_i(x_i)$ where $\theta_i(x_i)$ are two integration constants corresponding to the value of $\theta_i(x)$ at the points $x_i$. These integration constants can be determined by integrating \eqref{fermphaseq} along the trajectories $\Gamma_{1,2}$ to the positions $x_{1,2}$ respectively, i.e.
\begin{eqnarray}\label{fermintphase}
\theta_i(x_i)=\int_{\Gamma_i} (k_\mu+eA_\mu)\di x^\mu+\theta_0
\end{eqnarray}
 where $\theta_0$ is some arbitrary global phase just before the wavepacket was split up by the first beam splitter. Using the above we can rewrite \eqref{locsuperposwavefunctions} as
\begin{eqnarray}\label{waveaddition}
a\phi^{(1)}_A(x)+b\phi^{(2)}_A(x)\approx\varphi(x)\ee^{\ii\theta_1(x)}\left(a\psi_A^{(1)}(\Gamma_1)+ b\psi_A^{(2)}(\Gamma_2)\ee^{\ii\Delta\theta}\right)
\end{eqnarray}
where
\begin{eqnarray}
\Delta\theta = (k_\mu+e A_\mu)(x_1^\mu-x^\mu_2)+(\theta_2(x_2)-\theta_1(x_1)).\label{fermionphasedifference}
\end{eqnarray}
It is important to note that this phase difference is independent of $x^\mu\in D_1$ as all dependence on $x$ has been factored out in \eqref{waveaddition}. Furthermore, we note that the choice of $x_1$ and $x_2$ is arbitrary and the phase difference $\Delta\theta$ is also independent of this choice. To see this, consider a different choice of positions, $x_1'=x_1+\delta x_1$ and $x_2'=x_2+\delta x_2$ on $\Gamma_1$ and $\Gamma_2$. This results in a change in the integration constants \eqref{fermintphase} of $\theta_i(x_i)\to\theta_i(x_i)+(k_\mu+eA_\mu)\delta x_i^\mu$ which exactly cancels the change in the term $(k_\mu+eA_\mu)(x_1^\mu-x^\mu_2)$ in \eqref{fermionphasedifference}. Therefore $\Delta\theta$ is independent of the arbitrary positions $x_1$ and $x_2$.

Note that $\Delta\theta$ is not the phase difference determined empirically in a Mach--Zehnder type interference experiment. This is because the transported quantum states $\psi_A^{(1)}(\Gamma_1)$ and $\psi_A^{(2)}(\Gamma_2)$ can contain an additional phase difference induced from their specific evolutions on the Bloch sphere. This transport induced phase difference can be determined from \cite{Bengtsson}
\begin{eqnarray}\label{transphase}
\ee^{\ii\Delta\theta_\text{Trans}}=\frac{\langle\psi^{(1)}(\Gamma_1)|\psi^{(2)}(\Gamma_2)\rangle}{|\langle\psi^{(1)}(\Gamma_1)|\psi^{(2)}(\Gamma_2)\rangle|}.
\end{eqnarray}
The region $D_1$ is assumed to be small enough that $\psi^{(1)}_A$ and $\psi^{(2)}_A$ do not vary significantly with changes in $x_1$ and $x_2$. Thus, the {\em total} phase difference $\Delta\theta_\text{Tot}$, which is the quantity that we actually measure in a Mach--Zehnder experiment, is then given by
\begin{eqnarray*}
\Delta\theta_\text{Tot}=\Delta\theta+\Delta\theta_\text{Trans}.
\end{eqnarray*}
This total phase difference $\Delta\theta_\text{Tot}$ can be determined completely from the trajectories $\Gamma_1$ and $\Gamma_2$ and the spacetime geometry $g_{\mu\nu}$ using the transport equation \eqref{fermionTP}. In particular, the phase difference measured by some detector in $D_1$ is independent of the motion of that detector.

Lastly, if we restrict ourselves to measurements that do not probe the spatial profile we can neglect the factor $\varphi(x)\ee^{\ii\theta_1(x)}$  in \eqref{waveaddition}. All contributions to the phase difference are then contained in $\psi_A^{(\ii)}$ and $\Delta\theta$, and so at $D_1$ we are left with the two-dimensional quantum state
\begin{eqnarray*}
\ket{\psi}_\text{recomb}=a\psi_A^{(1)}(\Gamma_1)+ b\psi_A^{(2)}(\Gamma_2)\ee^{\ii\Delta\theta}.
\end{eqnarray*}
Therefore the assumptions that led to \eqref{waveaddition} established a formalism for determining the resulting qubit quantum state in region $D_1$ of a Mach--Zehnder type interferometer.

%%%%%%%%%%%%%%%%%%%%%%%%%%%%%
\subsubsection{Photons}
%%%%%%%%%%%%%%%%%%%%%%%%%%%%%

The derivation of the phase difference for photons follows essentially the same path as that for fermions. The starting point is to consider the wavepacket in the small region $D_1$
\begin{equation}
aA_{(1)}^I(x)+bA_{(2)}^I(x)=a\varphi_1(x)\psi^I_{(1)}(x)\ee^{\ii\theta_1(x)}+ b\varphi_2(x)\psi^I_{(2)}(x)\ee^{\ii\theta_2(x)}\label{superposwavefunctionsphotons}
\end{equation}
where $x$ is some local Lorentz coordinate system, and $a$ and $b$ are real-valued coefficients. The functions $\varphi_i(x)$, $\psi_{(i)}^{I}(x)$ and $\theta_i(x)$ ($i=1,2$) are defined in section \ref{photongeoapprox}. We then make use of the equations of the WKB approximation and the localization assumptions to simplify the expression \eqref{superposwavefunctionsphotons}. As in the case of fermions, this means that the envelopes $\varphi_1(x)$ and $\varphi_1(x)$ are rigidly transported along their respective trajectories and so they differ at most by a translation i.e. $\varphi_i(x)=\varphi(x-x_i)$.

Again, if we assume the change in the envelope is small
\begin{eqnarray*}
\frac{(x^\mu-x_i^\mu)\nabla_\mu\varphi(x)}{\varphi(x)}\ll1,\qquad i=1,2
\end{eqnarray*}
for all points $x\in\m D_1$, the translational difference in the envelopes can be neglected and can be factored out:
\begin{eqnarray*}
aA_{(1)}^I(x)+bA_{(2)}^I(x)\approx\varphi(x)\ee^{\ii\theta_1(x)}\left(a\psi^I_{(1)}(\Gamma_1)+ b\psi^I_{(2)}(\Gamma_2)\ee^{\ii\Delta\theta(x)}\right)
\end{eqnarray*}
where $\Delta\theta = \theta_{2}(x)-\theta_{1}(x)$. We then solve the partial differential equation $ \nabla_{\mu}\theta=k_{\mu}$ to determine
\begin{eqnarray}\label{photonphase}
\theta_i(x)=k_\mu(x^\mu-x^\mu_i)+\theta_{i}(x_{i})+\theta_0
\end{eqnarray}
where $\theta_{i}(x_{i}) = \int_{\Gamma_i} k_\mu\di x^\mu$ are again integration constants. Using that $k_{\mu}$ is null and that we are integrating along its integral curves, we have $\int_{\Gamma_i} k_\mu\di x^\mu\equiv0$. Thus, the only contribution to the phase difference is
\begin{eqnarray}
\Delta\theta  =k_\mu(x_1^\mu-x^\mu_2).\label{photonsphasedifference}
\end{eqnarray}
Again note that this phase difference is independent of the position $x^\mu\in D_1$ at which the phase difference is computed. We also have that the phase difference $\Delta\theta$ is independent of the choice of points  $x_1$ and $x_2$. This follows since a change $x^{\mu}_i\to x^{\prime\mu}_i = x^{\mu}_i+\delta x^{\mu}_i=  x^{\mu}_i+\epsilon_{i} k^{\mu}$ leaves $\Delta\theta$ invariant since $k_\mu\delta x_i^\mu=0$.

As in the fermionic case there is also a phase difference $\Delta\theta_\text{Trans}$ defined by \eqref{transphase} related to the transport along the trajectories. What is actually measured in a Mach--Zehnder interference experiment is then
\begin{eqnarray*}
\Delta\theta_\text{Tot}=\Delta\theta+\Delta\theta_\text{Trans}.
\end{eqnarray*}
Just as in the case for fermions we now neglect the spatial part and we end up with the final qubit quantum state at $D_1$
\begin{eqnarray*}
\ket{\psi}_\text{recomb}=a\psi^I_{(1)}(\Gamma_1)+ b\psi^I_{(2)}(\Gamma_2)\ee^{\ii\Delta\theta(x)}
\end{eqnarray*}
We have now obtained a formalism for describing interference experiments for photons solely in terms of two-dimensional quantum states.
%%%%%%%%%%%%%%%%%%%%%%%%%%%%%
\subsubsection{The recipe for adding qubit states}\label{recipe}
%%%%%%%%%%%%%%%%%%%%%%%%%%%%%%%
Above we have established a formalism for quantum interference phenomena for both fermions and photons in a Mach--Zehnder interference experiment. This description can be summarized by the following recipe for correctly adding the two quantum states:
\begin{enumerate}
\item Transport the quantum states $\psi_\sigma^{(1)}$ and $\psi_\sigma^{(2)}$ to the arbitrary positions $x_1$ and $x_2$ on the respective paths $\Gamma_1$ and $\Gamma_2$ in the recombination region $D_1$ using the appropriate transport equation, either \eqref{fermionTP} for fermions or \eqref{photonPT} for photons.
\item determine the integration constants $\theta_1$ and $\theta_2$. For fermions this is determined by equation \eqref{fermintphase}. For photons this is identically zero.
\item determine the phase difference $\Delta\theta$ using either \eqref{fermionphasedifference} or \eqref{photonsphasedifference}.
\item Finally, the two-dimensional quantum state in region $D_1$ is given by
\begin{eqnarray*}\ket{\psi}_\text{recomb}=a\psi_\sigma^{(1)}+ b\psi_\sigma^{(2)}\ee^{\ii\Delta\theta}.
\end{eqnarray*}
\end{enumerate}

%%%%%%%%%%%%%%%%%%%%%%%%%%%%%%%%%%%%%%%%%%%%%
\subsection{The physical interpretation of phase in terms of wave geometry}
%%%%%%%%%%%%%%%%%%%%%%%%%%%%%%%%%%%%%%%%%%%%%
We now provide an intuitive wave-geometric picture for the various terms in the phase difference $\Delta\theta$. To do this we will focus on the specific case of fermions. Note however that the essential picture is also applicable to photons and we will comment on photons when necessary. The phase difference for fermions \eqref{fermionphasedifference} is given by
\begin{align}
\Delta\theta=&(k_\mu+eA_\mu)\Delta x^\mu + \int_{\Gamma_2} (k_\mu+eA_\mu)\di x^\mu-\int_{\Gamma_1} (k_\mu+eA_\mu)\di x^\mu \nonumber\\
=& \oint_\Gamma k_\mu \di x^\mu+e\oint_\Gamma A_\mu\di x^\mu
\end{align}
where $\Delta x^\mu\equiv x_1^\mu-x_2^\mu$ and $\Gamma= \Gamma_2+\Gamma_{2\to 1}-\Gamma_1$, where $\Gamma_{2\to 1}$ denotes the straight path going from point $x^\mu_2$ to $x^\mu_1$. The second term in the integral accounts for an Aharonov--Bohm phase. Let us consider the first term. The various contributions to this term are
\begin{eqnarray}
\oint_{\Gamma} k_\mu\di x^\mu = \int_{\Gamma_2} k_\mu\di x^\mu-\int_{\Gamma_1} k_\mu\di x^\mu + k_\mu\Delta x^\mu. \label{massphase}
\end{eqnarray}
The first two terms in the decomposition can each be thought of as representing the accumulation of global phase  along each trajectory, while the third is related to the displacement of the wavepackets. We now show how to interpret these two contributions wave-geometrically.

%%%%%%%%%%%%%%%%%%%%%%%%%%%%%%%%%%%%%%%%%%%%%
\subsubsection{The internal phase shift}
%%%%%%%%%%%%%%%%%%%%%%%%%%%%%%%%%%%%%%%%%%%%%

The first two terms in \eqref{massphase} are integrals of the wavevector $k^\mu$ along the paths $\Gamma_i$, $i=1,2$. If we parameterize the paths with proper time $\frac{\di x^\mu}{\di\tau}=\frac{\hbar}{m}k^\mu$, the integrals become
\begin{eqnarray}\label{pathintphase}
\int_{\Gamma_i} k_\mu\di x^\mu= \int_{\Gamma_i}\di\tau \frac{mc^2}{\hbar}.
\end{eqnarray}
\begin{figure}[h]
a) $t_1$ \hspace{6cm} b) $t_2$\\
  \centering
\ifpdf
\includegraphics[height=25mm]{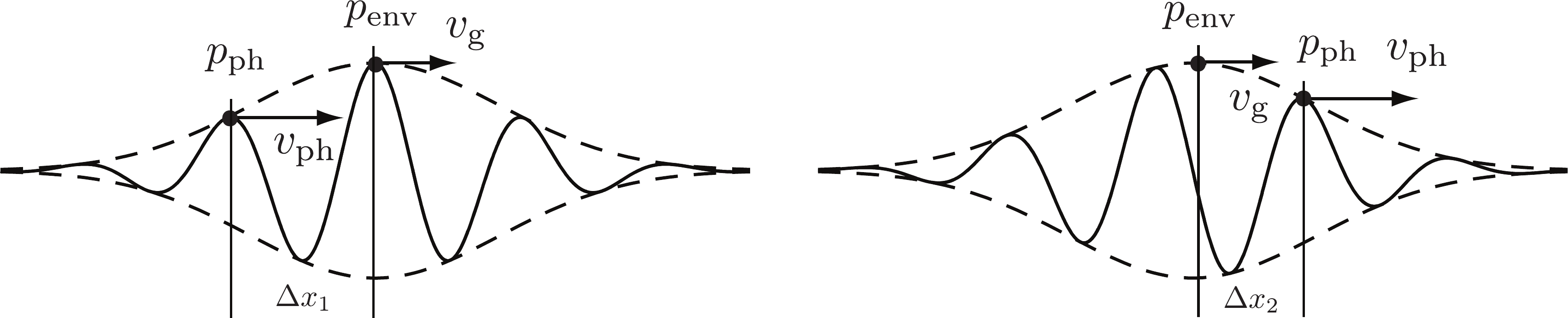}
\else
\includegraphics[height=25mm]{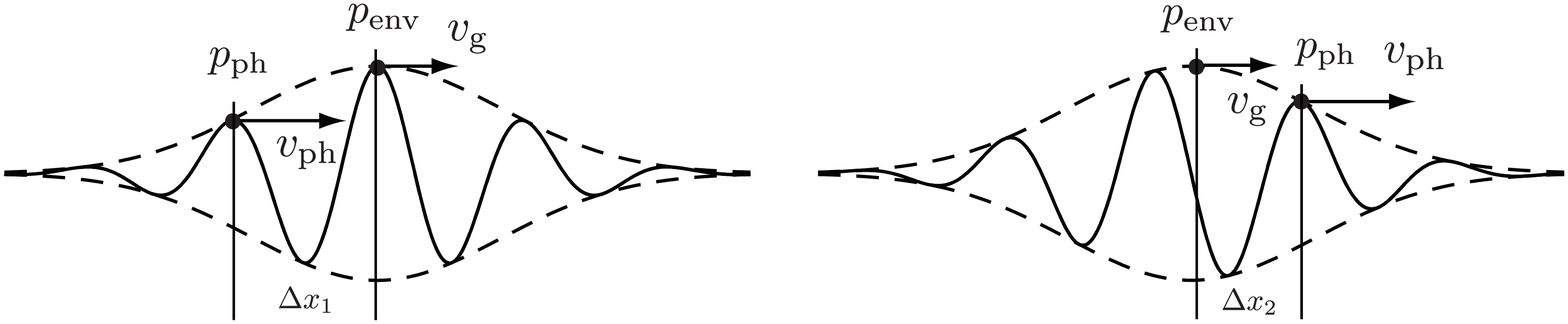}
\fi
\caption{\small An illustration of the accumulation of internal phase from a time $t_1$ (a) to a later time $t_2$ (b) along a trajectory. The internal phase $\theta_\text{int}=\Delta x_\text{int} /\lambdabar$ is determined by the difference in the offset $\Delta x_\text{int} = \Delta x_{2} - \Delta x_{1}$ of a point $p_\text{ph}$ of constant phase and a point $p_\text{env}$ of constant position on the envelope at the two times $t_1$ and $t_2$. For timelike packet velocities the phase velocity $v_\text{ph}$ is greater than the group velocity $v_\text{g}=c^2/v_\text{ph}<c$ so the internal phase is seen to accumulate along the trajectory.
\label{fig:inthaseoffsets}}
\end{figure}
This results in a phase discussed in \cite{Stodolsky,Alsing} and motivated from the relativistic path integral. We can also understand this term in a simple wave-geometric picture. Consider a point $p_{\text{env}}(\tau)$ defined by $\varphi(x)=\text{const}$ which is fixed on the rigidly moving envelope (see figure \ref{fig:inthaseoffsets}). With respect to some arbitrary reference frame with 4-velocity $n^\mu$, the velocity at which $p_{\text{env}}$ moves in this frame is called the {\em group velocity} $v_\text{g}$ (see e.g. \cite{Rindler}), defined by $u^0=e^0_\mu u^\mu=\gamma_{v_\text{g}}=(1-v_\text{g}^2/c^2)^{-1/2}$, and corresponds to the particle's velocity. Secondly, consider a fixed phase point $p_{\text{ph}}$ defined by $\theta(x)=\text{const}$. The speed at which this phase point moves is given by $v_\text{ph} \equiv c^2/v_{\text{g}}$ and is called the {\em phase velocity}. Thus, if $v_{\text{g}}<c$ the points of constant phase move with respect to the wavepacket. It is this difference in velocity that results in the accumulation of the above mentioned path integral phase \eqref{pathintphase}. To see this, first calculate how much distance $\delta x_\text{int}$ is gained by $p_{\text{ph}}$ relative to $p_{\text{env}}$ during some time interval $\di t$ measured in this reference frame. This is given by
\begin{eqnarray*}
\delta x_\text{int} = p_{\text{ph}}-p_{\text{env}} = \left(\frac{c^2}{v_{\text{g}}}-v_{\text{g}}\right)\di t.
\end{eqnarray*}
In order to see how many radians of phase this distance is equivalent to we divide by the reduced wavelength $\lambdabar\equiv \hbar/p =  \hbar/m v_\text{g}$;
\begin{eqnarray*}
\frac{\delta x_\text{int} }{\lambdabar}=\frac{\left(\frac{c^2}{v_{\text{g}}}-v_\text{g}\right)\di t}{\lambdabar}=\frac{\frac{c^2}{v_\text{g}}\gamma^{-2}\di t m\gamma v_\text{g}}{\hbar}=\frac{mc^2}{\hbar}\gamma^{-1}\di t=\frac{mc^2}{\hbar}\di\tau.
\end{eqnarray*}
During a finite period of time we have $\theta_\text{int}\equiv\int\di\tau mc^2/\hbar=\Delta x_\text{int}/\lambdabar$, which is nothing but the path integral phase. We can now interpret the path integral phase as how much the constant phase surfaces have shifted inside the wavepacket. We call this an {\em internal phase shift} $\theta_\text{int}$. When we add two wavepackets it is important to keep track of this phase shift as it may lead to destructive or constructive interference. $\theta_\text{int}$ calculated for each trajectory is simply the integration constants $\theta_{i}(x_{i})$ (e.g. \eqref{fermintphase}).

Recall that for photons there was no contribution to the phase from the path integral, i.e. the integration constants are $\theta_{i}(x_{i})=0$. From a wave-geometric picture this is due to the fact that the group and phase velocities are equal and therefore $\theta_\text{int} = 0$ .

%%%%%%%%%%%%%%%%%%%%%%%%%%%%%%%%%%%%%%%%%%%%%
\subsubsection{The displacement induced phase difference}
%%%%%%%%%%%%%%%%%%%%%%%%%%%%%%%%%%%%%%%%%%%%%

Let us now provide a wave-geometric interpretation for the third term in \eqref{massphase}, $k_\mu \Delta x^\mu$. First, for simplicity let $\Delta x^\mu = x_1^\mu-x^\mu_2$ be spacelike and orthogonal to some arbitrary unit timelike vector $n^\mu$, i.e. $\Delta x^\mu = h^\mu_{\ \nu} \Delta x^\nu$ where $h^\mu_{\ \nu} = \delta^\mu_\nu-n^\mu n_\nu$ projects onto the orthogonal space of $n^\mu$.\footnote{$\Delta x^\mu$ can always be made spacelike orthogonal by changing the arbitrary end points of the trajectories $\Gamma_{1}$ and $\Gamma_{2}$.} $k_\mu \Delta x^\mu$ then simplifies to:
\begin{eqnarray*}
|k_\mu \Delta x^\mu| =|k_\mu h^\mu_{\ \nu} \Delta x^\nu| = k_{\perp} \Delta x_\text{dis} |\cos(\alpha)|=\frac{\Delta x_\text{dis}}{\lambdabar}
\end{eqnarray*}
where $k_\perp = \sqrt{-h^{\mu\nu} k_\mu k_\nu}$ and $\Delta x_\text{dis} = \sqrt{-h_{\mu\nu} \Delta x^\mu\Delta x^\nu}$, and we have used that $|\cos(\alpha)|=1$ since the wavepackets are spatially displaced in the direction of motion, i.e. $k_{\perp}^\mu \propto \Delta x^\mu$.

\begin{figure}[h]
  \centering
  \ifpdf
    \includegraphics[height=40mm]{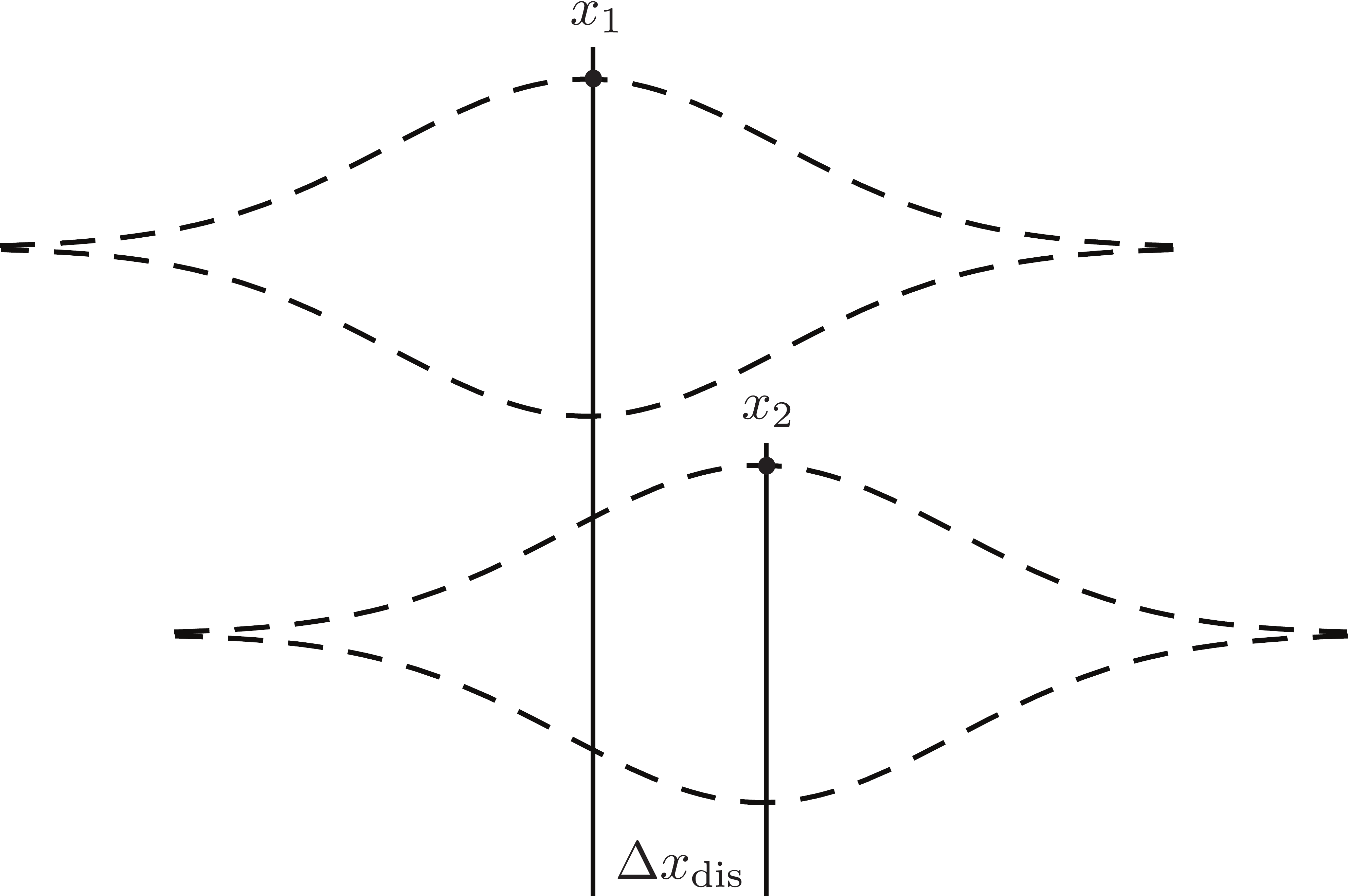}
    \else
  \includegraphics[height=40mm]{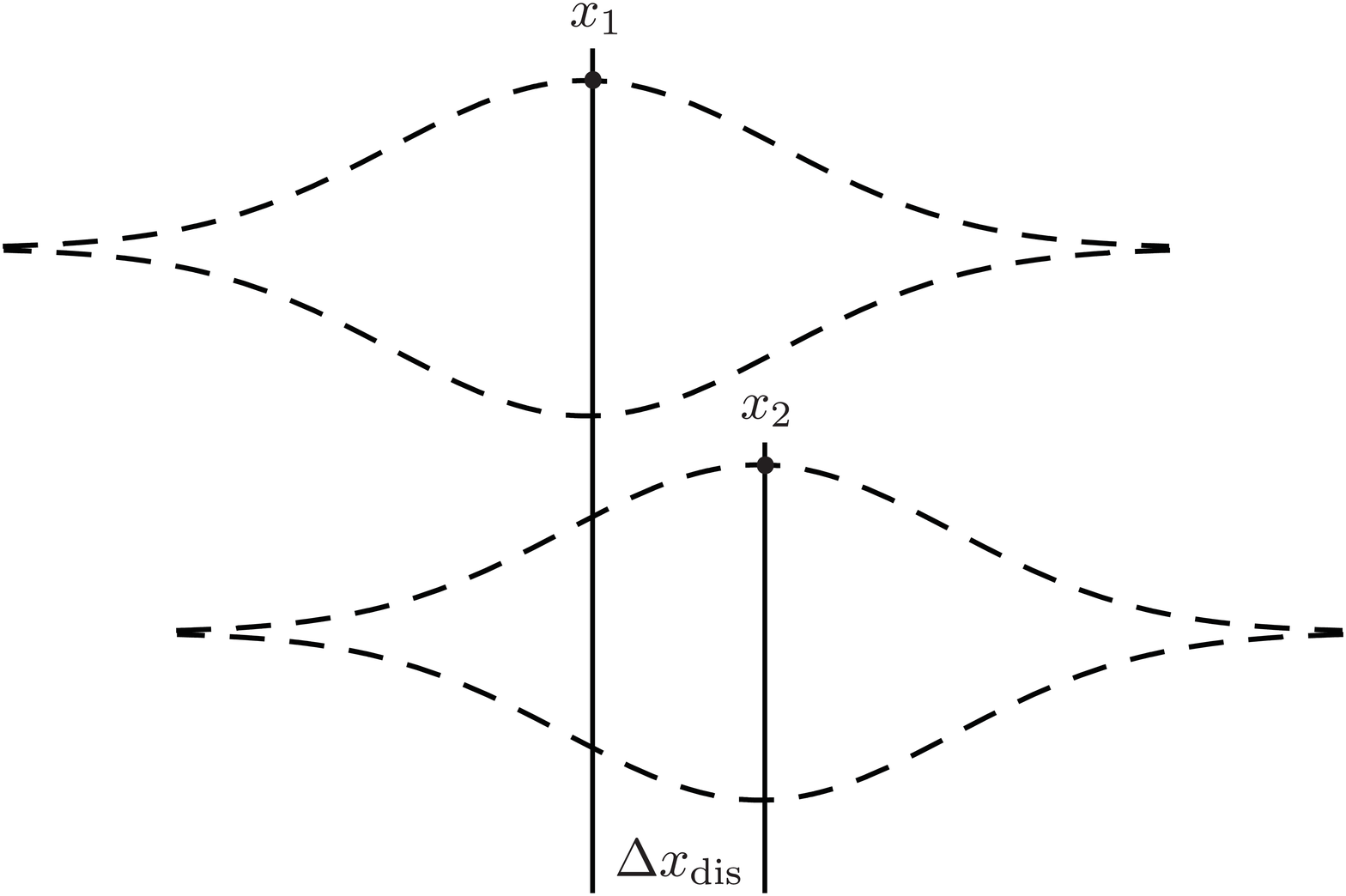}
  \fi
  \caption{\small Illustration of the recombination of two envelopes in the detector region $D_{1}$. The offset of the classical positions $x_1$, $x_2$ of two wavepackets with the same wavelength $\lambdabar$ produces a displacement phase of $\Delta\theta_\text{dis}=\Delta x_\text{dis}/\lambdabar$.
  \label{fig:Phaseoffsets}}
\end{figure}
This contribution to the phase difference, which is present for both fermions and photons, can therefore be interpreted as the two wavepackets being spatially displaced, as illustrated in figure \ref{fig:Phaseoffsets} and as argued in \cite{Mannheim}. Note that in order for this {\em displacement induced phase difference} $\Delta\theta_\text{dis}=k_\mu\Delta x^\mu$ to be experimentally, the variance in $\Delta x_\text{dis}$ over runs of an interference experiment must be significantly smaller than the wavelength $\lambdabar$. Furthermore, if $\Delta x_\text{dis} \sim \m L$  (see Fig.\ref{fig-spacetimeMZ}) then the interference effects will be drastically reduced and when $\Delta x_\text{dis} \geq \m L$ no interference phenomena will be present.

%%%%%%%%%%%%%%%%%%%%%%%%%%%%%%%%%%%%%%%%%%%%%%%%%%%%%%%%%
\subsubsection{Addition of quantum states in the wave-geometric picture}
%%%%%%%%%%%%%%%%%%%%%%%%%%%%%%%%%%%%%%%%%%%%%%%%%%%%%%%%%

\begin{figure}[h]
\centering
\small
\ifpdf
\includegraphics[height=48mm]{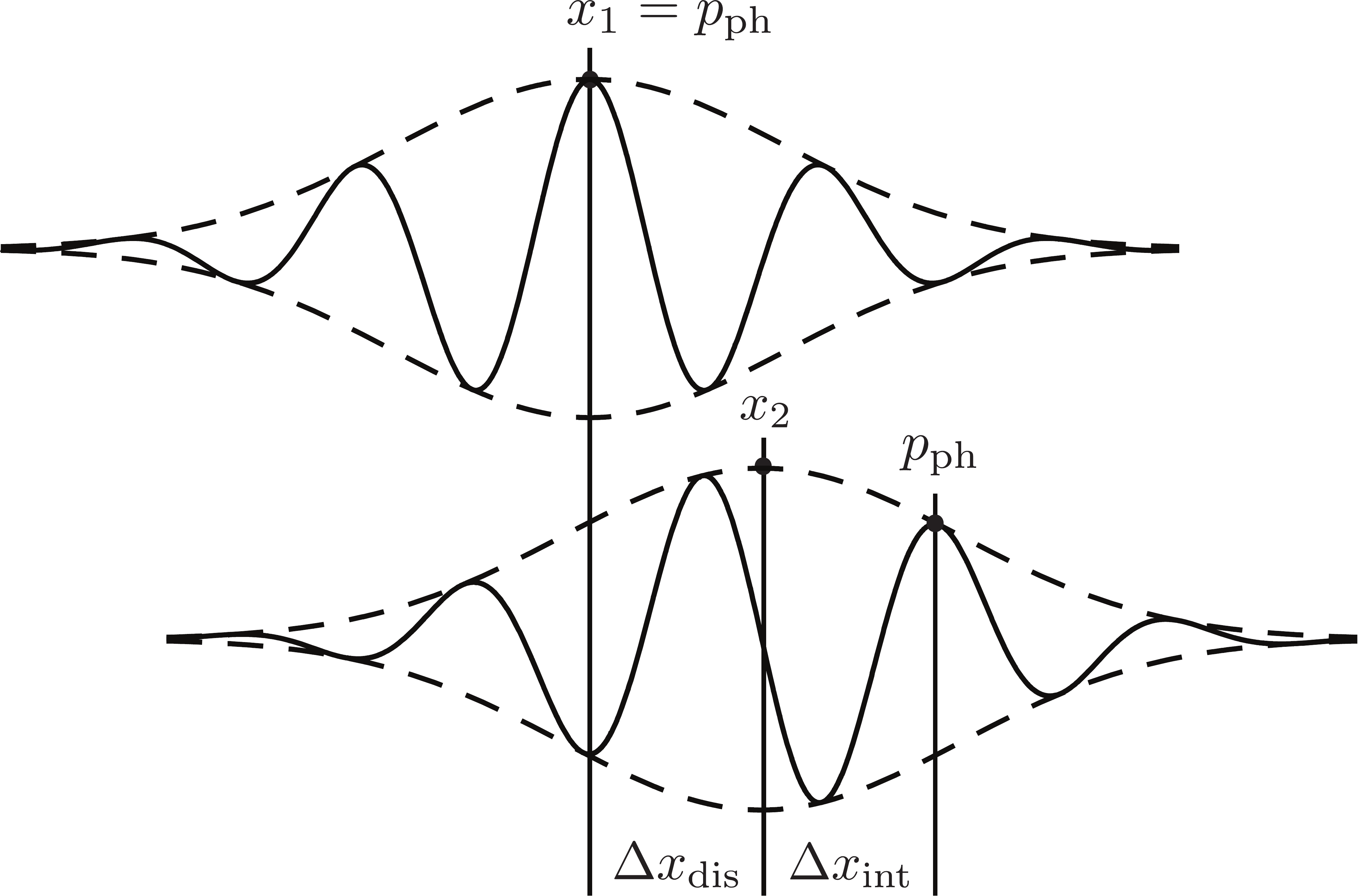}
\else
\includegraphics[height=48mm]{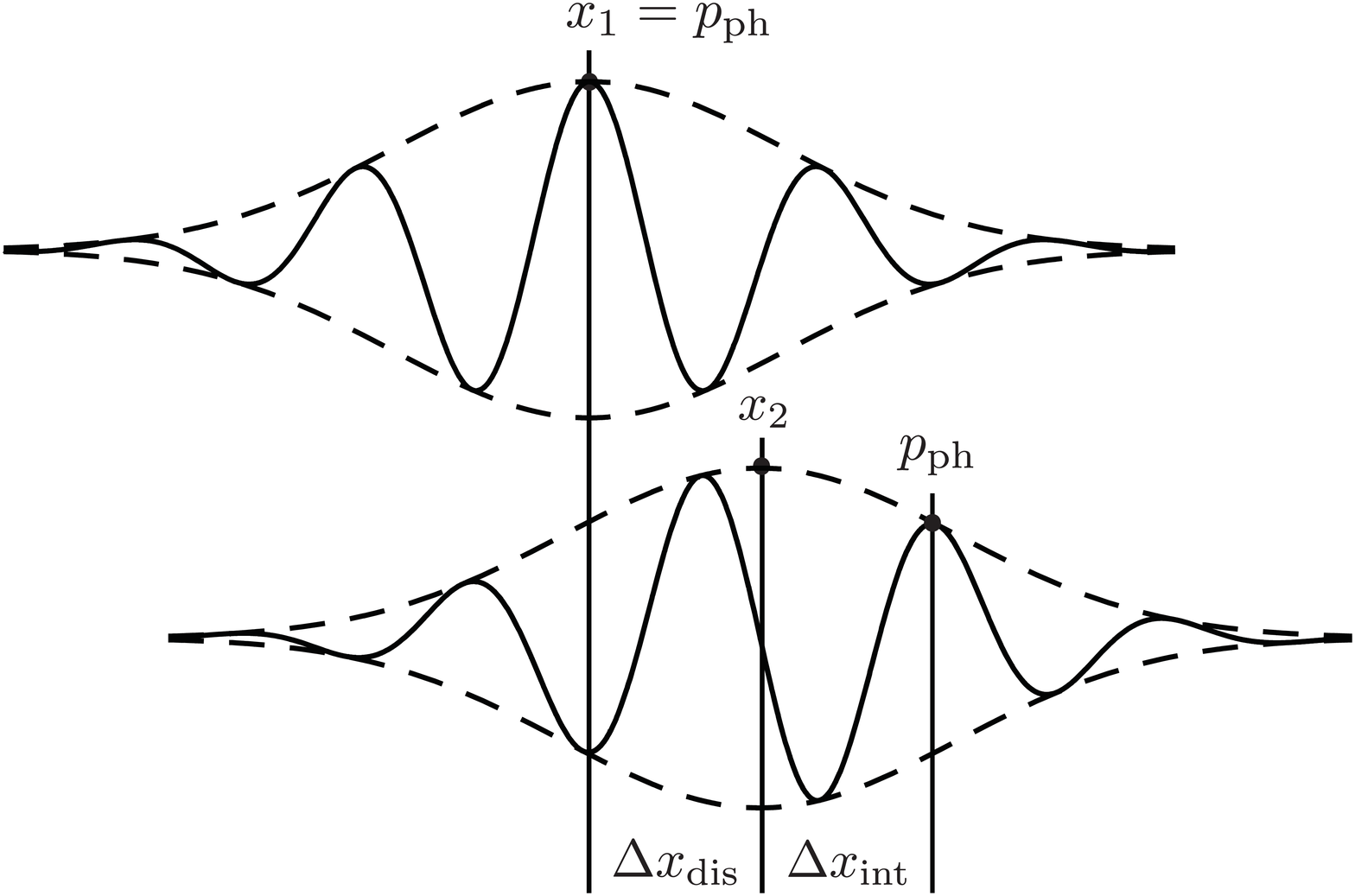}
\fi
\caption{From a wave-geometric point of view we can understand the phase difference $\Delta\theta$ as the sum of two contributions: the difference in the {\em internal phase} given by $\Delta\theta_\text{int}=\Delta x_\text{int}/\lambdabar$ and a phase difference $\Delta\theta_\text{dis}=\Delta x_\text{dis}/\lambdabar$ originating from the wavepackets being spatially displaced.\label{fig-spacewaveaddition}}
\end{figure}

The recipe for adding two quantum states \S\ref{recipe} can now readily be understood in terms of wave geometry. In the detector region $D_{1}$ we have two wavepackets whose envelopes, centred at $x_1$ and $x_2$, overlap but are slightly offset  (as in figure \ref{fig:Phaseoffsets}). Furthermore, each wavepacket has a rapidly oscillating phase that has evolved in a path-dependent way along each of the two distinct trajectories $\Gamma_{1}$ and $\Gamma_{2}$ (as in figure \ref{fig:inthaseoffsets}). These two effects produce respectively the displacement induced phase difference and the internal phase difference. We then add these wavepackets to obtain the total phase difference $\Delta\theta$. This is illustrated in figure \ref{fig-spacewaveaddition}. The total phase difference is again  $\Delta\theta_\text{Tot} = \Delta\theta_\text{Trans.}+ \Delta\theta$.

%%%%%%%%%%%%%%%%%%%%%%%%%%%%%%%%%%%%%%%%%%%%%%%%%%%%%%%%%%%%%
\subsection{An example: relativistic neutron interferometry}\label{neutronphase}
\begin{figure}[h]
\center
\ifpdf
\includegraphics[height=50mm]{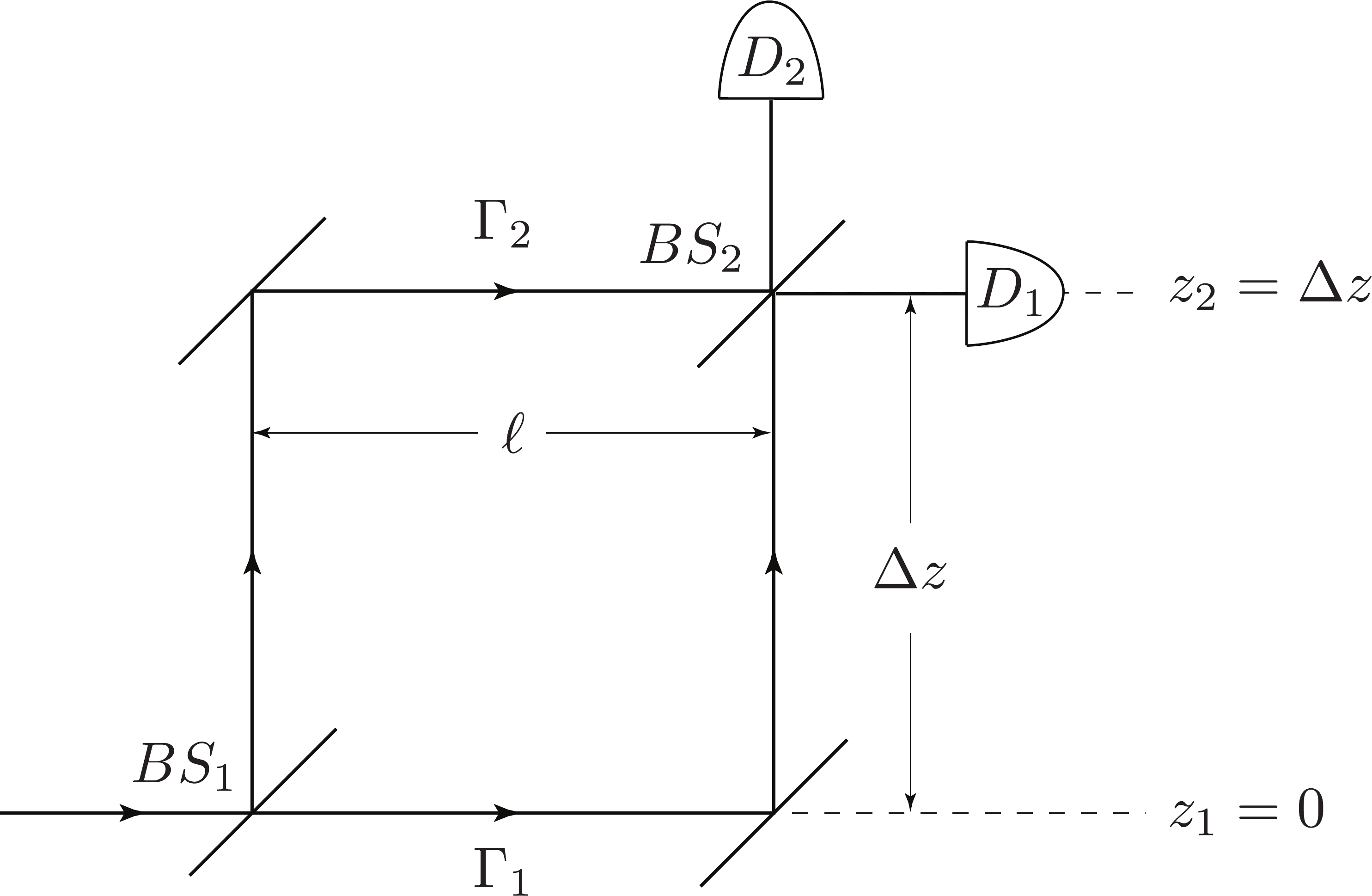}
\else
\includegraphics[height=50mm]{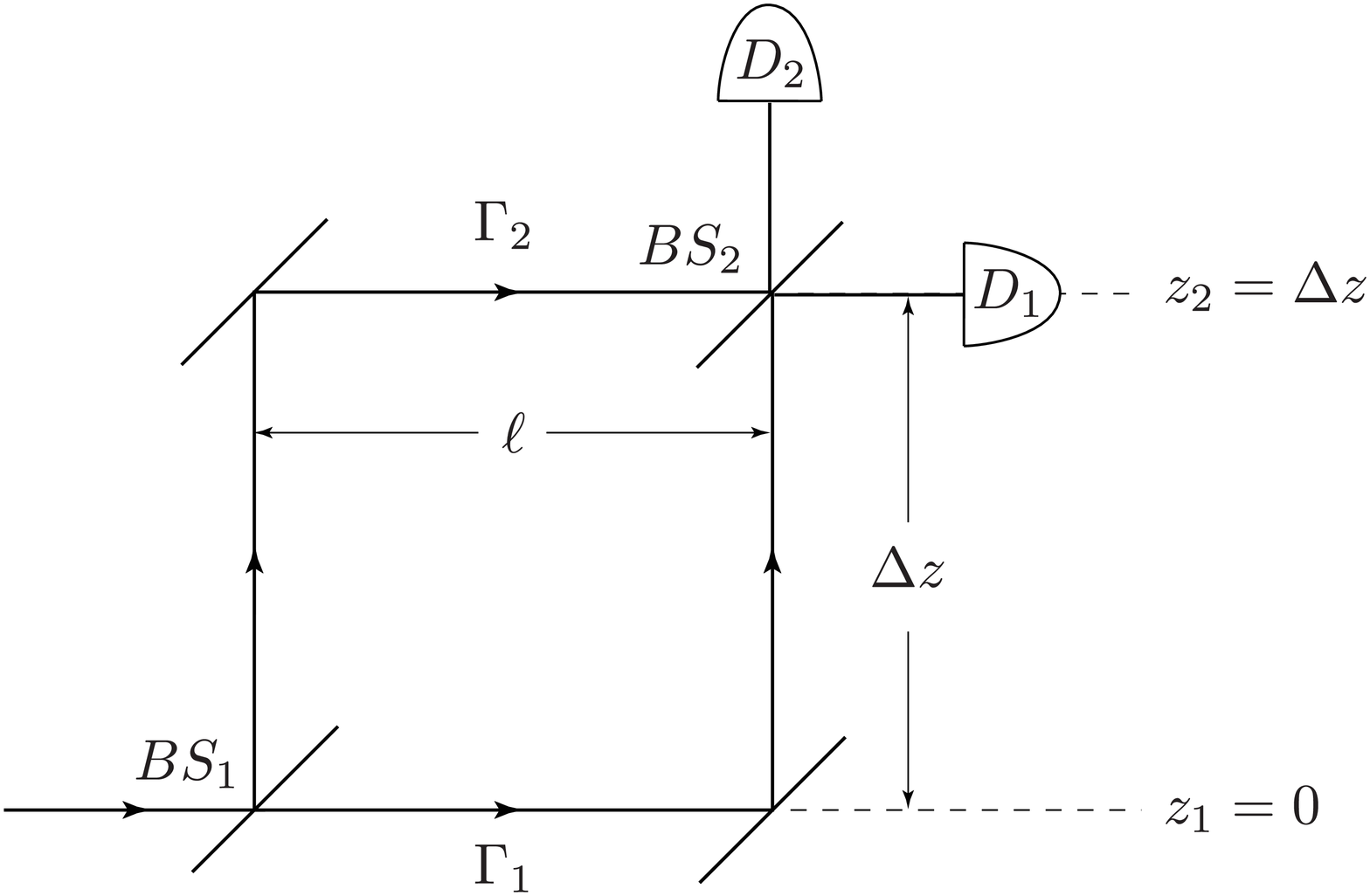}
\fi
\caption{A Schematic diagram of a Neutron interferometer used in the COW experiment. A neutron incident on the first beam splitter $BS_{1}$ is split into a spatial superposition travelling along two distinct paths $\Gamma_{1}$ and $\Gamma_{2}$. We find that $\Gamma_{2}$ accumulates a phase shift with respect to $\Gamma_{1}$ as it is higher in the gravitational field by $\Delta z$. \label{fig-neutroninterferometer}}
\end{figure}

%%%%%%%%%%%%%%%%%%%%%%%%%%%%%%%%%%%%%%%%%%%%%%%%%%%%%%%%%%%%%
As a concrete example for implementing the above recipe for calculating the phase difference we consider the gravitational neutron interferometry experiment illustrated in Fig.\ref{fig-neutroninterferometer}, known as the Colella--Overhauser--Werner (COW) experiment \cite{Colella}. The setup is geometrically identical to a Mach--Zehnder interferometer: The wavepacket is as usual split up into a spatial superposition and the respective wavepackets then travel along two distinct paths. The interferometer is oriented such that one path is higher up in the gravitational field relative to the other path. Essentially the two components of the spatial superposition have different speeds and experience two different gravitational potentials, which leads, in the recombination region, to a phase shift. Interference fringes have been observed (see e.g. \cite{Colella,Werner,Sakurai}) when the interferometer is rotated in the gravitational field, altering the difference in height of the paths.

In this section we are going to derive the phase difference for this experiment using the relativistic formalism  developed above. The spin of the neutrons is ignored and we treat them as scalar particles with no internal discrete degree of freedom. Therefore there is no need to use the transport equation \eqref{fermionTP}. The effects due to the spin could be included by computing \eqref{transphase}; however the corrections to the overall phase difference are minute, as noted in \cite{VarjuRyder}. From this analysis we will arrive at an exact relativistic result which contains, in certain limits, both approximate relativistic corrections to the COW experiment \cite{Anandan,AudretschLammerzahl,VarjuRyder} as well as the non-relativistic result \cite{Colella,Werner,Sakurai}.

One might represent the gravitational field for this experiment by the Schwarzschild metric. However, since the size of the experimental apparatus is less than a metre and hence small compared to the curvature scale, we can mimic gravity by simply going to an accelerated reference frame. This can be achieved by making use of the Rindler coordinates \cite{Rindler,MTW} in which the flat spacetime metric takes the form\footnote{Note that we could also consider rotating reference frames which would lead to the Sagnac effect \cite{Werner,Anandan,VarjuRyder}, but for simplicity we will stick to the Rindler metric.}
\begin{eqnarray*}
g_{\mu\nu}= \begin{pmatrix}(1+\frac{zg}{c^2})^{2}&0&0&0\\0&-1&0&0\\0&0&-1&0\\0&0&0&-1\end{pmatrix}.
\end{eqnarray*}

Since the Rindler metric is static (i.e. independent of $t$) we have a Killing vector $\eta^\mu=(1,0,0,0)$ and hence conserved energy $E\equiv p_\mu\eta^\mu=mc^2\gamma g_{00}$, where $\gamma\equiv(g_{00}-\frac{v^2}{c^2})^{-1/2}$ with $v$ the speed of the neutron as measured in the frame defined by the Killing vector $\eta^\mu$. We can now use this conserved energy to determine the speed $v_2$ of the neutron in the upper path given the speed in the lower path $v_1$, i.e. $E_1=mc^2\gamma_1 g_{00}(z_1)=mc^2\gamma_2 g_{00}(z_2)=E_2$. We take the lower path to be at height $z_1=0$, and so if the difference in height is $\Delta z$ we have the height of the top path being $z_2=\Delta z$. Therefore we have the relation $\gamma_2g_{00}(z_2)=\gamma_1$ since $g_{00}(z_1)=1$.

The easiest way to calculate the phase difference is by using the formula
\begin{equation}
\Delta\theta=\oint_{\Gamma} k_\mu\di x^\mu = \int_{\Gamma_2} k_\mu\di x^\mu-\int_{\Gamma_1} k_\mu\di x^\mu + k_\mu\Delta x^\mu\label{COWphase}
\end{equation}
where we have used \eqref{fermionphasedifference} and assumed that $A_{\mu}(x)$ is constant.
This formula contains two arbitrary spacetime points $x_1$ and $x_2$. Here we take these points to be where the trajectories $\Gamma_{1}$ and $\Gamma_{2}$, respectively, hit the second beam splitter $BS_{2}$. Since the spatial positions of these two events are the same in our Rindler coordinate system we have $\Delta x^\mu=(\Delta t,0,0,0)$, where $\Delta t$ is the difference in arrival time. This is given by
\begin{eqnarray*}
\Delta t=\ell\left(\frac{1}{v_1}-\frac{1}{v_2}\right)
\end{eqnarray*}
where $\ell$ is the length of the horizontal legs of the paths. The contribution of the phase difference from the third term is thus
\begin{eqnarray*}
k_\mu\Delta x^\mu=\frac{mc^2\ell}{\hbar}\left(\frac{g_{00}(\Delta z)\gamma_2}{v_1}-\frac{g_{00}(\Delta z)\gamma_2}{v_2}\right)=\frac{mc^2\ell}{\hbar}\left(\frac{\gamma_1}{v_1}-\frac{\gamma_1}{v_2}\right).
\end{eqnarray*}
Let us now turn to the first and second terms in \eqref{COWphase}, representing the internal phase shifts. Since the internal phases accumulated along the vertical components of each path are equal the quantity cancels in the calculation of the phase difference and so it is unnecessary to calculate them. The internal phase shifts of the upper and lower horizontal paths are given by
\begin{eqnarray*}
\theta^{(1)}_\text{int}=\frac{mc^2}{\hbar}\tau_1=\frac{mc^2\ell}{\hbar}\frac{1}{\gamma_1v_1}, \qquad\theta^{(2)}_\text{int}=\frac{mc^2}{\hbar}\tau_2=\frac{mc^2\ell}{\hbar} \frac{1}{\gamma_2v_2}.
\end{eqnarray*}
The phase difference is then given by
\begin{eqnarray*}
\Delta\theta=\theta^{(2)}_\text{int}-\theta^{(1)}_\text{int}+k_\mu\Delta x^\mu=\frac{mc^2\ell}{\hbar}\left(\frac{1}{\gamma_2v_2}-\frac{1}{\gamma_1v_1}+\frac{\gamma_1}{v_1}-\frac{\gamma_1}{v_2}\right)
\end{eqnarray*}
which simplifies to
\begin{eqnarray}\label{exactphasediff}
\Delta\theta=\frac{m\ell\gamma_1}{\hbar}\left(v_1-\frac{v_2}{(1+\frac{\Delta zg}{c^2})^2}\right)
\end{eqnarray}
where we can make the replacement $v_2=c\sqrt{g_{00}\left(1-g_{00}\gamma_1^{-2}\right)}$.

Eq.\eqref{exactphasediff} is the exact result for the gravitationally induced phase shift in the Rindler metric. This compares to various results in the literature \cite{Anandan,AudretschLammerzahl,Sakurai,Werner,VarjuRyder} for the gravitational effect in the COW experiment, which turn out to be approximations of \eqref{exactphasediff}. It is instructive to take various limits to demonstrate these connections.

Firstly, if we take the weak field limit, $\Delta z g/c^2\ll 1$ we obtain
\begin{eqnarray}
\Delta\theta\approx\frac{m\ell v_1\gamma_1}{\hbar}\left(1-\sqrt{1-2\frac{\Delta zg}{v_1^2}}\right).
\end{eqnarray}
If we furthermore take $\Delta zg/v_1^2\ll1$, which corresponds to assuming $|v_{1}-v_{2}|/v_1\ll1$, and consider only first order terms, we obtain the phase result in \cite{Anandan}. If we instead take the non-relativistic limit ($\gamma_1\approx1$) we obtain the result of \cite{AudretschLammerzahl}. Expanding this result to two orders of $\Delta z g/v_1^2$ gives
\begin{eqnarray}
\Delta\theta\approx\frac{m\ell}{\hbar}\left(\frac{\Delta zg}{v_1}+\frac{\Delta z^{2}g^{2}}{v_1^2}\right),\label{correction}
\end{eqnarray}
which indicates the `$g^2$' correction term derived in \cite{AudretschLammerzahl}. To leading order in  $\Delta zg/v_1^2$ the non-relativistic limit gives the standard theoretical prediction of the phase difference $\Delta\theta_\text{COW}$ observed in the COW experiment;
\begin{eqnarray}\label{standardphasediff}
\Delta\theta_\text{COW}=\frac{m\Delta z\ell g}{\hbar v_1}.
\end{eqnarray}
There is a reported small discrepancy between measurement and theory \cite{Colella}. However, the error introduced by neglecting corrections in $\Delta z g/v_1^2$ is too small to account for this discrepancy \cite{AudretschLammerzahl}.

The standard result \eqref{standardphasediff} is obtained using a path integral approach (see e.g. \cite{Sakurai,Werner}). The path integral method allows only for summation over paths which start and end at the same two spacetime points. However, the classical trajectories in this problem in fact do not arrive at the second beam splitter at the same time. Therefore, the standard expression, although a very good approximation in the specific case of the actual experiment under consideration, is not exact even non-relativistically.

%%%%%%%%%%%%%%%%%%%%%%%%%%%%%%%%%%%%%%%%%%%%%%%%%%%%%%%%%%%%%%%%%%%%%
\section{Elementary operations and measurement formalism\label{sec-unitarity}\label{sec-measurement}}
%%%%%%%%%%%%%%%%%%%%%%%%%%%%%%%%%%%%%%%%%%%%%%%%%%%%%%%%%%%%%%%%%%%%%

In order to develop quantum information theory in curved spacetimes we need to understand how elementary operations such as unitary transformations and state updating are represented within the reference frame covariant formalism of this paper. This section is dedicated to these issues. In addition we show how Hermitian observables are represented, how to calculate their expectation values, and how to construct explicitly a quantum observable given the measurement direction of a Stern--Gerlach device, or a polarizer.

%%%%%%%%%%%%%%%%%%%%%%%%%%%%%%%%%%%%%%%%%%%
\subsection{Fermions}
%%%%%%%%%%%%%%%%%%%%%%%%%%%%%%%%%%%%%%%%%%%

In this section we develop the notion of unitarity, observables and projectors for fermions. The notion of unitarity and observables is not straightforward for two reasons: (1) the inner product is velocity dependent and (2) Hilbert spaces associated with distinct points in spacetime must be thought of as separate. The formalism that we develop addresses these issues in a reference-frame-covariant way.

%%%%%%%%%%%%%%%%%%%%%%%%%%%%%%%%%%%%%%%%%%%
\subsubsection{Unitarity and Hermitian operators}
%%%%%%%%%%%%%%%%%%%%%%%%%%%%%%%%%%%%%%%%%%%

Unitarity is traditionally defined for automorphisms $U:\m H\mapsto\m H$, i.e. unitary maps take elements from one Hilbert space back to the same Hilbert space. The map $U$ is unitary if it satisfies
\begin{eqnarray}
\bk{U\phi}{U\psi}=\bk\phi\psi
\label{ordinaryunitary}
\end{eqnarray}
where the inner product is given by $\bk{\phi}{\psi}=\delta^{A'A}\bar\phi_{A'}\psi_A$. However, for our purposes this definition is too restrictive: we are interested in localized qubits transported along some spacetime trajectory $\Gamma$. The Hilbert spaces associated with the points along $\Gamma$ must be thought of as distinct and the therefore a map induced by the transport equation \eqref{fermionTP} cannot be thought of as a map from a Hilbert space to itself. Furthermore, the inner products for the Hilbert spaces depend on the respective 4-velocity. It is then clear that the transformation induced by the transport equation \eqref{fermionTP} is not going to be unitary according to \eqref{ordinaryunitary} as we are not dealing with automorphisms.

Consider therefore a map $U:\m H_1\rightarrow\m H_2$ where $\m H_1$ and $\m H_2$ are two Hilbert spaces on the trajectory $\Gamma$. That a quantum state belongs to $\m H_1$ is indicated by a subscript $\ket{\cdot}_1$ and similarly for $\m H_2$. The `generalized' definition of unitarity then becomes
\begin{eqnarray}
_2\bk{U\phi}{U\psi}_2\ =\ _1\bk\phi\psi_1\label{genuni}
\end{eqnarray}
using the inner product $_i\bk{\phi}{\psi}_i=\bar\phi^{(i)}_{A'}I_{u_i}^{A'A}\psi^{(i)}_A$, with $I^{A'A}_{u_i}\equiv u_I^{(i)}\bar\sigma^{IA'A}$, and $\phi^{(i)}_{A},\psi^{(i)}_{A}\in\m H_i$, $i=1,2$. If we adapt the tetrad such that $u^I=(1,0,0,0)$ along the trajectory we see that the inner product $I_u^{A'A}=u_I\bar\sigma^{IA'A}$ becomes the ordinary inner product $\delta^{A'A}$ which is independent of both position and momentum. We would therefore expect the inner product between two quantum states along some trajectory $\Gamma$ to be conserved. To see this let $\phi(\tau)$ and $\psi(\tau)$ represent two quantum states that are Fermi--Walker transported, according to equation \eqref{spinhalfFW} along $\Gamma$. Then we have
\begin{eqnarray}
\frac{\di}{\di\tau}\bk\phi\psi=\frac{D^{FW}}{D\tau}\bk\phi\psi=u_I\bar{\sigma}^{IA'A}\frac{D^{FW}\bar\phi_{A'}}{D\tau}\psi_A+u_I\bar{\sigma}^{IA'A}\bar\phi_{A'}\frac{D^{FW}\psi_A}{D\tau}=0 \quad\label{fermIPcons}
\end{eqnarray}
since the Fermi--Walker derivative of $u^I$ is zero by construction, and the inner product $\bk\cdot\cdot$ is defined using $I_{u(\tau)}^{A'A}$. Strictly speaking, the inner product should be labelled with $\tau$  (i.e. $_\tau\bk{\cdot}{\cdot}_\tau$) in order to indicate that we are dealing with different Hilbert spaces. However, for convenience we omit this cumbersome notation.

Let us now consider a more general evolution dictated by a Schr\"odinger equation
\begin{eqnarray}
\frac{D^{FW}\psi_{A}}{D\tau} = \frac{\di\psi_A}{\di\tau}-\ii\left(\frac12\frac{\di x^\mu}{\di\tau}\omega_{\mu IJ}+u_Ia_J\right)L^{IJ\ B}_{\ \ A}\psi_{B} = \ii A_A^{\ B}\psi_B\label{fermionevolution}
\end{eqnarray}
where $A_A^{\ B}$ represents some operator on $\psi_{B}$. Requiring the inner product to be preserved under the evolution implies that $A_A^{\ B}$ must for all $\phi,\psi\in\m H$ satisfy
\begin{eqnarray}
u_I\bar{\sigma}^{IA'A}\bar\phi_{A'}A_A^{\ B}\psi_B-u_I\bar{\sigma}^{IA'A}\bar{A}_{A'}^{\ B'}\bar\phi_{B'}\psi_A =0 \label{herm}
\end{eqnarray}
or equivalently $\bk{\phi}{A\psi}=\bk{A\phi}{\psi}$, which is nothing but the standard definition of a Hermitian operator. Since this must hold for all $\phi_A$ and $\psi_A$ we must have $I_u^{B'A}A_A^{\;B} = I_u^{A'B}\bar{A}_{A'}^{\;B'}$.

In spinor notation we can define $\m A^{A'A}\equiv A_B^{\ A}I_{u}^{A'B}$ which yields an equivalent definition of Hermiticity for the spinorial object $\m A^{A'A}$:
\begin{eqnarray*}
\bar{\m A}^{A'A} =\m A^{A'A}.
\end{eqnarray*}
An object $\m A^{A'A}$ satisfying this condition can be written as
\begin{eqnarray}
\m A^{A'A}=N_I\bar{\sigma}^{IA'A}\label{fermionhermitian}
\end{eqnarray}
for some real-valued coefficients $N_I$. We also have $A_A^{\ B}=I_{uAA'}\m A^{A'B}$ where $I_{uAA'}$ is the inverse of $I^{A'A}_u$ defined by $I_{uAA'}I_{u}^{A'B}=\delta_A^B$. The corresponding operator $A_{A}^{\ B}$ is then given by
\begin{eqnarray}
A_A^{\;B}&\equiv&I_{uAA'}\m A^{A'B}=u_I\sigma^I_{\;AB'}N_J\bar{\sigma}^{JB'B}=u_IN_J(\sigma^{[I}\bar{\sigma}^{J]}+ \sigma^{\{I}\bar{\sigma}^{J\}})_A^{\;B}\notag\\
&=&-2\ii u_IN_JL^{IJ\ B}_{\;\;A}+u_{I} N^{I}\delta_A^{\;B}\label{fermionoperator}
\end{eqnarray}
where ${L^{IJ}}_{A}^{\ B}$ are the left-handed $\mathfrak{sl}(2,\mathbb C)$ generators and the term in $\delta_A^{\ B}$ generates changes in global phase \footnote{Note that one could also have chosen the right-handed representation ${R^{IJ}}_{\ B'}^{A'}$ of the Lorentz group as this would yield the same result.}.

It should be noted that the operator $A_{A}^{\ B}$ does not `look' Hermitian when written out in matrix form. For example, $A_{1}^{\ 2}\neq\bar{A}_{2}^{\ 1}$.  Rather, it is only the object $\m A^{A'A}\equiv I_u^{A'B}A_{B}^{\;A}$ which looks Hermitian in matrix form, i.e. $\m A^{A'A}=\bar{\m A}^{AA'}$ \footnote{Spinor notation gives the relationship $\overline{\m A^{A'A}}\equiv\bar{\m A}^{AA'}\equiv\bar{\m A}^{A'A}$ between the conjugate and row-column transpose. }. The reason for this difference can be clearly seen by expressing $A_{A}^{\ B}$ in the rest frame of the qubit. In the particle rest frame, Hermitian operators are expressed as $\tilde{A}_A^{\ B}=N_I\sigma^0_{\ AA'}\bar\sigma^{IA'B}$, and in this case  $\tilde{A}_{1}^{\ 2}=\bar{\tilde A}_{2}^{\ 1}$. Thus, from an operator $\hat{\tilde A}$ which is Hermitian with respect to $\delta^{A'A}$ we can construct another operator $\hat A$ which is Hermitian with respect to $I^{A'A}_u$ simply by applying a boost, i.e. $A_A^{\ B}=\Lambda_A^{\ C}\tilde{A}_C^{\ D}\Lambda^{-1B}_D$, where $\Lambda$ is the spin-$\half$ representation of the Lorentz boost that takes $\delta^I_0$ to $u^I$.

A general inner product preserving evolution can therefore be understood as being composed of two pieces. One piece, the Fermi--Walker derivative, dictates how acceleration and the gravitational field affects the quantum state and has therefore a purely geometric character. The Fermi--Walker derivative maps elements between neighbouring Hilbert spaces $\m H_{(x,p)(\tau)}$ and $\m H_{(x,p)(\tau+\delta\tau)}$. The remaining term $A_A^{\ B}$ encodes possible non-geometric influences on the quantum state and is an automorphism $A:\m H\rightarrow\m H$. This second term is required to be Hermitian with respect to the inner product $I^{A'A}_u$.\footnote{Hermiticity can alternatively be defined in terms of the partial $\di/\di\tau$ or covariant $D/D\tau$ derivatives but in doing so we would have to modify the definition of a Hermitian operator.}

We have already seen an example of an evolution of the form \eqref{fermionevolution}. In the WKB limit of the minimally coupled Dirac equation we arrived at the transport equation \eqref{fermionTP}, where the Hermitian operator took on the form
\begin{eqnarray*}
B_A^{\ B}\equiv -\frac e{2m} B_{IJ}L^{IJ\ B}_{\ \ A}= -\frac e{2m} h_I^{\ K}h_J^{\ L}F_{KL}L^{IJ\ B}_{\ \ A}
\end{eqnarray*}
where $B_{IJ}\equiv h_I^{\ K}h_J^{\ L}F_{KL}$ is the magnetic field experienced by the particle. To see that the magnetic precession term $B_A^{\ \ B}$ is Hermitian with respect to the inner product, we expand the left side of the Hermiticity definition \eqref{herm}:
\begin{eqnarray*}
\bk{\phi}{\hat B\psi}-\bk{\hat B\phi}{\psi}=-\frac e{2m} \bar\phi\left(u_K\bar{\sigma}^KB_{IJ}\hat L^{IJ}-\bar B_{IJ}\hat{\bar{L}}^{IJ}u_K\bar{\sigma}^K\right)\psi.
\end{eqnarray*}
With $B_{IJ}$ real and making use of the identity $[\bar{\sigma}^K,\hat L^{IJ}]=\ii[\eta^{IJ}\bar{\sigma}^K-\eta^{JK}\bar{\sigma}^I]$ \cite{Bailin}, we have
\begin{align}
\bk\phi{\hat B\psi}-\bk{\hat B\phi}\psi&=-\frac e{2m} u_K B_{IJ}\bar\phi [\bar{\sigma}^K, \hat L^{IJ}]\psi\\&=-\ii\frac e{2m} \bar\phi\bar{\sigma}^K\psi[u_KB_I^{\ I}-B_K^{\ I}u_I]=0
\label{covhermitcommutator}
\end{align}
since $B_{IJ }u^{J}=0$ and $B_I^{\ I}=0$. The magnetic precession is thus a Hermitian automorphism with respect to the inner product $I_u^{A'A}$.

%%%%%%%%%%%%%%%%%%%%%%%%%%%%%%%%%%%%%%%%%%
\subsubsection{Observables and projective measurements\label{fermeas}}
%%%%%%%%%%%%%%%%%%%%%%%%%%%%%%%%%%%%%%%%%%

Observables are represented by Hermitian operators $A_A^{\ B}$, which will take the form indicated in \eqref{fermionoperator}. The covariant expression of the expectation value of the observable $A$ for a spinor $\psi_A$ is given by
\begin{eqnarray}
\langle\psi|A|\psi\rangle=\bar{\psi}_{A'}N_I\bar\sigma^{IA'A}\psi_A. \label{spacetimemeas}
\end{eqnarray}
Note that in \eqref{spacetimemeas} all indices have been contracted, indicating the expectation value is manifestly a Lorentz invariant scalar and could in principle represent an empirically accessible quantity.

In order to complete the measurement formalism we need to discuss how to determine the post-measurement quantum state. We do this for the simple case of projection-valued measures, however we can easily extend the formalism to generalized measurements. A Hermitian operator has a real eigenvalue spectrum and its normalized eigenstates $|\psi^{(k)}\rangle$ are orthogonal, i.e.
\begin{eqnarray*}
\langle\psi^{(k)}|\psi^{(l)}\rangle=I_{u}^{A'A}\bar{\psi}^{(k)}_{A'}\psi^{(l)}_A=\delta^{kl}.
\end{eqnarray*}
where $k,l=\pm$. The spectral decomposition of an observable $\hat{A}$  is $A_{A}^{\ B} = \sum_\pm\lambda_\pm P^{\pm B}_{\ A}$, where the $\lambda_\pm$ are the eigenvalues of $A_{A}^{\ B}$ and the $P^{\pm B}_{\ A}$ represents the corresponding projector onto the eigenstate $|\psi^{\pm}\rangle$. In spinor notation, the projectors are given by $P^{\pm B}_{\ A}=I_{u}^{A'B}\bar{\psi}^{\pm}_{A'}\psi^{\pm}_A$. A pair of projection operators $P^{\pm B}_{\ A}$ which, together  with the identity operator, span the space of Hermitian observables on $\m H$ can also be written as
\begin{eqnarray}\label{spinprojectors}
P^{\pm B}_{\ A}=\frac{1}{2}(\delta_A^{\ B}\mp 2\ii u_In_JL^{IJ\ B}_{\ \ A})
\end{eqnarray}
with  $n_In^I=1$ and $n_Iu^I=0$. These are the ordinary Bloch sphere projectors but written in a reference frame covariant way. One can then suspect that a measurement of spin along some unit direction can be represented by such projectors. This is indeed the case as we shall see now in the specific case of Stern--Gerlach measurements.

%%%%%%%%%%%%%%%%%%%%%%%%%%%%%%%%%%%%%%%%%%%%%%%%%%%%%%%%%%%%%%%%%%%%%
\subsubsection{The spin operator for a relativistic Stern--Gerlach measurement\label{sec-SternGerlach}}
%%%%%%%%%%%%%%%%%%%%%%%%%%%%%%%%%%%%%%%%%%%%%%%%%%%%%%%%%%%%%%%%%%%%%
To be able to extract empirical predictions from the above formalism we need to determine how $N^{I}$ in \eqref{fermionoperator} corresponds to the relevant parameters defining the experimental setup, e.g. the spatial orientation of a Stern--Gerlach magnet. In the literature there exist several proposals for relativistic spin operators and these have been studied for various reasons (see e.g. \cite{FoldyWouthuysen,HehlNi,Ryder98,Ryder99,Mashhoon95,Czachor,Ternotworol,Friis-relent}). In this section we are going to be concerned exclusively with constructing a spin observable associated with a relativistic Stern--Gerlach measurement. Notably, the spin operator that we obtain differs from other proposals and we will elaborate on this in a forthcoming paper. %(PTW12)

In order to obtain the correct relativistic spin observable it is necessary to understand in more detail the physical aspects of the measurement process. In a Stern--Gerlach spin measurement, a particle is passed though an inhomogeneous magnetic field. This causes the wavepacket to separate into two packets of orthogonal spin. A subsequent position measurement then records the outcome. To gain further insight, let us consider this measurement process in the fermion's rest frame where $e^\mu_t=u^\mu$.  In such a frame the stationary qubit is exposed to a magnetic field $B^i$ for a short period of time and it is clear that it is the direction $\frac1BB^i$ (with $B^2\equiv B^iB^j\delta_{ij}$) of the magnetic field that determines what component of the spin we are measuring \cite{PeresQM}.

If the qubit is moving non-relativistically with respect to the Stern--Gerlach device, the spatial direction $\frac1BB^i$ of the magnetic field approximately agrees with the orientation of the Stern--Gerlach device $m^{i}$. However, if the qubit is moving relativistically with respect to the apparatus these directions do not necessarily coincide, nor is their relationship straightforward. We will establish a relation between these two directions, and in doing so we will identify the correct spin observable for a relativistic Stern--Gerlach measurement. In particular the  relativistic spin operator/observable that we obtain depends on the spatial orientation $m^I$ of the apparatus and the 4-velocities $v^I$  and $u^I$ of the apparatus and qubit.

To proceed we first work out an expression for the electromagnetic field $F_{IJ}$ generated by the Stern--Gerlach apparatus. To do that we first introduce the magnetic field 4-vector $M^I=Mm^I$ where $M$ is the magnitude of the Stern--Gerlach magnetic field. We can now define the electromagnetic tensor as
\begin{eqnarray}
F_{IJ}\equiv-\epsilon_{_{LIJK}}v^LM^K.\label{EMfield}
\end{eqnarray}
In the rest frame of the apparatus it takes the form
\begin{eqnarray*}
F_{IJ}\stackrel{*}{=}\begin{pmatrix}0&0\\0&B_{ij}\end{pmatrix}
\end{eqnarray*}
so there is only a magnetic field, and no electric field, generated by the Stern--Gerlach apparatus in its own rest frame. In order to simplify the calculation the gradient $e^{\mu}_{I}\nabla_\mu M$ of the magnetic field strength is assumed to point in the same direction as the magnetic field itself, i.e. $e^{\mu}_{I}\nabla_\mu M\propto m_I$\footnote{Although this is not strictly possible as the magnetic field must satisfy $\nabla_{i}m^{i} = 0$ one can always choose a field which approximately has $e^{\mu}_{I}\nabla_\mu M\propto m_I$ locally \cite{Ballentine}.}.

We can now calculate the magnetic field 4-vector $B^I$ corresponding to the magnetic field as measured in the rest frame of the qubit. It is given by $B^I\equiv\half\epsilon^{LIJK}u_LF_{JK}$. Using \eqref{EMfield}, we obtain
\begin{eqnarray*}
B^I=-\half\epsilon^{LIJK}u_L \epsilon_{_{MJKN}}v^MM^N=M^I(v\cdot u)-v^I(M\cdot u)
\end{eqnarray*}
with $\epsilon^{LIJK}\epsilon_{_{MJKN}}=-2(\delta^L_M\delta^I_N-\delta^I_N\delta^L_M)$  and $\epsilon_{_{0123}}=1$ \cite[p87]{MTW}. For spin measurements, the 4-vector $n^I$ is now the normalized qubit rest-frame magnetic field, i.e.
\begin{eqnarray*}
n^I(m,u,v)\equiv\frac{B^I}{B}
\end{eqnarray*}
where $B\equiv\sqrt{-B^IB^J\eta_{_{IJ}}}$, and it is easy to check that $n\cdot u=0$. This expression $n^I$ becomes singular only for unphysical or trivial situations characterized by $u^I$ being null, or $M=0$.

Thus, given the spatial orientation of the Stern--Gerlach apparatus and the 4-velocities of the apparatus and qubit we obtain, using \eqref{fermionoperator}, a relativistic spin operator  given by
\begin{eqnarray*}
\m S_{A}^{\ B} = -2\ii u_I n_J(m,u,v)L^{IJ\ B}_{\ \ A}.
\end{eqnarray*}
The expectation values are calculated using \eqref{spacetimemeas} and the corresponding projectors are given by \eqref{spinprojectors}. We now have a fully relativistic and reference frame invariant measurement formalism for a Stern--Gerlach measurement.

%%%%%%%%%%%%%%%%%%%%%%%%%%%%%%%%%%%%%%%%%%
\subsection{Photons\label{sec-photonunitaritymmt}}
%%%%%%%%%%%%%%%%%%%%%%%%%%%%%%%%%%%%%%%%%%

In this section we develop the notion of unitarity, observables and projectors for photons. The definition of unitarity is in some sense simpler than for fermions as the inner product is not velocity dependent; the difficulty is only in handling the gauge degrees of freedom.

As we have already stated, the polarization state of a photon can be represented by a spatial complex 4-vector $\psi^I$ orthogonal to the null wavevector $k_\mu$. Furthermore, the corresponding quantum state of a photonic qubit with null velocity $u^I$ was identified as being a member of an equivalence class of polarization vectors $\psi^I\sim \psi^I+\upsilon u^I$ all orthogonal to $u^I$ (\S\ref{secphotonQS}). With the orthogonality condition and the gauge degree of freedom this space therefore reduced to a two-dimensional Hilbert space on which unitary and Hermitian operators act. We now develop the notion of unitarity, observables and projectors within this four-dimensional formalism.

%%%%%%%%%%%%%%%%%%%%%%%%%%%%%%%%
\subsubsection{Unitarity and Hermitian operators}
%%%%%%%%%%%%%%%%%%%%%%%%%%%%%%%%

Unitarity and Hermiticity are more straightforward with polarization vectors than spinors because the definition of unitarity $\bk{U\phi}{U\psi} = \bk{\phi}{\psi}$ is in terms of a standard inner product $\eta_{IJ}\bar\phi^I\psi^J$ where $\eta_{IJ}$ is constant. The requirement for unitarity again translates into requiring that the inner product between two polarization vectors is conserved along trajectories, i.e.
\begin{eqnarray}
\frac{\di \bk{\phi}{\psi}}{\di \lambda} = \frac{D \bk{\phi}{\psi}}{D \lambda} = \frac{D \bar \phi_{I}}{D \lambda}\psi^{I} + \bar\phi_{I} \frac{D \bar \psi^{I}}{D \lambda} = 0\label{photonderivIP}.
\end{eqnarray}

Consider now a Schr\"odinger evolution of the form
\begin{eqnarray}
\frac{D\psi^{I}}{D\lambda}=\beta u^{I} +\ii A^{I}_{\ J}\psi^{J}
\end{eqnarray}
which is more general than the transport equation \eqref{photonPT}. Substituting into \eqref{photonderivIP}, we get
\begin{eqnarray*}
\bar{A}^{\ J}_{I}\bar\phi_{J}\psi^{I} -\bar\phi_{I}A^{I}_{\ J}\psi^{J} = 0
\end{eqnarray*}
where we have used that $\psi^{I}u_{I} = u^{I}\phi_{I} = 0$. Requiring that this hold for all $\phi^{I}$ and $\psi^{I}$ we obtain the standard definition of Hermiticity; $\bar {A}^{\ \;I}_{J} = A^{I}_{\ J}$.

In addition to the above we also require that the gauge condition $u_I\psi^I=0$ be preserved. This implies that
\begin{eqnarray*}
\frac{\di(u_{I}\psi^{I})}{\di \lambda} = \frac{D (u_{I}\psi^{I})}{D \lambda} = \frac{D \psi^{I}}{D \lambda}u_{I} = A^{I}_{\ J}\psi^{J} u_{I} = 0.
\end{eqnarray*}
This condition ensures that Hermitian operators $A^{I}_{\ J}$ map polarization vectors into polarization vectors. Again this should hold for all $\psi^{J}$, so we have the condition $ A^{I}_{\ J} u_{I} \propto u_{J}$. The following two conditions suffice for characterizing a general Hermitian operator;
\begin{subequations}
\label{HK0HH}
\begin{align}
\bar{A}^{\ I}_{J} &= A^{I}_{\ J},\label{HH} \\
A^{I}_{\ J} u^{J} & \propto u^{I}. \label{HK0}
\end{align}
\end{subequations}

In order to determine the form of valid operators it is convenient to express the matrix $A^I_{\ J}$ in terms of a set of basis vectors $\{u^I,w^I,f_1^I,f_2^I\}$ which spans the full tangent space. Recall from Section \ref{sec-photonqsident} that a diad frame $f_A^I\equiv(f_1^I,f_2^I)$ defines a spacelike two-dimensional subspace orthogonal to two null vectors $u^I,w^I$ with $u^Iw_I=1$. $f_A^If^A_J=h^I_J=\delta^I_J-u^Iw_J-w^Iu_J$ is the metric on the spacelike subspace \cite{Poisson}. One can then define a sixteen-dimensional complex vector space spanned by the outer products of $\{u^I,w^I,f_1^I,f_2^I\}$ with the dual vectors $\{u_J,w_J,f_{1J},f_{2J}\}$. Components of an arbitrary matrix $A^{I}_{\ J}\in\langle B\otimes\bar B\rangle$ of this space can then be expanded in terms of these sixteen elements.

A valid map $A^I_{\ J}$ on polarization vectors must satisfy equations \eqref{HK0HH}. In terms of the sixteen basis elements, no terms in $w^I$ or $w_J$ can exist, since $w^Ik_I\neq0$. Any remaining terms that involve $u^{I}$ or $u_J$ are pure gauge and do not change the polarization vector. A hermitian operator is therefore, up to gauge, represented as
\begin{eqnarray*}
A^{I}_{\ J} = a f_1^I f^1_J + \beta f_1^I f^2_J + \bar\beta f_2^I f^1_J + b f_2^I f^2_J
\end{eqnarray*}
where the real numbers $a$ and $b$ and the complex number $\beta$ constitute the four remaining real degrees of freedom. The two conditions in \eqref{HK0HH} thus reduce a $4\times 4$ hermitian matrix to effectively a $2\times 2$ hermitian operator that acts on the transverse spacelike (polarization) degrees of freedom. Such an operator can be written in terms of the Pauli matrix basis as $A^{A}_{\ B} = C^a\sigma_{a\ B}^{\ A}$, where the Pauli matrices act on the two dimensional Jones vector. $C^a$ consisting of four coefficients $a=0,1,2,3$ does %FIXED
not transform as a 4-vector and therefore does not have any spatial significance. For the operator $A_A^{\;B}$ where  $a = 1,2$ or $3$ the eigenbasis corresponds to respectively the diagonal linear polarization basis, the circular polarization basis, and the horizontal--vertical polarization basis of the Jones vector. The relation between the four-dimensional and two-dimensional hermitian operators is $A^I_{\ J}=f^{I}_{A} A^{A}_{\ B} f_{J}^{B}$.

%%%%%%%%%%%%%%%%%%%%%%%%%%%%%%%%
\subsubsection{Observables and projective measurements\label{polarizationmeasurement}}
%%%%%%%%%%%%%%%%%%%%%%%%%%%%%%%%
The construction of observables and projectors is identical to that of fermions. Observables are represented by Hermitian operators $A^{I}_{\ J}$. Let $P_{(k)}^{I}$ represent the eigenvectors of $A^{I}_{\ J}$. The $P_{(k)}^{I}$ form an orthonormal basis with $\bar{P}_{(k)}^I{P_{(l)}}_I=\delta_{kl}$. The probability $p_{k}$ of getting outcome $\lambda_{k}$ is given in tetrad notation by
\begin{eqnarray}
p_k=|\bar P_{(k)}^I\psi_I|^2\label{eq-polarizeroverlap}.
\end{eqnarray}
The corresponding projector for an eigenvector $P^I_{(k)}$ is $P^I_{\; J} = P^I_{(k)}\bar{P}_{(k)J}$ and the post-measurement state, up to gauge, is given by  $\psi^I\to {\psi'}^{I} = P^{I}_{\ J} \psi^J$.

In this case we have a clean interpretation of projectors $P^I_{k}$ as polarizer vectors $P^I$: complex, spacelike normalized vectors orthogonal to photon velocity, $P^Iu_I=0$. Polarizer vectors correspond to the physical direction and parameters of an optical polarizer: A linear polarizer direction is of the form $\ee^{\ii\theta}P^I$ with $P^I$ real, and a circular polarizer is a complex vector $P^I=\frac{1}{\sqrt{2}}(P^I_1+\ii P^I_2)$ with $P_1^IP_2^J\eta_{IJ}=0$ and $\bar P^I_1P^J_1\eta_{IJ}=\bar P^I_2P^J_2\eta_{IJ}=1$. The probability of transmission of a polarization vector through a polarizer is simply the modulus square of the overlap of the polarization state with the polarizer vector \eqref{eq-polarizeroverlap}. Such an overlap clearly does not depend on the tetrad frame used, and indeed all tetrad indices are contracted in \eqref{eq-polarizeroverlap}. The probability $p$ is then manifestly a Lorentz scalar. It is easy to verify that the formalism is invariant under gauge transformations $\psi^I\rightarrow \psi^{I}+\upsilon u^I$ and $P^I\rightarrow P^I+\kappa u^I$. Thus, the probability $p$ is both gauge invariant and Lorentz invariant as should be the case. With this completed measurement formalism it is then not necessary to work with the  Jones vector or Wigner rotations, both of which involve working with the cumbersome  adaption procedure.

%%%%%%%%%%%%%%%%%%%%%%%%%%%%%%%%%%%%%%%%%%%%%%%%%%%%%%%%%%%%%%%%%%%%%
\section{Quantum entanglement}\label{secQIinCST}
%%%%%%%%%%%%%%%%%%%%%%%%%%%%%%%%%%%%%%%%%%%%%%%%%%%%%%%%%%%%%%%%%%%%%
Until now we have been concerned with the question of how the quantum state of some specific physical realization of a single qubit is altered by moving along some well-defined path in spacetime. We shall now show how this formalism can easily be extended to describe entanglement of multiple qubits.

%%%%%%%%%%%%%%%%%%%%%%%%%%%%%%%%%%%%%%%%%%%%%%%%%%%%%%%%%%%%%%%%%%%%%
\subsection{Bipartite states}
%%%%%%%%%%%%%%%%%%%%%%%%%%%%%%%%%%%%%%%%%%%%%%%%%%%%%%%%%%%%%%%%%%%%%
We have seen that it is necessary to associate a separate Hilbert space with each pair of position and momentum $(x^\mu,p_\mu)$. A single qubit moving along a specific path $x^\mu(\lambda)$ in spacetime will therefore have its state encoded in a sequence of distinct Hilbert spaces associated with the spacetime points along the path \footnote{Recall that while we can uniquely determine the 4-momentum $p^\mu=m\frac{\di x^\mu}{\di\tau}$ from the trajectory $x(\tau)$ in the case of massive fermions, the same is not true for photons. Due to the arbitrariness of the parametrization of the null trajectory $x(\lambda)$ we can only determine the null momentum up to a proportionality factor, i.e. $p^\mu\propto\frac{\di x^\mu}{\di\lambda}$ .}. The formalism that we have so far developed determines how to assign a quantum state to each distinct Hilbert space along the path along which we move the qubit. The one-parameter family of quantum states $|\psi(\lambda)\rangle$ is parameterized by some parameter $\lambda$ of the path $x(\lambda)$, and the sequence of Hilbert spaces associated with the path is $\mathcal{H}_{(x,p)(\lambda)}$. The quantum state $|\psi(\lambda)\rangle$ belongs to the specific Hilbert space $\mathcal{H}_{(x,p)(\lambda)}$.

Let us now consider how to generalize the formalism of this paper to the quantum state of two, possibly entangled, qubits in curved spacetime. Instead of one worldline we will now have two worldlines, $x_1(\lambda_1)$ and $x_2(\lambda_2)$, and consequently instead of one parameter $\lambda$ we now have two, $\lambda_1$ and $\lambda_2$. Corresponding to each value of $\lambda_1$ and $\lambda_2$ we have two spacetime points and two Hilbert spaces, $\mathcal{H}_{(x_1,p_1)(\lambda_1)}$ and $\mathcal{H}_{(x_2,p_2)(\lambda_2)}$. It is therefore clear that the quantum state describing the two qubits is mathematically described by a quantum state $|\psi(\lambda_1,\lambda_2)\rangle\in\mathcal{H}_{(x_1,p_1)(\lambda_1)}\otimes\mathcal{H}_{(x_2,p_2)(\lambda_2)}$ which belongs to the tensor product Hilbert space.

In order to calculate statistics (e.g. correlation functions) we need to provide an inner product for the tensor product Hilbert space $\mathcal{H}_{(x_1,p_1)(\lambda_1)}\otimes\mathcal{H}_{(x_2,p_2)(\lambda_2)}$. The natural choice is the inner product induced by the inner products for the individual Hilbert spaces.

In the case of fermions the natural choice for the parameterization $\lambda$ is the proper time $\tau$. The Hilbert spaces $\mathcal{H}_{(x_1,p_1)(\tau_1)}$ and $\mathcal{H}_{(x_2,p_2)(\tau_2)}$ have the inner products given by $I^{A'A}_1=u^1_{I}\bar{\sigma}^{IA'A}$  and $I^{A'A}_2=u^2_{I}\bar{\sigma}^{IA'A}$ where $u^1_I$ and $u^2_I$ are the respective 4-velocities. In our index notation a bipartite quantum state can be represented by an object with two spinor indices $\psi_{A_1 A_2}(x_1(\tau_1),x_2(\tau_2))$. Note however that the indices $A_1$ and $A_2$ relate to two distinct spinor spaces associated with two distinct points $x_1(\tau_1)$ and $x_2(\tau_2)$ and therefore cannot be contracted. The inner product between two bipartite quantum states $|\psi\rangle$ and $|\phi\rangle$ becomes
\begin{eqnarray*}
\bk{\psi}{\phi}_{p_1,p_2}=u^1_I\bar{\sigma}^{I A_1'A_1}u^2_J\bar{\sigma}^{JB_2'B_2}\bar{\psi}_{A_1'B_2'}\phi_{A_1B_2}.
\end{eqnarray*}
where $p_1$, $p_2$ are the momenta of the two qubits. In the case of photons the quantum state $\ket{\psi}$ can be represented by a polarization 4-vector $\psi^I(\lambda)$. A bipartite state $\ket{\psi}$ is then given by a two-index object $\psi^{I_1 I_2}(x_1(\lambda_1),x_2(\lambda_2))$, where $I_1$ and $I_2$ belong to two different tangent spaces and thus cannot be contracted. The requirement that the polarization vector be orthogonal to the null wavevector generalizes to  $u_{I_1}\psi^{I_1I_2} = 0= u_{I_2}\psi^{I_1I_2}$. The inner product between two bipartite quantum states  $|\psi\rangle$ and $|\phi\rangle$ becomes
\begin{eqnarray*}
\bk{\psi}{\phi}_{p_1,p_2} = \eta_{I_1 J_1} \eta_{I_2 J_2} \bar{\psi}^{I_1I_2} \phi^{J_1J_2} .
\end{eqnarray*}
We could also consider bipartite states $\ket{\phi}$ where one component is an electron and the other is a photon. Mathematically this would be represented as $\phi^{I_1}_{A_2}(x_1(\lambda), x_2(\tau))$ and the inner product can be constructed similarly.

Let us now turn to the evolution of bipartite quantum states. The physically available interactions of the qubits are given by local operations. Mathematically this means that the most general evolution of the state vector is given by two separate Schr\"odinger equations:
\begin{eqnarray}
\ii\frac{D^{T}}{D\lambda_1}|\psi(\lambda_1,\lambda_2)\rangle&=&\hat{A}_1(\lambda_1)\otimes \mathbb{I}|\psi(\lambda_1,\lambda_2)\rangle\label{twoSE1}\\
\ii\frac{D^{T}}{D\lambda_2}|\psi(\lambda_1,\lambda_2)\rangle&=&\mathbb{I}\otimes\hat{ A}_2(\lambda_2)|\psi(\lambda_1,\lambda_2)\rangle\label{twoSE2}
\end{eqnarray}
where $\hat{A}_1(\lambda_1)$ and $\hat{A}_2(\lambda_2)$ are possible local Hermitian operators (as defined in \S\ref{sec-unitarity}) acting on the Hilbert spaces $\mathcal{H}_{(x_1,p_2)(\lambda_1)}$ and $\mathcal{H}_{(x_2,p_2)(\lambda_2)}$. $D^{T}/D\lambda$ denotes the transport law, i.e. the Fermi--Walker transport for fermions or the parallel transport for photons.

This mathematical description of the evolution of the quantum state is not standard since we have two Schr\"odinger equations rather than one. However, if we introduce an arbitrary foliation $t(x)$ these two equations can be combined into one Schr\"odinger equation. First we express the parameters as functions of the foliation $\lambda_1=\lambda_1(t)$ and $\lambda_2=\lambda_2(t)$. This allows us to write the quantum state as only depending on one time parameter: $|\psi(t)\rangle=|\psi(\lambda_1(t),\lambda_2(t))\rangle$. The evolution of the quantum state now takes a more familiar form
\begin{eqnarray*}
\ii\frac{D^{T}}{Dt}|\psi(t)\rangle&\equiv&\ii\left(\frac{\di\lambda_1}{\di t} \frac{D^{T}}{D\lambda_1}+\frac{\di\lambda_2}{\di t} \frac{D^{T}}{D\lambda_2}\right)|\psi(\lambda_1(t),\lambda_2(t))\rangle\\
&=&\left(\frac{\di\lambda_1}{\di t}\hat{A}_1\otimes\mathbb{I}+ \frac{\di\lambda_2}{\di t}\mathbb{I}\otimes\hat{A}_2\right)| \psi(\lambda_1(t),\lambda_2(t))\rangle\\&=&\hat{A}|\psi\rangle
\end{eqnarray*}
where $\hat{A}\equiv\frac{\di\lambda_1}{\di t}\hat{ A}_1\otimes\mathbb{I}+\frac{\di\lambda_2}{\di t}\mathbb{I}\otimes\hat{A}_2$ is the total Hamiltonian acting on the full state. For this single Schr\"odinger equation to hold for all paths $x^\mu_1(\lambda_1)$ and $x^\mu_2(\lambda_2)$, and all choices of foliation $t(x)$, and so for all values of $\frac{\di\lambda_1}{\di t}$ and $\frac{\di\lambda_2}{\di t}$, it is necessary that both equations \eqref{twoSE1} and \eqref{twoSE2} hold. Thus, the two mathematical descriptions of the evolution of the quantum state are equivalent when only local operations enter in the evolution.

If the Hamiltonian is not a local one, i.e. not of the form $\hat{A}=a\hat{A}_1\otimes\mathbb{I}+b\mathbb{I}\otimes\hat{A}_2$, then it is not possible to cast it into the previous form %FIXED
with two independent evolution equations and it is also necessary to introduce a preferred foliation. However, if all interactions are local the introduction of an arbitrary foliation is not necessary.

The generalization to multipartite states is straightforward. Furthermore, if we are dealing with identical particles the wavefunction should be symmetrized or antisymmetrized with respect to the particle label, depending on whether we are dealing with bosons or fermions. This will correctly reproduce the Pauli exclusion phenomenon and the Hong--Ou--Mandel bunching phenomenon \cite{HOM}. %fixed
%
%%%%%%%%%%%%%%%%%%%%%%%%%%%%%%%%%%%%%%%%%%%%%%%%%%%%%%%%%%%%%%%%%%%%%
\subsection{State updating and the absence of simultaneity}
%%%%%%%%%%%%%%%%%%%%%%%%%%%%%%%%%%%%%%%%%%%%%%%%%%%%%%%%%%%%%%%%%%%%%
Let us now discuss the issue of state updating for entangled states. Let $\Gamma_1$ and $\Gamma_2$ be two spacetime trajectories along which two qubits are being transported. Furthermore let $x_1(\lambda_1)$ and $x_2(\lambda_2)$ each represent a distinct point on the corresponding trajectory.  Consider now that the two qubits are in the entangled state
\begin{eqnarray*}
\ket{\psi(\lambda_1,\lambda_2)} = \frac{1}{\sqrt 2}\left(\ket{+,\lambda_1}\ket{-,\lambda_2}-\ket{-,\lambda_1}\ket{+,\lambda_2}\right).
\end{eqnarray*}
If a measurement on qubit 1 is carried out at the spacetime point $x_1(\lambda_1)$ with outcome `$+$', the bipartite state has to be updated as follows:
\begin{eqnarray*}
\ket{\psi(\lambda_1,\lambda_2)} \to \ket{+,\lambda_1}\ket{-,\lambda_2}.
\end{eqnarray*}
Note that the value of $\lambda_2$ is left completely arbitrary after the state update. It should therefore be clear that even though the state update is associated with a distinct spacetime point $x_1(\lambda_1)$ on the trajectory $\Gamma_1$ no such point can be identified for $\Gamma_2$. In other words, as long as the local unitary evolution for particle 2 is well-defined, the state updating can be thought of as occurring at any point $x_2(\lambda_2)$ along $\Gamma_2$. The point $x_2(\lambda_2)$ could be in the past, elsewhere, or in the future of $x_1(\lambda_1)$. The reason for this freedom in state updating is that a projection operator on particle 1 commutes with any local unitary operator acting on particle 2.

%%%%%%%%%%%%%%%%%%%%%%%%%%%%%%%%%%%%%%%%%%%%%%%%%%%%
\subsection{An example: quantum teleportation}\label{teleportation}
%%%%%%%%%%%%%%%%%%%%%%%%%%%%%%%%%%%%%%%%%%%%%%%%%%%%
As an example of entanglement and state updating let us look at quantum teleportation in a curved spacetime. Although the mathematics is virtually the same as for the non-relativistic treatment, the interpretation is more delicate. In particular, in a curved spacetime the claim that the input state is in some sense the ``same" as the output state seems to lack a well-defined mathematical meaning. However, as we are going to see, in order to carry out the standard teleportation protocol the parties involved must first establish a shared basis in which the entangled state takes on a definite and known form. For example, Alice and Bob could choose the singlet state. This will be called the `canonical' form of the entangled state. Once this shared basis has been established, Alice and Bob have a well-defined convention for comparing quantum states associated with these different points in spacetime. The problems associated with quantum teleportation in curved spacetime are therefore similar to the problems associated with teleportation in flat spacetime when the maximally entangled state is unknown \cite{BRS,Rudolph,BRSdialogue,BRST}.

Consider then three qubits moving along three distinct trajectories $\Gamma_1$, $\Gamma_2$, and $\Gamma_3$. For concreteness assume that the qubits are physically realized as the spins of massive fermions. The tripartite state is then given by
\begin{eqnarray}
\ket{\Upsilon;\lambda_1,\lambda_2,\lambda_3}\in \mathcal{H}_{(x_1,p_1)(\lambda_1)}\otimes\mathcal{H}_{(x_2,p_2) (\lambda_2)}\otimes\mathcal{H}_{(x_3,p_3)(\lambda_3)} \nonumber
\end{eqnarray}
or, written in our index notation, $\Upsilon_{A_1A_2A_3}(\lambda_1,\lambda_2,\lambda_3)$.

In order to proceed we define a basis for each one of the three Hilbert spaces $\mathcal{H}_{(x_1,p_1)(\lambda_1)}$, $\mathcal{H}_{(x_2,p_2)(\lambda_2)}$, and $\mathcal{H}_{(x_3,p_3)(\lambda_3)}$ and {\it for all points} along the trajectories $\Gamma_1$, $\Gamma_2$, and $\Gamma_3$. The three pairs of basis vectors are assumed to be orthonormal, and to evolve according to the local unitary evolution (e.g. by pure gravitational evolution given by the Fermi--Walker transport \eqref{spinhalfFW}). Therefore, once we have fixed the basis for particle $i$ at one point $x_i(\lambda_i)$ on the trajectory $\Gamma_i$, the basis is uniquely fixed everywhere else along the trajectory, at least where the local unitary evolution is well-defined. It follows that a state can be expressed as a linear combination of these basis states with the components independent of $\lambda_i$, $i=1,2,3$.

We denote these three pairs of (one-parameter families of) orthonormal states as $\phi^{(i)}_{A_i}(\lambda_i)$ and $\psi^{(i)}_{A_i}(\lambda_i)$, with $i=1,2,3$. The orthonormality conditions are explicitly given by
\begin{eqnarray*}
\bk{\phi^{(i)};\lambda_i}{\psi^{(i)};\lambda_i}&=&u^{(i)}_{I_i}\bar{\sigma}^{I_iA'_iA_i} \bar{\phi}^{(i)}_{A'_i}(\lambda_i)\psi^{(i)}_{A_i}(\lambda_i)=0\\
\bk{\phi^{(i)};\lambda_i}{\phi^{(i)};\lambda_i}&=&u^{(i)}_{I_i}\bar{\sigma}^{I_iA'_iA_i} \bar{\phi}^{(i)}_{A'_i}(\lambda_i)\phi^{(i)}_{A_i}(\lambda_i)=1\\
\bk{\psi^{(i)};\lambda_i}{\psi^{(i)};\lambda_i}&=&u^{(i)}_{I_i}\bar{\sigma}^{I_iA'_iA_i} \bar{\psi}^{(i)}_{A'_i}(\lambda_i)\psi^{(i)}_{A_i}(\lambda_i)=1
\end{eqnarray*}
with $i=1,2,3$.

Consider now the specific tripartite three-parameter family of states
\begin{equation}
\Upsilon_{A_1A_2A_3}(\lambda_1,\lambda_2,\lambda_3)= \frac{1}{\sqrt{2}}\left(\alpha\phi^{(1)}_{A_1}(\lambda_1)+ \beta\psi^{(1)}_{A_1}(\lambda_1)\right)\left(\phi^{(2)}_{A_2}(\lambda_2) \phi^{(3)}_{A_3}(\lambda_3)+\psi^{(2)}_{A_2}(\lambda_2)\psi^{(3)}_{A_3} (\lambda_3)\right)\label{tripartite}
\end{equation}
where $\alpha$ and $\beta$ are independent of $\lambda_1$ (since evolution is entirely in the basis vectors) and $|\alpha|^2+|\beta|^2=1$. The maximally entangled state (involving particles $2$ and $3$) only has coefficients equal to $0$ or $1$ which are trivially independent of the parameters $\lambda_2$ and $\lambda_3$. We will call the maximally entangled state in equation \eqref{tripartite} the {\em canonical form}. Other choices of this canonical form are possible but the teleportation protocol used below (see \eqref{Bobunitaries}) will then change accordingly.

As in the flat spacetime description we now proceed to rewrite the state in the Bell basis for the Hilbert space $\mathcal{H}_{(x_1,p_1)(\lambda_1)}\otimes\mathcal{H}_{(x_2,p_2)(\lambda_2)}$:
\begin{eqnarray*}
\Phi^\pm_{A_1A_2}(\lambda_1,\lambda_2)&\equiv&\frac{1}{\sqrt{2}}\left(\phi^{(1)}_{A_1} (\lambda_1)\phi^{(2)}_{A_2}(\lambda_2)\pm\psi^{(1)}_{A_1}(\lambda_1)\psi^{(2)}_{A_2} (\lambda_2)\right)\\
\Psi^\pm_{A_1A_2}(\lambda_1,\lambda_2)&\equiv&\frac{1}{\sqrt{2}}\left(\phi^{(1)}_{A_1} (\lambda_1)\psi^{(2)}_{A_2}(\lambda_2)\pm\psi^{(1)}_{A_1}(\lambda_1)\phi^{(2)}_{A_2} (\lambda_2)\right).
\end{eqnarray*}
In these new bases the state $\ket{\Upsilon}$ reads
\begin{multline*}
\Upsilon_{A_1A_2A_3}(\lambda_1,\lambda_2,\lambda_3)=\frac{1}{2}\big(\Phi^+_{A_1A_2} \left(\alpha\phi_{A_3}+\beta\psi_{A_3}\right)+\Phi^-_{A_1A_2}\left(\alpha\phi_{A_3}- \beta\psi_{A_3}\right)\\ +\Psi^+_{A_1A_2}\left(\beta\phi_{A_3}+\alpha\psi_{A_3}\right)+\Psi^-_{A_1A_2} \left(-\beta\phi_{A_3}+\alpha\psi_{A_3}\right)\big).
\end{multline*}
Alice now performs a Bell basis measurement on the particles $1$ and $2$. It is important to note that the specific physical measurement operation that Alice needs to carry out depends on the bases $(\phi^{(1)}_{A_1},\psi^{(1)}_{A_1})$ and $(\phi^{(2)}_{A_2},\psi^{(2)}_{A_2})$. The basis $(\phi^{(2)}_{A_2},\psi^{(2)}_{A_2})$ is determined by the maximally entangled state. If Alice does not know the maximally entangled state she will not be able to do the correct Bell basis measurement.

The outcome of the Bell basis measurement ($\Phi^+$, $\Phi^-$, $\Psi^+$, or $\Psi^-$) is then communicated to  Bob's side: Bob performs the local unitary operation $U$ (which is assumed to act on the state but not on the basis) given by
\begin{eqnarray}\label{Bobunitaries}
U=\left\{\begin{array}{c}\hat\one\ \ \text{if}\ \ \Phi^+\\\hat\sigma_z\ \ \text{if}\ \ \Phi^-\\\hat\sigma_x\ \ \text{if}\ \ \Psi^+ \\\ii\hat\sigma_y\ \ \text{if}\ \ \Psi^-\end{array}\right.
\end{eqnarray}
where $\hat\sigma_x,\hat\sigma_y,\hat\sigma_z$ take the usual form when expressed in the local orthonormal basis $(\phi^{(3)}_{A_3},\psi^{(3)}_{A_3})$. Thus, as was the case for Alice, in order for Bob to know which specific physical operation to carry out, the basis has to be specified. Since the basis is determined by the maximally entangled state, Bob must know the entangled state in order to carry out his operations. Note that if the basis in which Bob applies the operation is incorrect, the state that Bob obtains at the end differs depending on the outcome of the Bell measurement. In the case where Bob implements the operation in the correct basis, Bob's state is given by $\alpha\phi_{A_3}+\beta\psi_{A_3}$.

We should now ask whether we can sensibly view this protocol as a `teleportation' of a quantum state from Alice to Bob in the sense that Bob received the {\em same} state as Alice sent. This hinges on there being a meaningful way of comparing quantum states associated with distinct spacetime points. However, as we have already stressed, if spacetime is curved, sameness of quantum states cannot be established uniquely by parallel transporting one qubit to the other as this would depend on the specific path along which we transport the qubit.

On the other hand, the maximally entangled state, by determining the bases for it to take the canonical form, defines a shared spinor basis for Alice and Bob. If we change the basis, the maximally entangled state would of course change accordingly and would no longer take on the canonical form. Given a shared basis we have a well-defined way of comparing quantum states and in particular a well-defined way to claim that they are the same or not. Thus, when we have a maximally entangled state, there {\it is} a natural way of comparing quantum states associated with distinct spacetime points. It is in using this convention for comparing quantum states that we can claim that Bob did indeed receive the same quantum state, and therefore we can say that the state was in this sense teleported.\footnote{We note that all examples of experimentally produced entangled qubit pairs are produced in localized spatial regions and distributed to the parties. The components of the entangled state will then undergo local unitary evolution along each trajectory.}

In the case of fermions it is also easy to see that the maximally entangled state will also establish a shared reference frame, i.e. a shared tetrad. This comes about because from the left-handed spinor $\psi_A$ by means of which the quantum state is expressed we can construct the null {\it Bloch 4-vector} $b^I=\bar\sigma^{IA'A}\bar\psi_{A'}\psi_A$. A maximally entangled state can therefore be loosely understood geometrically as a kind of `non-local connection'.

%%%%%%%%%%%%%%%%%%%%%%%%%%%%%%%%%%%%%%%%%%%%%%%%%%%%%%%%%%%%%%%%%%%%%%%%%%%%
\section{Conclusion, discussion, and outlook}\label{sec8}
%%%%%%%%%%%%%%%%%%%%%%%%%%%%%%%%%%%%%%%%%%%%%%%%%%%%%%%%%%%%%%%%%%%%%%%%%%%%%

Recently there has been increased interest in exploring relativistic quantum information theory in the context of phenomena from quantum field theory such as the Unruh effect and particle number ambiguity \cite{Alsing-DiracUnruh,Fuentes,Martinez}. In contrast, this paper explored relativistic quantum information in the regime where such effects are negligible and restricted attention to localized qubits for which the particle number ambiguity is circumvented. A localized qubit is understood in this paper to be any object that can effectively be described by a position and momentum $(x,p)$ and some two-component quantum state $\ket{\psi}$. We obtained a description of localized qubits in curved spacetimes, with the qubits physically realized as the spin of a massive fermion, and the polarization of a photon.

The original motivation for this  research was to develop a formalism for answering a simple experimental question: if we move a spatially localized qubit, initially in a state $|\psi_1\rangle$ at spacetime point $x_1$, along some classical spacetime path $\Gamma$ to another point $x_2$, what will the final quantum state $|\psi_2\rangle$ be? Rather than working directly with Wigner representations our starting point in answering this question was the one-particle excitations of the quantum fields that describe these physical systems. The one-particle excitations in curved spacetime satisfy respectively the Dirac equation minimally coupled to the electromagnetic field, and Maxwell's equations in vacuum. From these fields we were able to isolate a two-component quantum state and a corresponding Hilbert space.

In the case of fermions, the equation governing the transport of the spin of a fermion consisted of  a spin-$\half$ version of the Fermi--Walker derivative and a magnetic precession term, expressed in a non-orthonormal Hilbert space basis.  This result was expected since an electron can be regarded as a spin-$\half$ gyroscope, and the precession of a classical gyroscope along accelerated trajectories obeys the Fermi--Walker equation. By introducing an orthonormal Hilbert space basis, which physically corresponds to representing the spinor in  the particle's rest frame,  we reproduced the transport equation obtained in \cite{TerashimaUeda03,ASK} which made use of Wigner representations.

We showed by applying the WKB approximation to vacuum Maxwell equations that the polarization vector of a photon is parallel transported along geodesics and that this corresponds to a Wigner rotation. Furthermore, this rotation is proportional to the spin-1 connection term $\omega_{\mu12}$ when we consider a reference frame where the photon 3-velocity is along the $z$-axis. In this way the effect of spacetime geometry on the quantum state was easily identified.

We worked with faithful finite-dimensional but {\em non-unitary} representations of the Lorentz group, specifically a two-component left-handed spinor $\psi_A$, and polarization 4-vector $\psi^I$. Nevertheless, by identifying a suitable inner product we obtained a unitary quantum formalism. The advantage of working with non-unitary representations is that the objects which encode the quantum state transform covariantly under actions of the Lorentz group. As a result the transport equations are manifestly Lorentz covariant, in addition to taking on a simple form. The connection to the Wigner formalism was obtained by choosing a reference frame adapted to the particle's 4-velocity, which is a reason why the Wigner formalism is not manifestly Lorentz covariant.

In order to make empirical predictions we need a way to extract probabilities for outcomes from the formalism. Such a measurement formalism was developed for both fermions and photons. The predicted probabilities of outcomes of experiments were shown to be manifestly Lorentz invariant and thus reference frame invariant, resulting in a relativistically invariant measurement formalism. We also derived the specific Hermitian operator corresponding to a Stern--Gerlach measurement, by providing a physical model of this measurement process. In this way a unique spin operator can be identified given the spatial orientation and velocity of the Stern--Gerlach apparatus and the velocity of the particle. Notably this operator does not agree with previous competing proposals \cite{Czachor,Ternotworol,Friis-relent}.

A second advantage of working in terms of the Dirac and Maxwell fields instead of the Wigner representations is that global phases and quantum interference come out automatically from the WKB approximation. By considering spacetime Mach--Zehnder interference experiment we arrived at a general relativistic formula for calculating the gravitationally induced phase difference.  In the specific case of gravitational neutron interferometry we reproduced the existing formulae for the gravitationally induced phase difference as various limits of our formula. Our overall approach, however, provides a general, unified, and straightforward way of calculating phases and interference for any situation.

Finally we generalized this formalism to the treatment of multipartite states, entanglement, and teleportation, thereby extending the formalism to include all the basic elements of quantum information theory.

The Lorentz group played a primary role in the construction of qubits in curved spacetime. This role can be understood in terms of how gravity acts on physical objects. When an object is moved along some path $x^\mu(\lambda)$ in spacetime it passes through a sequence of tangent spaces. These are connected by infinitesimal Lorentz transformations that are determined from the trajectory and the gravitational field (i.e. the connection 1-form $\omega_{\mu\ J}^{\ I}$). That is, apart from a possible global phase, this sequence of infinitesimal Lorentz transformations determines how an object is affected by the gravitational field. In particular, if the object has internal degrees of freedom that transform under the Lorentz group, we can determine the effect of gravity on the state of these internal degrees of freedom. For example, this is the explanation for the presence of spin connection terms in the fermion Fermi--Walker transport. It is also an explanation for why the photon Wigner rotation is simply a rotation of the linear polarization and so respects the helicity of the photon: no Lorentz boost can change frames sufficiently to change the helicity of a photon.

In this paper we focused on just two physical realizations which constituted non-trivial representations of the Lorentz group. We can nevertheless contemplate other realizations such as composite two-level systems. In order to understand how gravity acts on the qubit state the same general approach applies: One needs to provide a mathematical model of the physical system. Once a model is established one can in principle determine how (if at all) the quantum state transforms under a Lorentz transformation. In addition to this there are other possible gravitational influences on the quantum state such as gravitationally induced phases.

As a concrete example of a physical realization that would behave very differently to the elementary realizations treated in this paper, consider a two-level system where the two levels are energy eigenstates $\ket{E_1}$ and $\ket{E_2}$. From ordinary non-relativistic quantum mechanics we know that the total state $\ket{\psi}=a\ket{E_1}+b\ket{E_2}$ will undergo the evolution
\begin{eqnarray*}
\ket{\psi(t)}=a\ee^{\ii E_1t/\hbar}\ket{E_1}+b\ee^{\ii E_2t/\hbar}\ket{E_2}.
\end{eqnarray*}
If the composite object is much smaller than the curvature scale and the acceleration is sufficiently gentle to not destroy it we can obtain a fully general relativistic generalization by simply replacing the Newtonian time $t$ with the proper time $\tau$:
\begin{eqnarray*}
\ket{\psi(\tau)}=a\ee^{\ii E_1 \tau/\hbar}\ket{E_1}+b\ee^{\ii E_2 \tau/\hbar}\ket{E_2}.
\end{eqnarray*}
Thus, because of the energy difference we develop a relative phase between the two energy levels which is proportional to the proper time of the trajectory. In principle we can make use of such a two-level system to measure the proper time of a spacetime trajectory.  Since proper time is path dependent we see that the transport of the quantum state is also path dependent, and we can also contemplate possible interference experiments.

The formalism presented in this paper provides a basis for quantum information theory of localized qubits in curved spacetime. One theoretical application of this is to extend the applicability of clock synchronization \cite{Josza,Giovannetti01,deBurgh} and reference frame sharing \cite{Rudolph,BRSdialogue} to curved spacetime, and provide an interesting physical scenario for the study of symmetries in quantum mechanics \cite{JonesWiseman,Spekkens}.

This formalism also has potential measurement applications. For example, in the last ten or fifteen years there has been interest in precision measurement of the effects of general relativity. Most recently there has been the experimental confirmation of the predicted frame-dragging effect by Gravity Probe B \cite{gravityprobeB}.\footnote{See also \cite{Brodutch11,BDT11} for a recent theoretical analysis of polarization rotation due to gravity.} On the other hand, in the same period there has been an increased interest in quantum precision measurements using techniques from quantum information theory \cite{Childs,Dariano,Preskill,Giovannetti04,Chiribella05,Xiang2011}. The formalism of this paper provides a bridge between these developments, providing a solid foundation for considering the effects of gravity on quantum states, and for considering the design of precision measurements of these effects. Importantly, by the use of entangled states one can in principle significantly increase the precision of such a measurement. For localized qubits in curved spacetimes as defined in this paper, the effect of gravity enters as classical parameters in the unitary evolution of the quantum state. Therefore, one should be able to use these same quantum information theory techniques to increase the precision in measurements of the gravitational field. It is plausible that such an amalgamation of the transformation of the discrete degrees of freedom of a quantum state and the phase accumulation, by increasing the degrees of freedom to be measured, will increase the sensitivity with which possible future sophisticated precision quantum measurements can measure gravitational effects.

For example, spacetime torsion is generally believed, even if non-zero, to be too small to measure with present day empirical methods \cite{Bergmann}. One problem is that torsion, as it is conventionally introduced in Einstein--Cartan theory, does not have any propagating degrees of freedom. Thus, the torsion in a spacetime region is non-zero if and only if the spin density is non-zero there. Experiments to measure torsion thus require objects to pass  {\em though} a material with non-zero spin density to accumulate an effect, while accounting for standard interactions. Needless to say, measuring torsion is then very difficult. However, electrons decouple from matter in the high energy WKB limit and effectively only feel the gravitational field including torsion. Thus, the spin of high energy fermions might carry information about the spacetime torsion.

Finally, on the more speculative side it would perhaps also be interesting to model closed timelike curves \cite{Deutsch,Ralph,Bartlett} within this formalism. It is possible that the relativistic approach to state evolution and bipartite states provides rules or constraints for the manipulation of quantum information in closed timelike curves.

To summarize: in this paper we have provided a complete account of the transport and measurement of localized qubits, realized as elementary fermions or photons, in curved spacetime. The manifest Lorentz covariance of the formalism allows for a relativistic treatment of qubits, with a perhaps more straightforward interpretation than approaches based on Wigner representations. The treatment of multipartite states, entanglement and interferometry provides a basis for quantum information theory of localized qubits in curved spacetime.

\section*{Acknowledgements}
We are indebted to Stephen Bartlett, Daniel Terno, Bruce Yabsley, Don Melrose, Jorma Louko, Florian Girelli, Gerard Milburn and other participants at the conference RQI4 for several stimulating and helpful discussions. We thank Emma Nimmo for help with diagrams. This research was supported by the Perimeter Institute--Australia Foundations (PIAF) program, and the Australian Research Council grant DP0880860.

%%%%%%%%%%%%%%%%%%%%%%%%%%%%%%%%%%%%%%%%%%%%%%%%%%%%%%%%%%%%%%%%
%%%%%%%%%%%%%%%%%%%%%%%%%%%%%%%%%%%%%%%%%%%%%%%%%%%%%%%%%%%%%%%%%

\appendix
\section*{Appendix}
\setcounter{section}{0}

%%%%%%%%%%%%%%%%%%%%%%%%%%%%%%%%%%%%%%%%%%%%%%%%%%%%%%%%%

%%%%%%%%%%%%%%%%%%%%%%%%%%%%%%%%%%%%%%%%%%%%%%%%%%%%%%%%%
\section{Spinors and $SL(2,\mathbb{C})$}\label{secspinornotation}
%%%%%%%%%%%%%%%%%%%%%%%%%%%%%%%%%%%%%%%%%%%%%%%%%%%%%%%%%

In our analysis of qubits in curved spacetime it will be necessary to introduce some notation for describing spinors. A spinor is a two-component complex vector $\phi_{A}$, where $A=1,2$ labels the spinor components, living in a two-dimensional complex vector space $W$. We are going to be using spinors as objects that transform under  $SL(2,\mathbb{C})$, which forms a double cover of $SO^{+}(1,3)$. Hence, $W$ carries a spin-$\half$ representation of the Lorentz group. The treatment of spinors in this section begins abstractly, and ends with details specific to Dirac spinors.

%%%%%%%%%%%%%%%%%%%%%%%%%%%%%%%%%%%%%%%%%%%%
\subsection{Complex vector spaces}\label{complexvectorspace}
%%%%%%%%%%%%%%%%%%%%%%%%%%%%%%%%%%%%%%%%%%%%

Mathematically, spinors are vectors in a complex two-dimensional vector space $W$. We denote elements of $W$ by $\phi_A$. Just as in the case of tangent vectors in differential geometry, we can consider the space $W^*$ of linear functions $\psi: W\mapsto \mathbb{C}$, i.e. $\psi(\alpha\phi_1+\beta\phi_2)=\alpha\psi(\phi_1)+\beta\psi(\phi_2)$. Objects belonging to $W^*$, which is called the {\em dual space} of $W$, is written with the index as a superscript, i.e. $\psi^A\in W^*$.

Since our vector space is a complex vector space it is also possible to consider the space $\overline{W}^*$ of all {\em antilinear} maps $\chi:W\mapsto\mathbb{C}$, i.e. all maps $\chi$ such that $\chi(\alpha\phi_{1}+\beta\phi_{2})=\bar{\alpha}\chi(\phi_{1})+\bar{\beta}\chi(\phi_{2})$. A member of that space, called the {\em conjugate dual space} of $W$,  is written as $\chi^{A'}\in\overline{W}^{*}$. The prime on the index distinguishes these vectors from the dual vectors.

Finally we can consider the space $\overline{W}$ dual to $\overline{W}^*$, which is identified as the {\em conjugate space} of $W$. Members of this space are denoted as $\xi_{A'}$.

In summary, because we are dealing with a complex vector space in quantum mechanics rather than a real one as in ordinary differential geometry we have four rather than two spaces:
\begin{itemize}
\item the space $W$ itself: $\phi_A\in W$;
\item the space $W^*$ dual to $W$: $\psi^A\in W^*$;
\item the space $\overline{W}^*$ conjugate dual to $W$: $\chi^{A'}\in\overline{W}^*$;
\item the space $\overline{W}$ dual to $\overline{W}^*$: $\xi_{A'}\in\overline{W}$.
\end{itemize}
%
%%%%%%%%%%%%%%%%%%%%%%%%%%%%%%%%%%%%%%%%%%%%
\subsubsection{Spinor index manipulation\label{sec-geometricquantum}}
%%%%%%%%%%%%%%%%%%%%%%%%%%%%%%%%%%%%%%%%%%%%
There are several rules regarding the various spinor manipulations that are required when considering spinors in spacetime. Specifically, we would like to mathematically represent the operations of  complex conjugation, summing indices, and raising and lowering indices. The operation of raising and lowering indices will require additional structure which we will address later.

Firstly the operation of complex conjugation: In spinor notion the operation of complex conjugation will turn a vector in $W$ into a vector in $\overline{W}$. The complex conjugation of $\phi_{A}$ is represented as
$$
\overline{\phi_{A}} = \overline{\phi}_{A'}.
$$

We will also need to know how to contract two indices. We can only contract  when one index appears as a superscript and the other as a subscript, and only when the indices are either both primed or both unprimed, i.e. $\phi_A\psi^A$ and $\xi_{A'} \chi^{A'}$ are allowed contractions. Contraction of a primed index with an unprimed one, e.g. $\phi_A\chi^{A'}$, is not allowed.

The reader familiar with two-component spinors \cite{Wald,Bailin,Penrose} will recognize the index notation (with primed or unprimed indices) presented is commonly used in treatments of spinors. It should be noted however that this structure has little to do with the Lorentz group or its universal covering group $SL(2,\mathbb{C})$. Rather, this structure is there as soon as we are dealing with complex vector spaces and is unrelated to what kind of symmetry group we are considering. We will now consider the symmetry given by the Lorentz group.

%%%%%%%%%%%%%%%%%%%%%%%%%%%%%%%%%%%%%%%%%%%
\subsection{$SL(2,\mathbb{C})$ and the spin-$\half$ Lorentz group\label{spinhalfLG}}
%%%%%%%%%%%%%%%%%%%%%%%%%%%%%%%%%%%%%%%%%%%
The Lie group $SL(2,\mathbb{C})$ is defined to consist of $2\times2$ complex-valued matrices $\m L_A^{\ B}$ with unit determinant which mathematically translates into
\begin{eqnarray*}
\frac{1}{2}\epsilon_{CD}\epsilon^{AB}\m L_A^{\ C}\m L_B^{\ D}=1
\end{eqnarray*}
where $\epsilon^{AB}$ is the antisymmetric Levi--Civita symbol defined by $\epsilon^{12}=1$ and $\epsilon^{AB}=-\epsilon^{BA}$ and similarly for $\epsilon_{AB}$. It follows immediately from the definition of $SL(2,\mathbb{C})$ that the Levi--Civita symbol is invariant under actions of this group. If we use the Levi--Civita symbols to raise and lower indices it is important due to their antisymmetry to stick to a certain convention, more precisely: whether we raise with the first or second index. See e.g. \cite{Bailin} or \cite{Wald} for competing conventions.

The generators $G^{IJ}$ in the corresponding Lie algebra $\mathfrak{sl}(2,\mathbb{C})$ is defined by (matrix indices suppressed)
\begin{eqnarray*}
[G^{IJ},G^{KL}]=\ii\left(\eta^{JK}G^{IL}-\eta^{IK}G^{JL}-\eta^{JL}G^{IK}+\eta^{IL}G^{JK}\right)
\end{eqnarray*}
and coincides with the Lorentz $\mathfrak{so}(1,3)$ algebra. In fact, $SL(2,\mathbb{C})$ is the double cover of $SO^+(1,3)$ and is therefore a spin-$\half$ representation of the Lorentz group. Note also that the indices $I,J,K,L=0,1,2,3$ labelling the generators of the group are in fact tetrad indices. The Dirac $4\times4$ representation of this algebra is given by
\begin{eqnarray*}
S^{IJ}=\frac{\ii}{4}[\gamma^I,\gamma^J]
\end{eqnarray*}
where the $\gamma^I$ are the usual $4\times4$ Dirac $\gamma$-matrices. This representation is reducible, which can easily be seen if we make use of the Weyl representation of the Dirac matrices
\begin{eqnarray*}
\gamma^I=\begin{pmatrix}0&\sigma^I_{\ AA'}\\ \bar{\sigma}^{IA'A}&0\end{pmatrix}
\end{eqnarray*}
in which the generators become
\begin{eqnarray*}
S^{IJ} = \frac \ii4 \left[\gamma^I,\gamma^{J}\right]=\begin{pmatrix}(L^{IJ})_{A}^{\ \ B}&0\\ 0&(R^{IJ})^{A'}_{\ \ B'}\end{pmatrix}
\end{eqnarray*}
where
\begin{eqnarray*}
&(L^{IJ})_{A}^{\ \ B} =  \frac \ii4\left(\sigma^I_{\ AA'} \bar{\sigma}^{JA'B} - \sigma^J_{\ AA'}\bar{\sigma}^{IA'B} \right)\\
&(R^{IJ})^{A'}_{\ \ B'} = \frac \ii4\left(\bar{\sigma}^{IA'A}\sigma^J_{\ AB'} - \bar{\sigma}^{JA'A} \sigma^I_{\ AB'}\right).
\end{eqnarray*}
In this way the Dirac $4\times4$ representation decomposes into a left- and right-handed representation. Since primed and unprimed indices are different kinds of indices the ordering does not matter. However, if we want the spinors $\sigma^I_{\ AA'}$ and $\bar{\sigma}^{JA'A}$ to be the usual Pauli matrices it is necessary to have the primed/unprimed index as a row/column for $\bar{\sigma}^{JA'A}$ and {\it vice versa} for $\sigma^I_{\ AA'}$ \cite{Bailin}. Furthermore, $\sigma^I_{\ AA'}$ and $\bar{\sigma}^{JA'A}$ are in fact the same spinor object if we use $\epsilon_{AB}$ and $\bar\epsilon_{A'B'}$ to raise and lower the indices. Nevertheless, it is convenient for our purposes to keep the bar since that allows for a compact index-free notation $\sigma^I=(1,\sigma^i)$, $\bar{\sigma}^I=(1,-\sigma^i)$, where the $\sigma^i$ are (in matrix form) the usual Pauli matrices.

The Dirac spinor can now be understood as a composite object:
\begin{equation}
\Psi = \begin{pmatrix}\phi_{A} \\ \chi^{A'}\end{pmatrix}\label{eq-Diracspinor}
\end{equation}
where $\phi_A$ and $\chi^{A'}$ are left- and right-handed spinors respectively. In this paper we take the left-handed component as encoding the quantum state. However, we could equally well have worked with the right-handed component as the result turns out to be the same.

Although $\epsilon_{AB}$ and $\bar\epsilon_{A'B'}$ are the only invariant objects under the actions of the group  $SL(2,\mathbb{C})$, the hybrid object $\bar{\sigma}^{IA'A}$ plays a distinguished role because it is invariant under the combined actions of the spin-1 and spin-$\half$ Lorentz transformations, that is
\begin{eqnarray*}
\bar{\sigma}^{IA'A}\rightarrow\Lambda^I_{\ J}\Lambda^{A}_{\ B}\bar\Lambda^{A'}_{\ B'} \bar{\sigma}^{JB'B}=\bar{\sigma}^{IA'A}
\end{eqnarray*}
where $\Lambda^I_{\ J}$ is an arbitrary Lorentz transformation and $\Lambda^A_{\ B}$ and $\bar\Lambda^{A'}_{\ B'}$ are the corresponding spin-$\half$ Lorentz boosts. $\Lambda^A_{\ B}$ is the left-handed  and $\bar\Lambda^{\ \;A'}_{B'}\ (=\bar\Lambda^{-1\, A'}_{\quad\ \ B'})$ the right-handed representation of $SL(2,\mathbb{C})$.

The connection between $SO(1,3)$ vectors in spacetime and $SL(2,\mathbb{C})$ spinors is established with the linear map $\bar{\sigma}^{IA'A}$, a hybrid object with both spinor and tetrad indices \cite{Wald}. The relation between a spacetime vector $\phi^I$ and a spinor $\phi_{A}$ is given by
\begin{eqnarray*}
\phi^{I} = \bar\sigma^{IA'A}\bar{\phi}_{A'}\phi_{A}.
\end{eqnarray*}
This relation can be thought of as the spacetime extension of the relation between $SO(3)$ vectors and $SU(2)$ spinors, i.e. this object is the Bloch 4-vector. This is in fact a null vector, and we can say that $\sigma^{IA'A}$ provides a map from the spinor space to the future null light cone.

%%%%%%%%%%%%%%%%%%%%%%%%%%%%%%%%%%%%%%%%%%%%%%%%%%%%%%%%%%%%
\subsection{The geometric structure of the inner product}\label{appIP}
%%%%%%%%%%%%%%%%%%%%%%%%%%%%%%%%%%%%%%%%%%%%%%%%%%%%%%%%%

In order to turn the complex vector space $W$ into a proper Hilbert space we need to introduce a positive definite sesquilinear inner product. A sesquilinear form is linear in the second argument, antilinear in the first, and takes two complex vectors $\phi_A,\psi_A \in W$ as arguments. The antilinearity in the first argument means that $\phi_A$ must come with a complex conjugation and the linearity in the second argument means that $\psi_A$ comes without complex conjugation. In order to produce a complex number we now have to sum over the indices. So we should have something looking like $\langle\phi|\psi\rangle=\sum\bar{\phi}_{A'}\psi_A$. However, we are not allowed to carry out this summation: both the indices appear as subscripts and in addition one comes primed and the other unprimed. The only way to get around this is to introduce some geometric object with index structure $I^{A'A}\in\bar{W}^*\otimes W^*$. The inner product then becomes
\begin{eqnarray*}
\langle\phi|\psi\rangle=I^{A'A}\bar{\phi}_{A'}\psi_A.
\end{eqnarray*}
In order to guarantee positive definiteness, the inner product structure $I^{A'A}$ should have only positive eigenvalues.

Now that we have defined an inner product structure we can state how the spinor index notation is related to Dirac bra-ket notation used in standard quantum theory. We can readily make the identifications
\begin{eqnarray*}
|\phi\rangle\sim\phi_A\qquad\langle\phi|\sim I^{A'A}\bar{\phi}_{A'}.
\end{eqnarray*}

In non-relativistic quantum theory, one would choose the inner product as $I^{A'A}=\delta^{A'A}$ where $\delta^{A'A}$ is the Kronecker delta. However,  a different structure arises from the inner product of the Dirac field in the WKB limit. To see this we begin with the conserved current $j^{\mu} =\bar\Psi(x)\hat\gamma^\mu\Psi(x)$. The net `flow' of this current through an arbitrary hypersurface forms the Dirac inner product, and is a conserved quantity. Now consider the Dirac inner product  between two 4-spinor fields $\Psi_1(x)$, $\Psi_2(x)$ in the Weyl representation \eqref{eq-Diracspinor}. We have
\begin{equation}
\int \bar\Psi_1(x)\hat\gamma^\mu\Psi_2(x)\   \di\Sigma_{\mu}=\int \bar{\sigma}^\mu_{\ A'A}\bar\chi^{A}_1(x)\chi^{A'}_2(x)+\bar{\sigma}^{\mu AA'}\bar\phi^1_{B'}(x)\phi^2_{A} (x) \ \di\Sigma_{\mu}
\label{covinproddemo}
\end{equation}
where the integration is over an arbitrary spacelike hypersurface $\Sigma$. If $n^{\mu}$ is the unit vector field normal to the hypersurface and $\di \Sigma$ is the induced volume element, we write $\di \Sigma^{\mu}= n^{\mu} \di \Sigma$. \eqref{covinproddemo} is further simplified by making use of the equations of motion $m\chi^{A'}=\ii\bar{\sigma}^{\mu A'A}D_\mu\phi_A$, where the covariant derivative reduces to $D_\mu\phi_A\approx k_\mu\phi_A$ in the WKB approximation. In this approximation we obtain
\begin{equation}
\int \bar\Psi_1(x)\hat\gamma^\mu\Psi_2(x)\   \di\Sigma_{\mu}\approx \int u^{1}_{\alpha} u^{2}_{\beta}\bar{\sigma}^{\alpha B'A} {\sigma}^\mu_{\ AA'}\bar{\sigma}^{\beta A'B}\bar\phi^{1}_{B'}(x)\phi^{2}_{B}(x)+\bar{\sigma}^{\mu AA'}\bar\phi^1_{A'}(x)\phi^2_{A} (x) \ \; \di\Sigma_{\mu}.
\end{equation}
If we further assume that $k^{1}_{\alpha}=k^{2}_{\alpha}$, i.e. the 4-momentum of the fields $\Psi_1$ and $\Psi_2$ in the WKB limit coincide, the inner product can be further simplified to
\begin{eqnarray}
\int \bar\Psi_1(x)\hat\gamma^\mu\Psi_2(x)\   \di\Sigma_{\mu} = \int 2 I_{u}^{A'A}\bar\varphi^{1}_{A'}(x)\varphi^{2}_{A} (x) u^{\mu}\ \di\Sigma_{\mu}
\label{covIPfrom DIP}
\end{eqnarray}
where we have made use of the identity \cite[Eqn (2.52) p16]{DHM2010}
\begin{equation}
\bar{\sigma}^{\alpha B'A}\sigma^\mu_{\ AA'}\bar{\sigma}^{\beta A'B} = g^{\alpha\mu}\bar{\sigma}^{\beta B'B}-g^{\alpha\beta}\bar{\sigma}^{\mu B'B}+g^{\beta\mu}\bar{\sigma}^{\alpha B'B}+\ii \epsilon^{\alpha\mu\beta\gamma}\bar{\sigma}^{\ B'B}_{\gamma}\label{posspinIP}.
\end{equation}
We therefore see that the inner product for the Weyl 2-spinor $I_{u}^{A'A}$, which we obtained in the WKB approximation, naturally emerges from the inner product of the Dirac field as the object contracting the spinor indices $A'$, $A$ at each point $x$ of the fields. \footnote{Note that the factor of 2 arises from differences in defining normalization: the Dirac spinor is normalized by $\Psi(x)^\dag\Psi(x)\equiv1$ with $\Psi=(\phi,\chi)$, but the Weyl 2-spinor is normalized by $\phi(x)^\dag\phi(x)\equiv1$.}

%%%%%%%%%%%%%%%%%%%%%%%%%%%%%%%%%%%%%%%%%%%%%%%%%%%%%%%%%%%%%%%%%%%%%%%%
\section{Jerk and non-geodesic motion}\label{jerk}
%%%%%%%%%%%%%%%%%%%%%%%%%%%%%%%%%%%%%%%%%%%%%%%%%%%%%%%%%%%%%%%%%%%%%%%

We have seen that the transport of qubits as massive fermions is governed by the spin-$\half$ Fermi--Walker transport equation (\S\ref{secFermion}). One might then expect that the transport of qubits as polarization of photons along non-geodesic null-trajectories should similarly be governed by a kind of Fermi--Walker transport. Transport of polarization vectors for these {\it non-geodesic} null trajectories was developed by Castagnino \cite{Castagnino,Jantzen,SamuelNityananda}. However, these two proposals are mathematically distinct, and it is not clear to us which one is the correct one. Furthermore, both of these proposals involve the `jerk'  along the path, i.e. the time derivative of the acceleration, making the transport equation for non-geodesic paths looks rather unpleasant.

It is easy to show that any transport of a polarization vector along a null path must involve three or more derivatives of the trajectory $x^\mu(\lambda)$, i.e. involve one or more derivative of the acceleration $a^\mu(\lambda)$. From linearity and the requirement that the transport reduces to the parallel transport for geodesics we deduce that the transport must have the form $\frac{D^\text{NFW}\phi^I}{D\lambda}=\frac{D\phi^I}{D\lambda}+T^I_{\ J}\phi^J=0$. We now show that no such choice of $T^I_{\ J}$ containing only the 4-velocity and the acceleration exists that preserves the orthogonality $\phi^Iu_I=0$ between the 4-velocity and the polarization vector. We have that
\begin{eqnarray*}
0=\frac{\di}{\di\lambda}(\phi^Iu_I)=\frac{D}{D\lambda}(\phi^Iu_I)=\frac{D\phi^I}{D\lambda} u_I+\phi^Ia_I=-T^I_{\ J}u_I\phi^J+\phi^Ia_I.
\end{eqnarray*}
However, if we now assume that the transport contains at most the second derivative of $x^\mu(\lambda)$ (i.e. the velocity $u^I$ and the acceleration $a^I$) we deduce that $T^I_{\ J}=\alpha u^Iu_J+\beta u^Ia_J+\gamma a^Iu_J+\delta a^Ia_J$. But since $u^Ia_I\equiv0$ we see that $T^I_{\ J}u_I\phi^J\equiv0$ and we have thus deduced that $\phi^Ia_I=0$ for all trajectories and all polarization vectors which is false. Therefore, we have a contradiction and we have to conclude that $T^I_{\ J}$ contains one or more derivatives of the acceleration $a^I$.

It is, however, not clear that it is appropriate to study transport of polarization vectors along  non-geodesic null trajectories. Physically, non-geodesic paths of photons can only be achieved in the presence of a medium, in which case the photon trajectories will be timelike. In our approach a physically motivated way to obtain non-geodesic trajectories would be to introduce a medium in Maxwell's equations through which the photon propagates. Nevertheless, even without explicitly including a medium, it is easy to include optical elements such as mirrors, prisms, and other unitary transformations as long as their effect on polarization can be considered separately to the effect of transport through curved spacetime.

%%%%%%%%%%%%%%%%%%%%%%%%%%%%%%%%%%%%%%%%%%%%%%%%%%%%%%%%%%%%%%%%
%%%%%%%%%%%%%%%%%%%%%%%%%%%%%%%%%%%%%%%%%%%%%%%%%%%%%%%%%%%%%%%%%
\bibliography{RQI}
%\end{indented}
\end{document}